\documentclass[pdflatex,sn-mathphys-num]{sn-jnl}%

\usepackage{graphicx}%
\usepackage{multirow}%
\usepackage{amsmath,amssymb,amsfonts}%
\usepackage{amsthm}%
\usepackage{mathrsfs}%
\usepackage[title]{appendix}%
\usepackage{xcolor}%
\usepackage{textcomp}%
\usepackage{manyfoot}%
\usepackage{booktabs}%
\usepackage{algorithm}%
\usepackage{algorithmicx}%
\usepackage{algpseudocode}%
\usepackage{listings}%
\let\orcidlogo\relax
\usepackage{orcidlink}

\usepackage{subcaption}
\usepackage{epstopdf}
\theoremstyle{thmstyleone}

\theoremstyle{thmstyletwo}%

\theoremstyle{thmstylethree}%

\begin{document}

\title[Article Title]{Assessment of the Gradient Jump Penalisation in Large-Eddy Simulations of Turbulence}

\author*[1]{\fnm{Shiyu} \sur{Du} \orcidlink{0000-0002-9256-2304}}  \email{shiyud@kth.se}

\author[2]{\fnm{Manuel} \sur{M\"unsch} \orcidlink{0009-0009-4570-0864}}

\author[3]{\fnm{Niclas} \sur{Jansson} \orcidlink{0000-0002-5020-1631}}

\author[1,2]{\fnm{Philipp} \sur{Schlatter} \orcidlink{0000-0001-9627-5903}}

\affil*[1]{\orgdiv{Department of Engineering Mechanics}, \orgname{KTH Royal Institute of Technology},  \orgaddress{\postcode{SE-100 44} \city{Stockholm},  \country{Sweden}}}

\affil[2]{\orgdiv{Institute of Fluid Mechanics (LSTM)}, \orgname{Friedrich--Alexander Universit\"at Erlangen--N\"urnberg (FAU)}, \orgaddress{\postcode{DE-91058} \city{Erlangen}, \country{Germany}}}

\affil[3]{\orgdiv{PDC Center for High Performance Computing},  \orgname{KTH Royal Institute of Technology}, \orgaddress{\postcode {SE-100 44} \city{Stockholm},  \country{Sweden}}}

\abstract{This research investigates the efficacy of the gradient jump penalisation (GJP) in large eddy simulations (LES) when coupled with active subgrid-scale (SGS) models. GJP is a stabilisation method tailored for the continuous Galerkin spectral element method, aiming at mitigating non-physical oscillations induced by discontinuous velocity gradients across element interfaces. We demonstrate that GJP effectively smoothens fields from LES without a salient impact on flow dynamics for the Taylor--Green vortex (TGV) at $Re=1600$, periodic hill flows at bulk Reynolds numbers $Re_b=10595$ and $37000$, as well as turbulent channel flow at $Re_{\tau} \approx 550$. 
In the TGV case, the application of GJP results in decreased fluctuations at only high wavenumbers compared to simulations without GJP. The periodic hill flow simulations indicate the  applicability of GJP in wall-resolved LES (WRLES) involving curved geometries, though it tends to dissipate some of the finer details in the solution. 
Finally, in the analysis of the canonical turbulent channel flow cases, GJP leads to a higher resolved turbulent kinetic energy than simulations without GJP and direct numerical simulations. GJP's mechanism is identified as providing enhanced dissipation at high wavenumbers but accompanied with insufficient dissipation at low wavenumbers, leading to a pronounced spectral cut-off. Non-physical oscillations on element interfaces are reflected as spikes in the power spectral density. By evaluating the sharpness of the strongest spike, GJP is shown to smoothen the spectra, however without completely removing the gradient jumps at low computational resolution.}


\maketitle

\section{Introduction}
The spectral element method (SEM) has been extensively used to perform high-fidelity simulations of turbulent flows, mainly motivated by two factors. One is its excellence in parallel computing since most operations are performed within a single reference element; while the other one is the higher accuracy under the same degree of freedom compared with some low-order numerical schemes due to its high-order nature. Although SEM based on the continuous Galerkin (CG) method has already proven to be an invaluable tool for well-resolved direct numerical simulation (DNS), using it on coarser grids typical for large eddy simulation (LES) is known to cause more-or-less well-documented difficulties, for example in~\cite{Fischer2001} and~\cite{Massaro2023}. In particular, the solution usually develops high-frequency oscillations (``wiggles'') inside the elements close to the element interfaces and significant discontinuities in its gradient across neighbouring elements, see~\cite{Chatterjee2017, Huang_2022, Mukha2024}. These issues are particularly relevant for LES since the resolution is generally ``poor" for a smooth solution and there lacks a mechanism providing sufficient dissipation by CG-SEM itself. In addition, most subgrid-scale (SGS) models employ the velocity-gradient tensor in some way. Thus, for LES using SGS models, gradient jumps directly propagate into model predictions. In an effort to alleviate such issues, during the past decades, several approaches have been proposed to mitigate the wiggles, such as directly utilizing the high-pass-filtered field~\cite{Schlatter2004, Schlatter2005_thesis}, applying hierarchical viscosity in the modal space, which is known as spectral vanishing viscosity (SVV)~\cite{Kirby2006}, and gradient jump penalisation (GJP)~\cite{Moura2022GJP}.

Some studies have already investigated GJP in implicit LES (iLES) in detail, with the aim of improving comparably well-resolved simulations. For example, the dispersion-diffusion behaviour of GJP on eigen-solutions was analysed in detail and instantaneous snapshots were compared with SVV in~\cite{Moura2022GJP}; and the integrated results of GJP were analysed and compared to solution jump penalisation in~\cite{Kou2023} to a Taylor--Green vortex flow. However, to our knowledge, the literature did not contain any study on their interactions with SGS models and their performance at a typical LES resolution which requires active SGS models instead of only relying on numerical schemes for sufficient dissipation. In addition, those studies had flow simulations at relatively low polynomial orders, \textit{e.g.} $P=3$ in~\cite{Kou2023} and $P=3,5$ in~\cite{Moura2022GJP}, where the solution was constructed on a smoother polynomial basis with potentially larger numerical dissipation.

The purpose of this work is to study the effect of the GJP on a coarse resolution, \emph{i.e.,} typical wall-resolved LES (WRLES) resolution, when coupled with active SGS models under a higher polynomial order. We implement GJP in the SEM solver Neko~\cite{Neko} for hexahedral elements and investigate the influence of the penalty term on the Taylor--Green vortex (TGV) flow at $Re=1600$ with isotropic meshes, turbulent flows over a periodic hill at $Re_b = 10595$ and $37000$ with curved meshes both visually and statistically, and further the canonical case of a turbulent channel flow at $Re_\tau = 550$ in terms of visualisations, statistics and spectra.

The remainder of this article is organised as follows. Section \ref{sec:methodology} introduces the governing equations, the SGS models adopted in this study and the details of GJP. This is followed by the LES of a TGV flow at $Re=1600$ performed with and without GJP in Section \ref{sec:tgv} to demonstrate the effect of wiggle-dissipation in ideal isotropic meshes. Section \ref{sec:periodic hill} applies GJP in the WRLES of periodic hill flow at $Re_b = 10595$ and $37000$. This is followed by the impact of GJP in a WRLES of fully developed turbulent channel flow at $Re_\tau\approx550$ in Section \ref{sec:turb_channel_550}. Finally, some conclusive remarks and future works are provided in Section \ref{sec:conclusion}.

\section{Methodology}\label{sec:methodology}
    \subsection{Governing equations} \label{sec:gov_eqn}
    The governing equations for incompressible flows in Cartesian coordinates are
    \begin{eqnarray}
        \frac{\partial u_i}{\partial t} + u_j \frac{\partial u_i}{\partial x_j} 
                                      &=& - \frac{1}{\rho}\frac{\partial p}{\partial x_i} + \frac{\partial \tau_{ij}}{\partial x_j}, \;(i=1,2,3) \label{eq:ns_tau}\\
        \frac{\partial u_i}{\partial x_i} &=& 0, \label{eq:div_free}
    \end{eqnarray}
     where $u_i$ and $p$ are the velocity and pressure fields resolved by the computational grid points. $\tau_{ij}$ is the stress tensor combining the viscous and SGS stresses from unresolved eddies. The SGS stresses are modelled by the eddy viscosity ansatz, which assumes a direct relation between the SGS stresses and the strain, similar as for the melocular stresses \cite{Boussinesq_1877}, leading to
    \begin{equation} \label{eq:tau_ij}
        \tau_{ij} = 2(\nu + \nu_{\rm sgs}) S_{ij} \ ,       
    \end{equation}
    with the strain-rate tensor
    \begin{equation}
    S_{ij} := \frac{1}{2}(\frac{\partial u_i}{\partial x_j} + \frac{\partial u_j}{\partial x_i}) \ .
    \end{equation}
    $\nu$ is the kinematic viscosity which is constant in this work; while $\nu_{\rm sgs}$ is the eddy viscosity which is a field evolving with time and space.
    The modelling of $\nu_{\rm sgs}$ is further discussed in Section \ref{sec:sgs_model}.
    
    \subsection{Subgrid-scale models} \label{sec:sgs_model}
        In the CG-SEM community, there is a tendency to avoid the use of explicit SGS models and leave the absent dissipation to numerical schemes such as high-pass filtering in~\cite{Schlatter2004, Schlatter2005_thesis, Gillyns2022} and SVV in~\cite{Manzanero2020} and~\cite{Wang2021}. However, in some simulations motivated by atmospheric flow studies such as~\cite{Mukha2024}, using explicit SGS models in CG-SEM is still unavoidable since one usually approaches such high Reynolds numbers that those aforementioned numerical techniques are not capable to compensate for the lack of the dissipation from flow physics, which can also be demonstrated by the wiggly fields in~\cite{Huang_2022} and also the spiky statistics of the case using high-order filtering as a forcing term in~\cite{Mukha2024}. Thus, we consider in this work three algebraic eddy-viscosity  models: Smagorinsky model, Sigma model and Vreman model.

        \subsubsection{Smagorinsky model} \label{sec:Smagorinsky}
        The Smagorinsky model was proposed in 1963~\cite{Smagorinsky1963} and has the following formulation,
        \begin{eqnarray} \label{eq:Smagorinsky_model}
            \nu_{\rm sgs} &=& (C_S \Delta)^2 \mathcal{|S|}\ , \\
            \mathcal{|S|} &=& (2S_{ij}S_{ij})^{1/2}\ .
        \end{eqnarray}
        Here we set the constant $C_S$ to $1.7$ as originally proposed by Lilly \cite{Lilly1966} for homogeneous isotropic turbulence. $\Delta$ is a length scale related to the filter width which will be discussed later in Section \ref{sec:filter_width}.
        
        \subsubsection{Sigma model} \label{sec:Sigma}
        The Sigma model proposed by Nicoud and Ducros ~\cite{Nicoud2011} is based on the singular values $\sigma_1, \sigma_2, \sigma_3$ ($\sigma_1 \ge \sigma_2 \ge \sigma_3 \ge 0$) of the velocity gradient tensor $g_{ij} := \partial u_i / \partial x_j$.
        \begin{eqnarray} \label{eq:Sigma_model}
            \nu_{\rm sgs} &=& (C_{\sigma} \Delta)^2 \mathcal{D}_{\sigma}, \\
            \mathcal{D}_{\sigma} &=& \frac{\sigma_3 (\sigma_1 - \sigma_2) (\sigma_2 - \sigma_3)}{\sigma_1 ^ 2}.
        \end{eqnarray}
        Here, the single model constant $C_{\sigma}$ is set to $1.35$ as proposed in~\cite{Nicoud2011} as a reference value.

        \subsubsection{Vreman model} \label{sec:Vreman}
        Vreman model was proposed by~\cite{Vreman2004}. The formulation of Vreman model is directly based on the velocity gradient tensor $g_{ij}$.
        \begin{eqnarray} \label{eq:Vreman_model}
            \nu_{\rm sgs} &=& C_{\rm Vreman} \Delta^2 \sqrt{\frac{B_\beta}{g_{ij}g_{ij}}}, \\
            B_\beta &=& \beta_{11}\beta_{22} - \beta_{12}^2 + \beta_{11}\beta_{33} - \beta_{13}^2 + \beta_{22}\beta_{33} - \beta_{23}^2, \\
            \beta_{ij} &=& g_{mi} g_{mj}.
        \end{eqnarray}
        Also in this case, since the focus of this contribution is on reproducibility, the model constant $C_{\rm Vreman}$ is set to the reference value of $0.07$ as recommended initially for the case of homogeneous isotropic turbulence \cite{Vreman2004}.

        \subsubsection{Filter width evaluation}\label{sec:filter_width}
        All three SGS models aforementioned require a length scale, which is typically related to the filter width, which is based on the grid space if no explicit filter is employed. In the SEM community, the right choice of this $\Delta$ has been the cause of a number of discussions, since the field is rendered by piecewise smooth polynomials, the implicitness of the grid filter and the non-equidistant distribution of collocation points. However, one could still related $\Delta$ with either the local evaluation related to quadrature points such as the estimation in~\cite{Chatterjee2017} or an element-wise evaluation related to the element size and the polynomial order, such as the estimation based on average element volume shared by each node in~\cite{Ntoukas2025} or based on the maximum node spacing in~\cite{Mukha2024}. Since in this work we mainly focused on the effect of GJP coupled with active SGS models instead of a perfect LES, we would adopt those choices for $\Delta$ without further discussion of which one gives a better dissipation property. To be more specific, for the TGV case and the turbulent channel flow case, we adopt the average element volume approach as turbulence on all under-resolved directions are homogeneous or developing towards homogeneity and the elements are consequently uniformly distributed; while for the periodic hill case, in contrast, we adopt the local pointwise approach since the under-resolved streamwise direction is heterogeneous due to the domain geometry.
    
    \subsection{Gradient jump penalisation} \label{sec:gjp}

    From the formulation shown in Section \ref{sec:Sigma} and \ref{sec:Vreman}, all three SGS models rely on an accurate estimation of the velocity gradient tensor, which would be jeopardised from the natural discontinuous gradient of CG and the non-physical oscillations appearing at element interfaces observed in several previous pioneer studies such as~\cite{Chatterjee2017} and~\cite{Mukha2024}. In this work, we apply GJP introduced by~\cite{Moura2022GJP} which was originally proposed to regularise non-physical oscillations in iLES. In~\cite{Moura2022GJP}, the formulation of GJP is introduced in detail, which we summarise here. For a 1D advection problem of scalar $T$ 
    \begin{equation}\label{eq:scalar}
        \frac{\partial T}{\partial t} + u \frac{\partial T}{\partial x} 
                                      = 0,
    \end{equation}
    where $u$ is the advection velocity, the weak formulation with the spectral element discretisation within an element $\Omega_e$ from an entire domain $\Omega$ is
    \begin{equation}\label{eq:weak_scalar}
        \int_{\Omega_e} \phi_i \frac{\partial T}{\partial t} dx
       +\int_{\Omega_e} \phi_i u \frac{\partial T}{\partial x} dx
       = 0,
    \end{equation}
    where $\phi_i$ is the polynomial of $i$-th order. The readers are recommended to refer to Appendix \ref{appendix:A} for a detailed demonstration of the formulation. The GJP is implemented by adding one more right-hand side (RHS) term to Equation~(\ref{eq:weak_scalar}) resulting in
    \begin{equation}\label{eq:GJP_RHS_1d}
        \int_{\Omega_e} \phi_i \frac{\partial T}{\partial t} dx
       +\int_{\Omega_e} \phi_i u \frac{\partial T}{\partial x} dx = -\tau \langle |u \cdot n| h_{\Omega_e}^2 G(T) \frac{\partial \phi_i}{\partial n}\rangle,
    \end{equation}
    where $\tau$ is a parameter which is optimised in~\cite{Moura2022GJP} to be $\tau = 0.8(P+1)^{-4}$ with $P$ to be the polynomial order, $n$ is the face normal which in this 1D case to be either $-1$ or $1$, $h_{\Omega_e}$ is the element size, and
    \begin{equation}
        \langle\rangle := \int_{\partial \Omega_e \backslash \partial \Omega} ds.
    \end{equation}
    There is one more term which needs more elaboration, \emph{i.e.,} the gradient jump term over an element interface $F$
    \begin{equation}\label{eq:gradient_jump}
        G(T)|_F := \nabla T |_{\partial \Omega^1_e \cap F} \cdot n_1
                  +\nabla T |_{\partial \Omega^2_e \cap F} \cdot n_2,
    \end{equation}
    where the sub/superscript $1$ and $2$ refers the quantity belonging to the elements on either side of $F$.

    For the 3D advection problem, the RHS of Equation~(\ref{eq:GJP_RHS_1d}) would become 
    \begin{equation}\label{eq:GJP_RHS}
        -\tau \langle |u_m \cdot n_m| h_{\Omega_e}^2 G(T) \phi_{t_1, i} \phi_{t_2, j} \frac{\partial \phi_{n, k}}{\partial n}\rangle,
    \end{equation}
    where $|u_m \cdot n_m|$ stands for the absolute volumetric flux, and $\phi_{t_1, i}, \phi_{t_2, j}$ are the $i$-th and $j$-th order polynomials tangential to the facet included in the integral and $\phi_{n, k}$ is the $k$-th order polynomial normal to the facet. The readers are directed to Appendix \ref{appendix:B} for a detailed demonstration. The final term $\frac{\partial \phi_{n, k}}{\partial n}$ could be evaluated separately by a derivative in the computational domain and its mapping into the physical domain illustrated as follows. For the convenience in the later illustrations, let us name the physical space of an element to be $(x,y,z)$ and the mapped computational space which is symbolised as $(\xi_1,\xi_2,\xi_3)$ ranged between $[-1,1]$. Then $\frac{\partial \phi_{n, k}}{\partial n}$ on the facet whose normal is mapped into $\xi_1$ at $\xi_1=1$ could be decomposed by the chain rule into
    \begin{equation}\label{eq:poly_1st_deriv}
       \left. \frac{\partial \phi_{n, k}}{\partial n}\right|_{n\rightarrow \xi_1^+} = \left.\frac{\partial \phi_k(\xi)}{\partial \xi}\right|_{\xi=1} \left(\frac{\partial \xi_1}{\partial x} + \frac{\partial \xi_1}{\partial y} + \frac{\partial \xi_1}{\partial z}\right).
    \end{equation}
    
    The first part could be evaluated by the polynomial property and independent of mesh; while the second part is from the mesh. Similar evaluation could be made for the facet whose normal is mapped to $\xi_1^-$, $\xi_2^\pm$ or $\xi_3^\pm$ by replacing $\xi_1^+$ and $\xi_1$ in Equation (\ref{eq:poly_1st_deriv}).
    
\begin{figure}[ht]
        \centering
        \begin{subfigure}{0.49\linewidth}
            \centering
            \includegraphics[width=\linewidth]{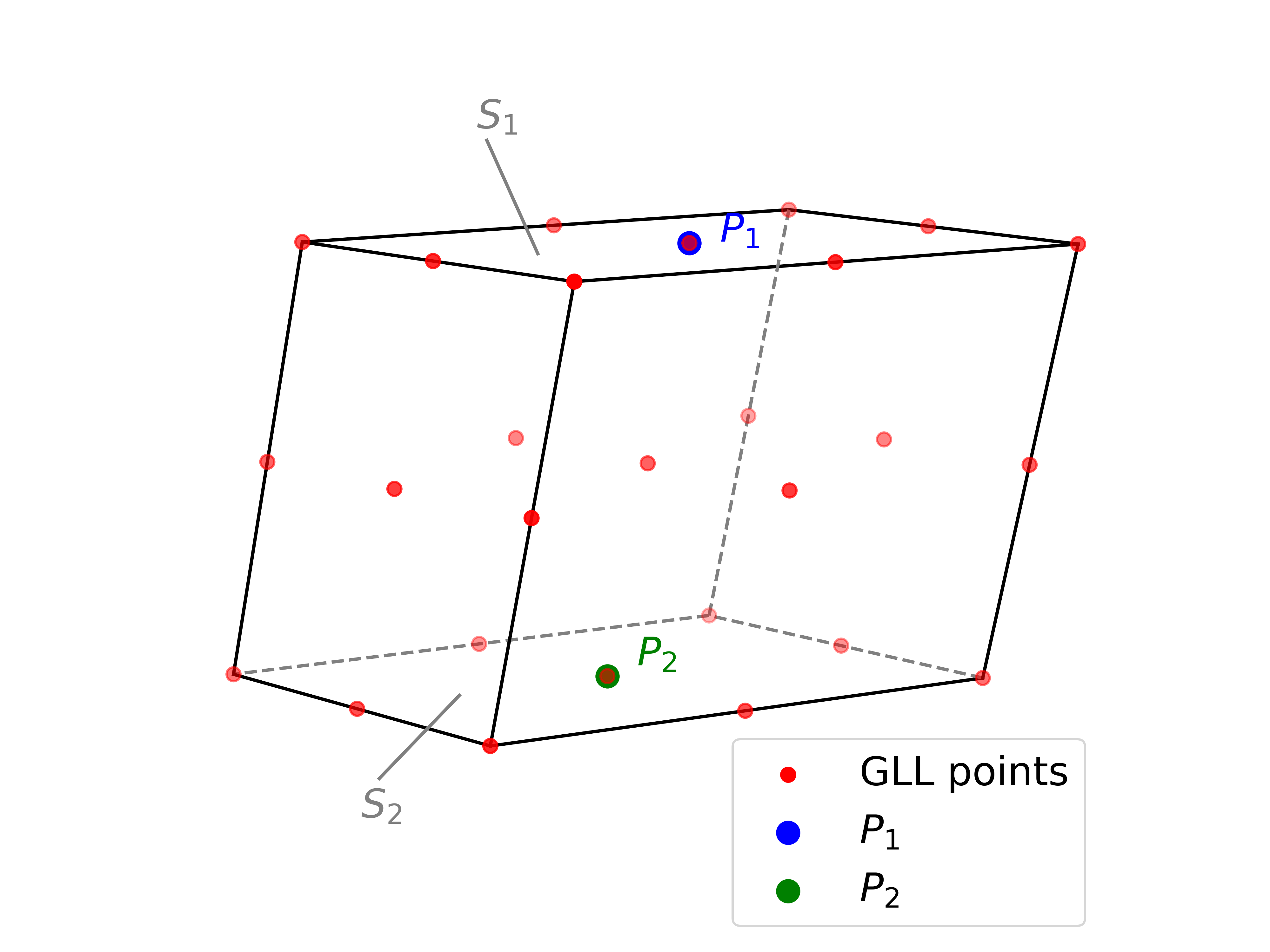}
            \caption{$S_1$, $S_2$, $P_1$ and $P_2$}
            \label{fig:hex_ele_gll}
        \end{subfigure}
        \begin{subfigure}{0.49\linewidth}
            \centering
            \includegraphics[width=\linewidth]{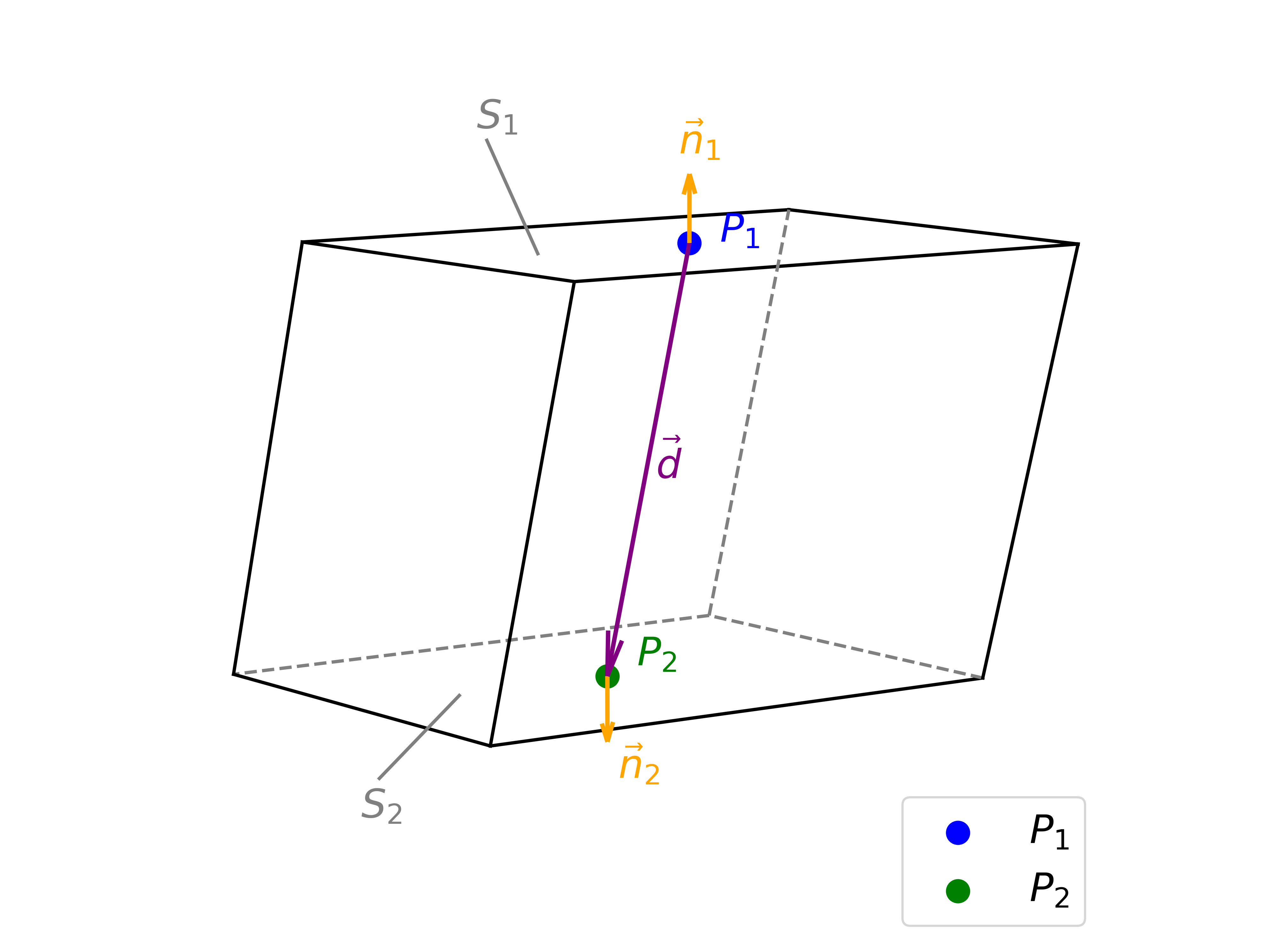}
            \caption{$\vec{d}$, $\vec{n}_1$ and $\vec{n}_2$}
            \label{fig:facet_distance_P1P2}
        \end{subfigure}
        \caption{An illustration of the determination of $h_{\Omega_e}\vert_{P_1,P_2}$ for a hexahedral element with $P=2$}
        \label{fig:gjp_facet_distance}
    \end{figure}

    In this work, elements are hexahedral instead of tetrahedral as was adopted in~\cite{Moura2022GJP}, which makes it not straightforward to evaluate the mesh size $h_{\Omega_e}$ as the $4$ vertices on the same facet might not lie in in the same plane. Here we give the GLL-point-associated definition as follows:
    \begin{enumerate}
        \item we associate $h_{\Omega_e}$ with each certain GLL-point pair that are the counterparts of each other on two opposite facets (here we denote the facets as $S_1$ and $S_2$), as illustrated in Figure \ref{fig:hex_ele_gll};
        \item $h_{\Omega_e}\vert_{P_1,P_2}$, which is associated with the GLL-point $P_1$ on facet $S_1$ and its counterpart point $P_2$ on facet $S_2$ is defined by the following steps and illustrated in Figure \ref{fig:facet_distance_P1P2},:
            \begin{enumerate}
                \item define $\vec{d}$ as the vector pointing from $P_1$ to $P_2$;
                \item name the normal vector of $S_1$ on $P_1$ as $\vec{n}_1$ and the normal vector of $S_2$ on $P_2$ as $\vec{n}_2$.
                \item $h_{\Omega_e}\vert_{P_1,P_2}$ associated with $P_1$ and $P_2$ is defined as the average value of the projected length of $\vec{d}$ on $\vec{n}_1$ and $\vec{n}_2$ respectively,
                \begin{equation}
                    h_{\Omega_e}\vert_{P_1,P_2} := \frac{1}{2}\left(\frac{\vert\vec{d} \cdot \vec{n}_2\vert}{\vert\vec{n}_2\vert} + \frac{\vert\vec{d} \cdot \vec{n}_1\vert}{\vert\vec{n}_1\vert}\right).
                \end{equation}
            \end{enumerate}
    \end{enumerate}

    Applying the weak form procedure as illustrated in Equation (\ref{eq:weak_scalar}) to Equation (\ref{eq:ns_tau}) for a 3D senario, one could write out the GJP term for momentum equations as an additional forcing term on the right hand side of the weak form equation similar to Equation (\ref{eq:GJP_RHS}) but changing scalar $T$ to velocity components $u_i$
    \begin{equation}\label{eq:GJP_u}
        -\tau \langle |u_m \cdot n_m| h_{\Omega_e}^2 G(u_i) \phi_{t_1} \phi_{t_2} \frac{\partial \phi_{n}}{\partial n}\rangle.
    \end{equation}
    Noticing that there is a parameter $\tau$ controlling the magnitude of the penalty term, from the parameter study in~\cite{Moura2022GJP}, a scaling law is proposed
    \begin{eqnarray}\label{eq:GJP_tau}
        \tau &=& a(P+1)^{-4} \; {\rm for} \; P>1, \\
        a &=& 0.8.
    \end{eqnarray}
    In the following part of this work, we will show the Taylor--Green vortex case and the periodic-hill flow case without GJP ($a=0.0$) and with GJP with $a=0.8$, and vary the value of $a$ from $0.0$ up to $1.2$ for the channel flow case, covering no penalisation case, under-penalised cases and over-penalised cases to look the effects of the penalty term on LES using explicit SGS models in detail.

\section{Taylor--Green vortex}\label{sec:tgv}
\begin{figure}[ht]
        \centering
        \begin{subfigure}{0.4\linewidth}
            \centering
            \includegraphics[width=\linewidth]{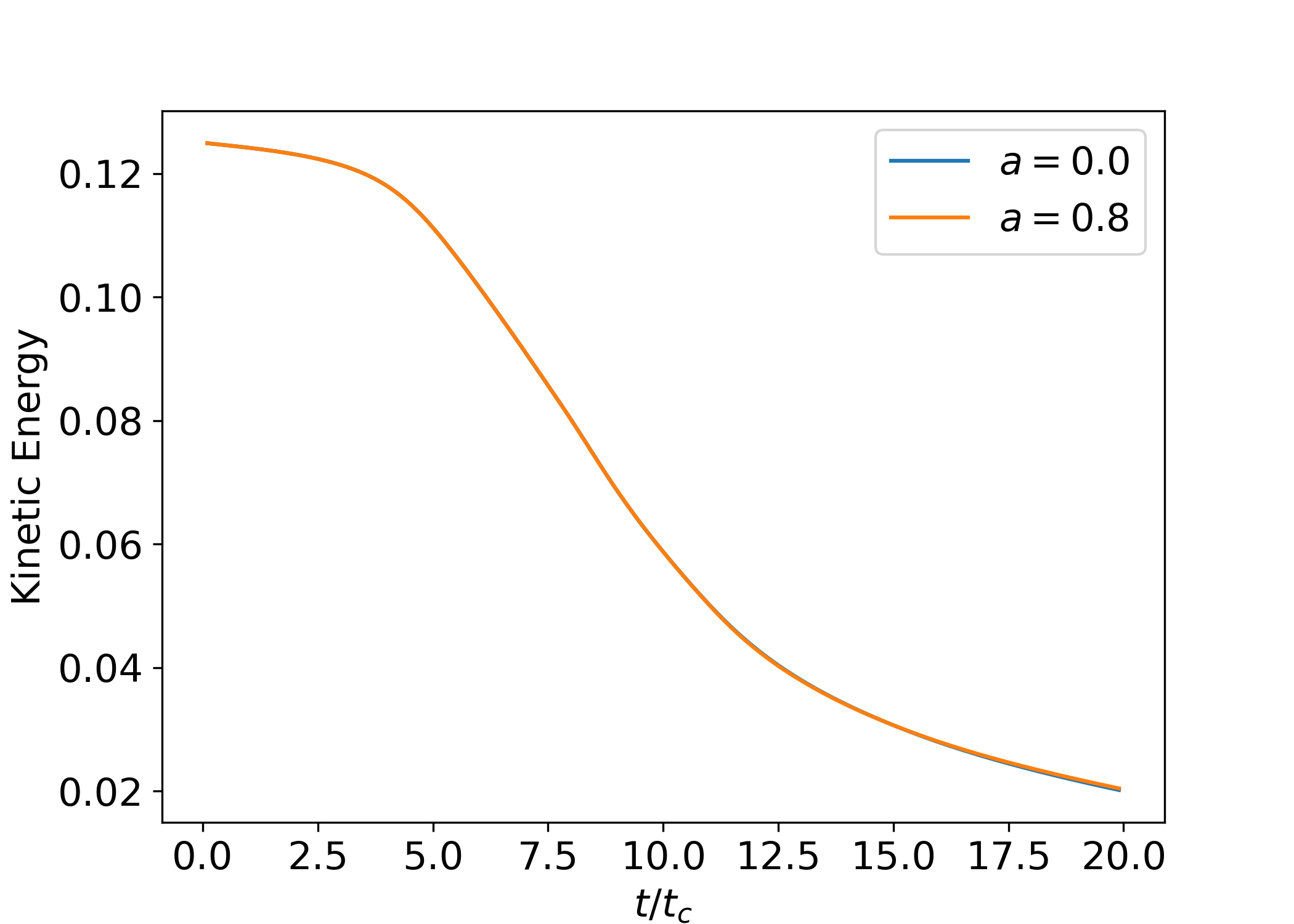}
            \caption{Volume-averaged kinetic energy}
            \label{fig:tgv_KE}
        \end{subfigure}
        \begin{subfigure}{0.4\linewidth}
            \centering
            \includegraphics[width=\linewidth]{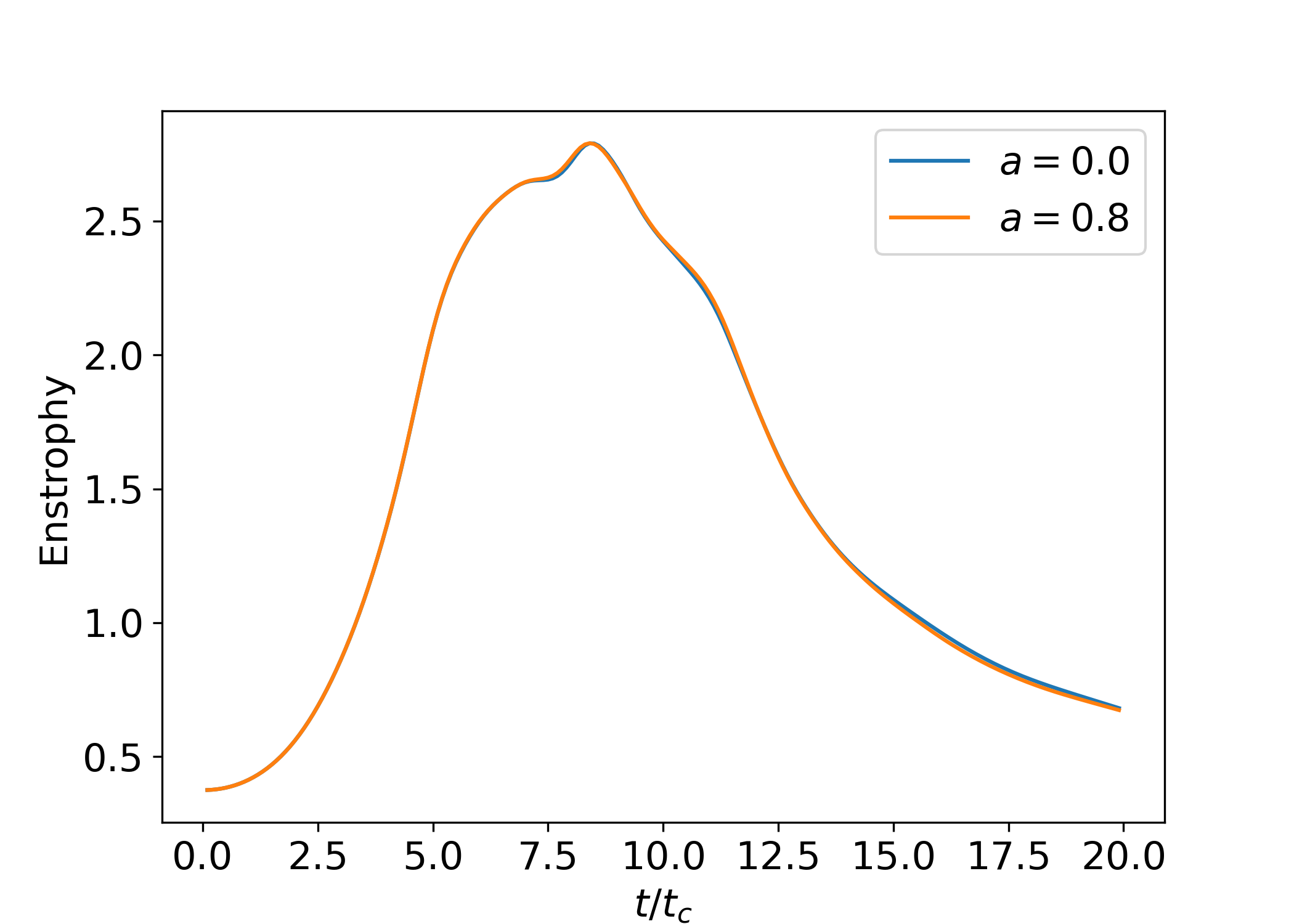}
            \caption{Volume-averaged enstrophy}
            \label{fig:tgv_Enst}
        \end{subfigure}
        \begin{subfigure}{0.4\linewidth}
            \centering
            \includegraphics[width=\linewidth]{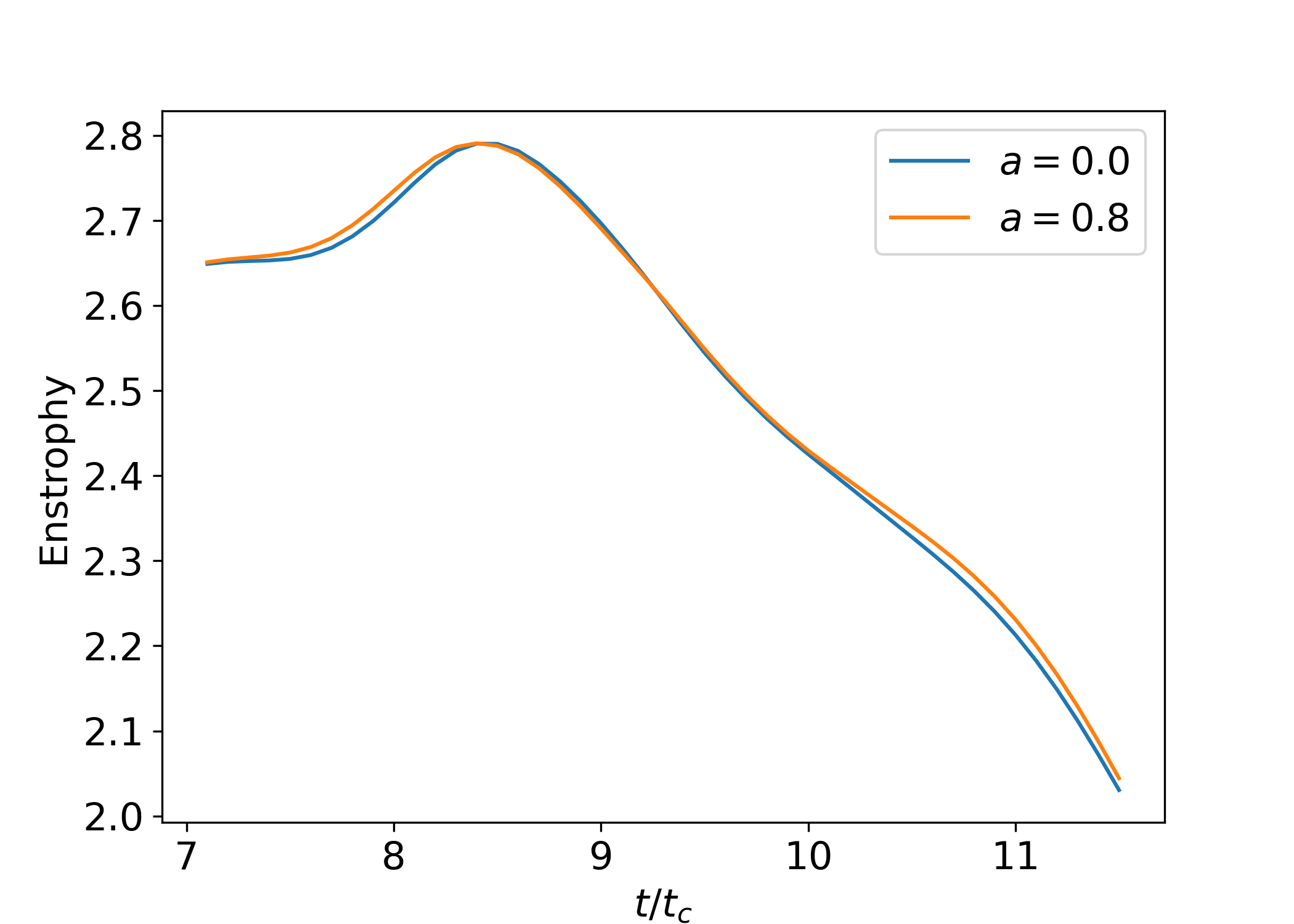}
            \caption{Zoomed panel b $t/t_c=7-11.5$}
            \label{fig:tgv_Enst_t7}
        \end{subfigure}\begin{subfigure}{0.4\linewidth}
            \centering
            \includegraphics[width=\linewidth]{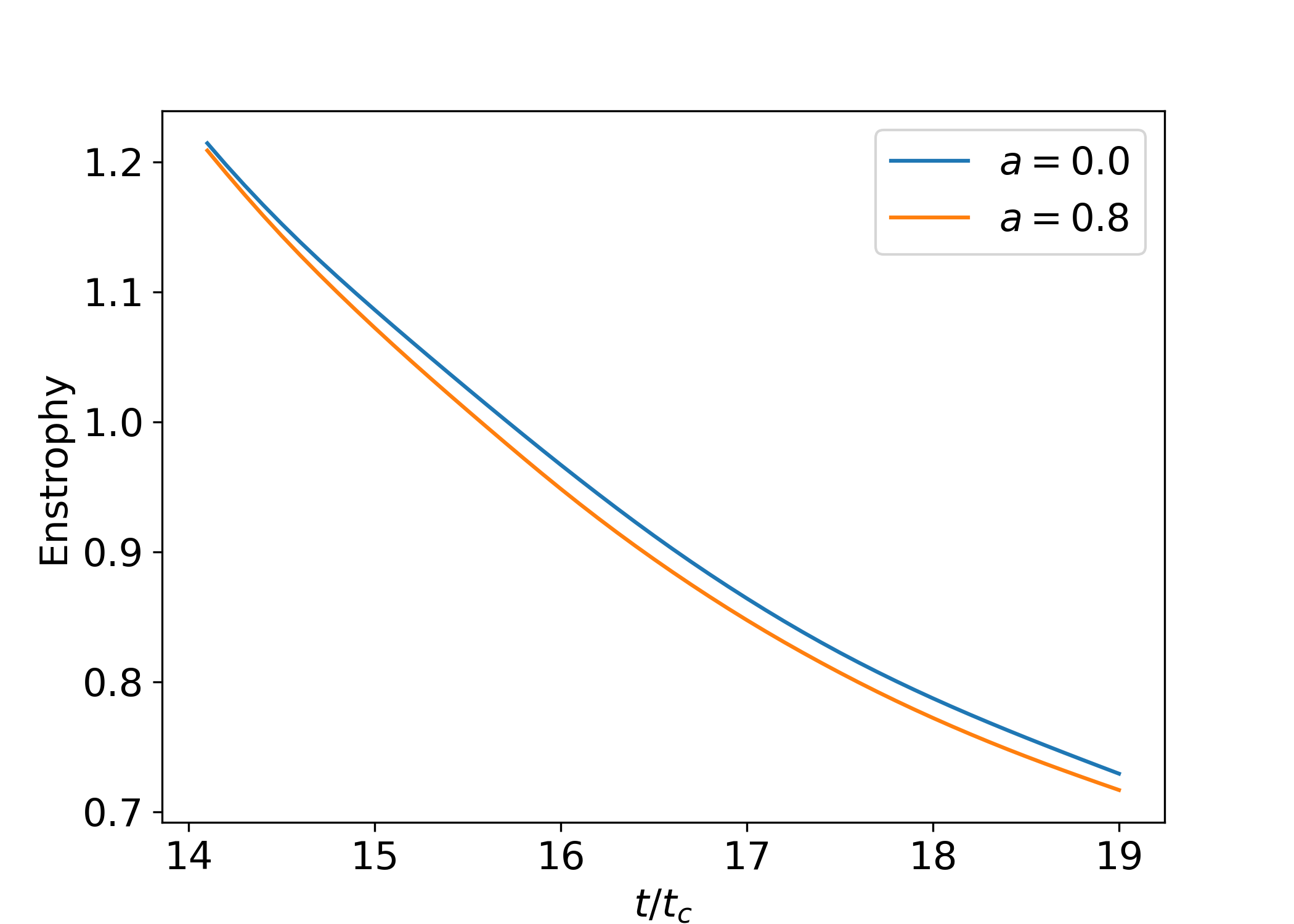}
            \caption{Zoomed panel b $t/t_c=14-19$}
            \label{fig:tgv_Enst_t14}
        \end{subfigure}
        \caption{Volume-averaged kinetic energy and enstrophy, obtained using the classical Smagorisnky model with $C_S=0.17$ and $\Delta=(V_{\Omega_e}/(P+1)^3)^{1/3}$.}
        \label{fig:tgv_KE_Enst}
    \end{figure}

To start with isotropic meshes, an LES of Taylor--Green vortex (TGV) at $Re = 1600$ is conducted on a uniform mesh $7 \times 7 \times 7$ elements with polynomial order $P=9$ to have the similar resolution as the coarse grid in \cite{Yilmaz2015} if modulated by average GLL-spacing. The TGV case uses a periodic domain $[-L\pi, L\pi]^3$ with the initial condition
    \begin{eqnarray}
        U(t=0) &=& V_0 \sin(\frac{x}{L})\cos(\frac{y}{L})\cos(\frac{z}{L}),\\
        V(t=0) &=& -V_0\cos(\frac{x}{L})\sin(\frac{y}{L})\cos(\frac{z}{L}),\\
        W(t=0) &=& 0.
    \end{eqnarray}
    The time of the simulation is scaled by $t_c = L/V_0$ and we use the classical Smagorinsky model with the filter width $\Delta$ defined as the averaged approach $\Delta = (V_{\Omega_e}/(P+1)^3)^{1/3}$ where $V_{\Omega_e}$ is the element volume, given the flow finally decays into isotropic homogeneous turbulence. We run two simulations without GJP (noted as "$a=0.0$") and with GJP where $a=0.8$ and plot the volume-averaged kinetic energy and enstrophy in Figure \ref{fig:tgv_KE_Enst}. As one can see, the kinetic energy goes through a decaying process while the enstrophy peaks at $t/t_c \approx 9$. This process corresponds to the well known physics in a TGV flow: starting from an unstable laminar initial condition, experiencing a transition from laminar to turbulent and proceeding towards a turbulent flow with an isotropic decay.
    
\begin{figure}[ht]
        \centering
        \begin{subfigure}{0.7\linewidth}
            \centering
            \includegraphics[width=\linewidth]{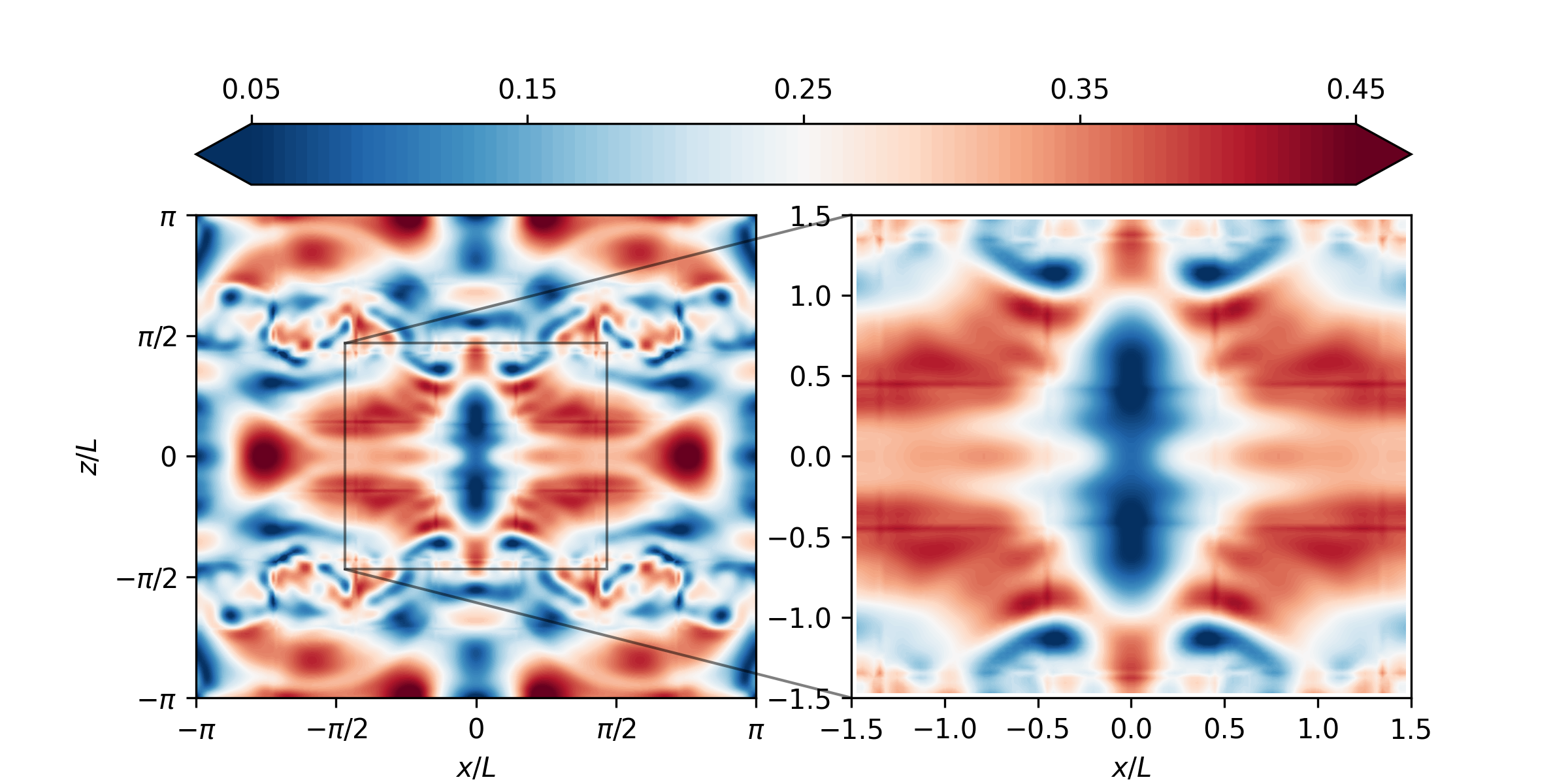}
            \caption{$a=0.0$}
            \label{fig:tgv_velmag_y0_t17}
        \end{subfigure}
        \begin{subfigure}{0.7\linewidth}
            \centering
            \includegraphics[width=\linewidth]{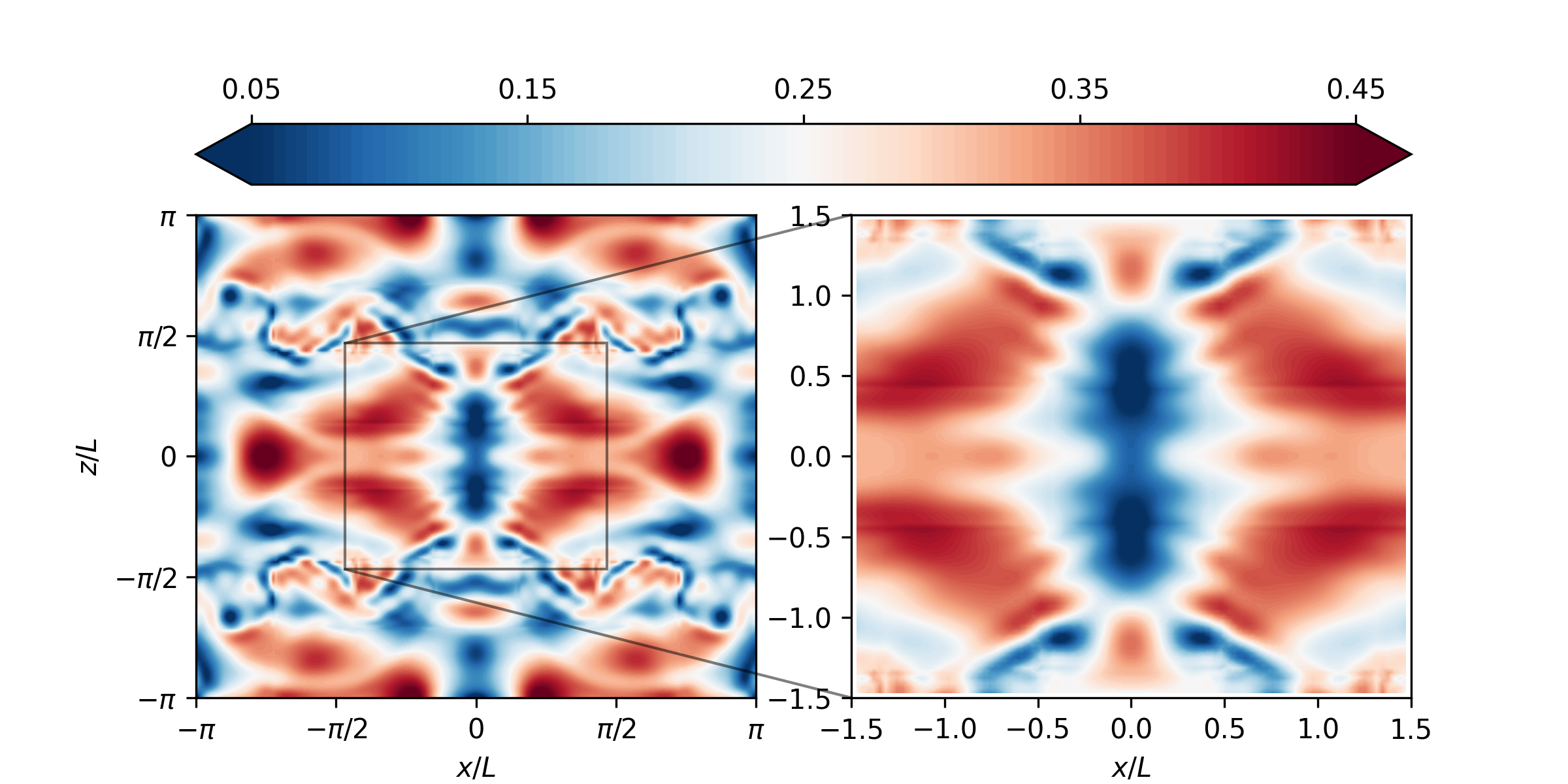}
            \caption{$a=0.8$}
            \label{fig:tgv_velmag_a08_y0_t17}
        \end{subfigure}
        \caption{Turbulent kinetic energy  $\sqrt{U^2+V^2+W^2}/V_0$ extracted in the plane at $y=-L\pi$ and $t/t_c=17$.}
        \label{fig:tgv_velmag_t17}
    \end{figure}

    From Figure \ref{fig:tgv_Enst_t7} and \ref{fig:tgv_Enst_t14}, small discrepancies of two cases are observed at $t/t_c \approx 8$, $11$ and $t/t_c>14$. To see it in detail, Figure \ref{fig:tgv_velmag_t17} shows the velocity magnitude field at $t/t_c = 17$ when the flow is much later than the transitional stage. Though one can still see the elements interfaces through the field due to the lack of $C^1$ continuity, the non-physical oscillations on element interfaces in Figure \ref{fig:tgv_velmag_y0_t17} in the center of the plane are dissipated to a large degree by applying GJP in Figure \ref{fig:tgv_velmag_a08_y0_t17}. 

\begin{figure}[ht]
        \centering
        \begin{subfigure}{0.49\linewidth}
            \centering
            \includegraphics[width=\linewidth]{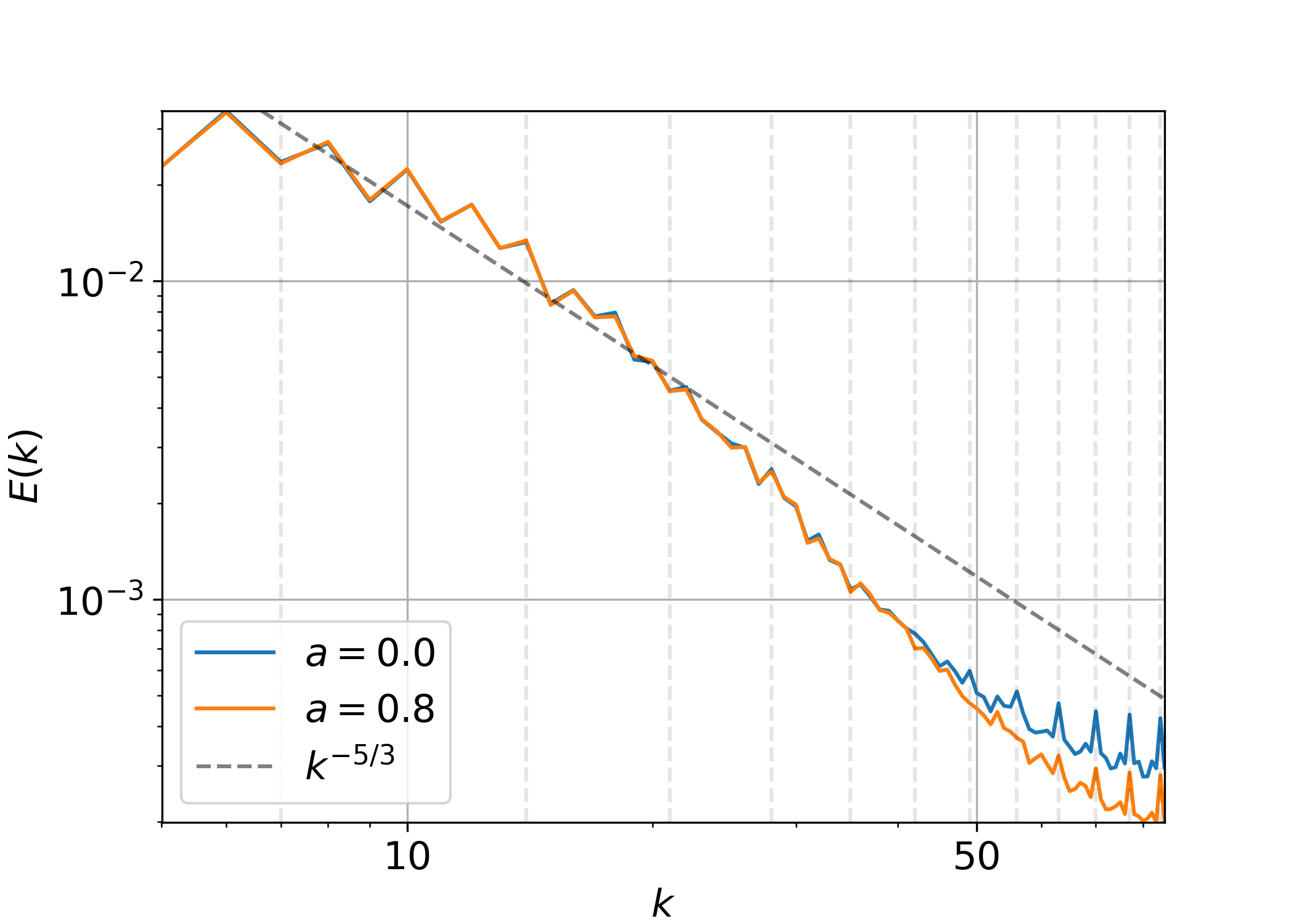}
            \caption{$t=11$}
            \label{fig:tgv_spectrum_t11}
        \end{subfigure}
        \begin{subfigure}{0.49\linewidth}
            \centering
            \includegraphics[width=\linewidth]{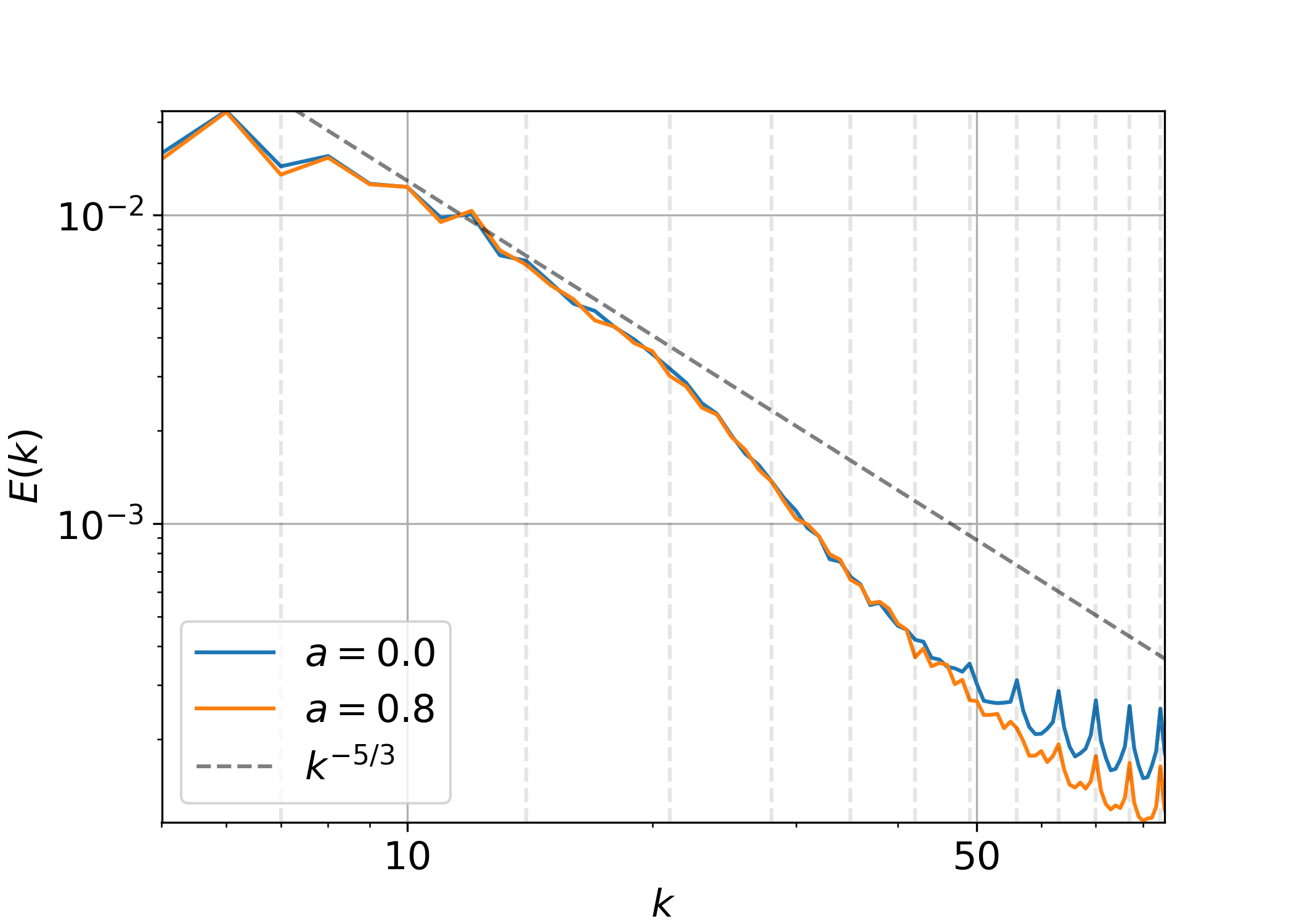}
            \caption{$t=17$}
            \label{fig:tgv_spectrum_t17}
        \end{subfigure}
        \caption{Spatial energy spectrum at two time instants, with superimposed $-5/3$ power law. }
        \label{fig:tgv_spectrum}
    \end{figure}

    In order to quantitatively assess how the flow is smoothened, we analyse the 1D spectrum of the kinetic energy by averaging over the spectrum on 3 Cartesian directions at time instant $t/t_c = 11$ and $17$ in Figure \ref{fig:tgv_spectrum}. In general we can see the case with GJP has the same energy with the one without GJP at low wavenumbers. Besides, since the spectrum is computed by performing discrete Fourier transform on $x,y$ and $z$ direction and then averaging over three directions, so wiggles are expected to be reflected in the spectrum in some way. Given the fact that the Fourier transform of a sharp periodic signal consists of spikes in the frequency domain located at the fundamental frequency and its higher harmonics, one should expect the wiggles to be reflected as clear spikes in wavenumber space. In Figure \ref{fig:tgv_spectrum}, those critical wavenumber are marked out by grey dash lines with the lowest to be $7$ corresponding to $7$ elements in each direction. Noting that the flow also shows large amplitude waves at low wavenumbers, we cannot see such spikes for low wavenumber such as $k=7$, $14$, $21$, etc. While for high wavenumbers starting from the $6$th harmonics where the flow dynamics does not contain high energy, the wiggles prevail and show spikes on harmonics of the element spacing, see the $a=0.0$ curve in Figure \ref{fig:tgv_spectrum} where GJP is not activated. By applying GJP, we can see for both time instants that  spike starts to show up from the higher $8$th harmonics which is of even lower energy. Moreover, the spikes are smoother than the $a=0.0$ case with lower peak and valley values.

    To summarize, GJP smoothen the wiggles in the TGV flow in a way such that the energy associated with high wavenumbers are generally lowered, while the energy associated with stronger low wavenumbers are preserved. Therefore, the overall energy, as shown in Fig.\ \ref{fig:tgv_KE_Enst} is not affected by GJP. 

\clearpage
\section{Periodic-hill flow}\label{sec:periodic hill}
In order to investigate the effect of GJP on a WRLES, we first perform the simulation of the flow over periodic hills, which belongs to the class of well established benchmark cases. It is characterised by curved surfaces leading to flow separation behind the first hill, a recirculation zone and re-attachment of the flow in front of the second hill. Further details and data for comparison at bulk Reynolds numbers from 5,600 to 37000 can be found within the ERCOFTAC Best Practice Guidelines \cite{Rapp2009Phill} and a accompanying publication \cite{BREUER2009433} . Here, simulations are conducted with the Sigma and Vreman model with and without GJP for Reynolds numbers of $10595$ and $37000$.
  
\subsection{Case setup}

\begin{figure}[ht]
        \centering
        \begin{subfigure}{0.49\linewidth}
            \centering
            \includegraphics[width=\linewidth]{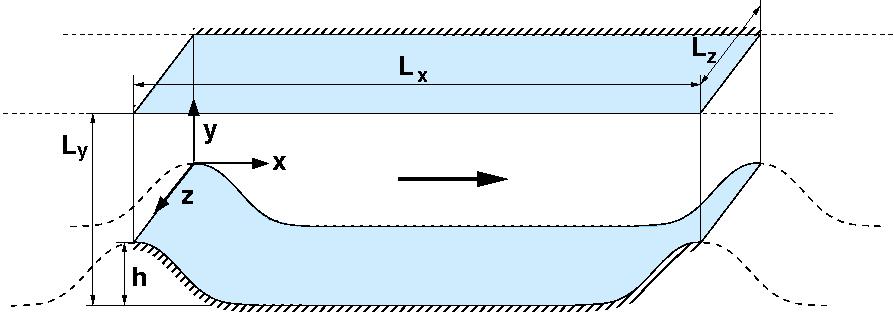}
            \caption{Geometry definitions}
            \label{fig:phill_hill3d}
        \end{subfigure}
        \begin{subfigure}{0.49\linewidth}
            \centering
             \includegraphics[width=\linewidth, clip]{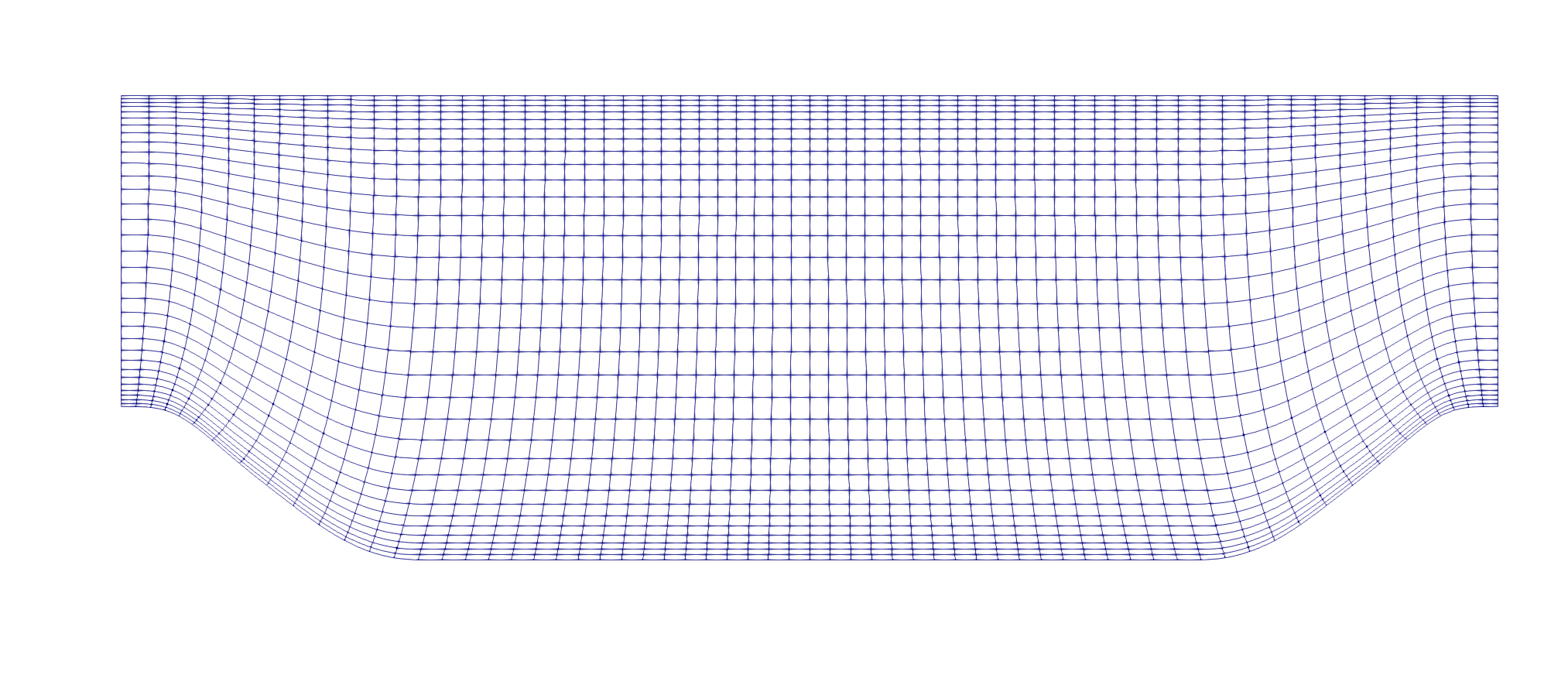}
            \vspace{-0.8cm}
            \caption{Mesh}
            \label{fig:phill_mesh}
        \end{subfigure}
        \caption{(a) Sketch of the geometry definitions taken from \cite{Rapp2009Phill} and (b) cross-section (x-y) of the mesh showing the elements for the LES case at $Re_b=37000$.}
        \label{fig:phill_case_setup}
    \end{figure}

The geometry and definitions of the periodic hill case are shown in Figure \ref{fig:phill_hill3d}. The domain length is $L_x = 9.0h$, the height is given with $L_y=3.036h$ and the span extension is equal to $L_z=4.5h$ where $h$ defines the height of the hills. On the lower and the upper wall, \emph{i.e.,} the blue surfaces in Figure \ref{fig:phill_hill3d}, no-slip boundary conditions are applied. For the remaining surfaces periodic boundary conditions are set, whereas a mass flux based forcing is applied in the streamwise $x$-direction to keep the Reynolds number $Re_b$ constant. Here, the Reynolds number is defined by the bulk velocity $U_b$ at the top of the hill, the kinematic viscosity of the fluid and the height $h$: $Re_b=\frac{U_{b} h}{\nu}$.

    \begin{figure}[ht]
        \centering
        \begin{subfigure}{0.32\linewidth}
            \centering
            \includegraphics[width=\linewidth]{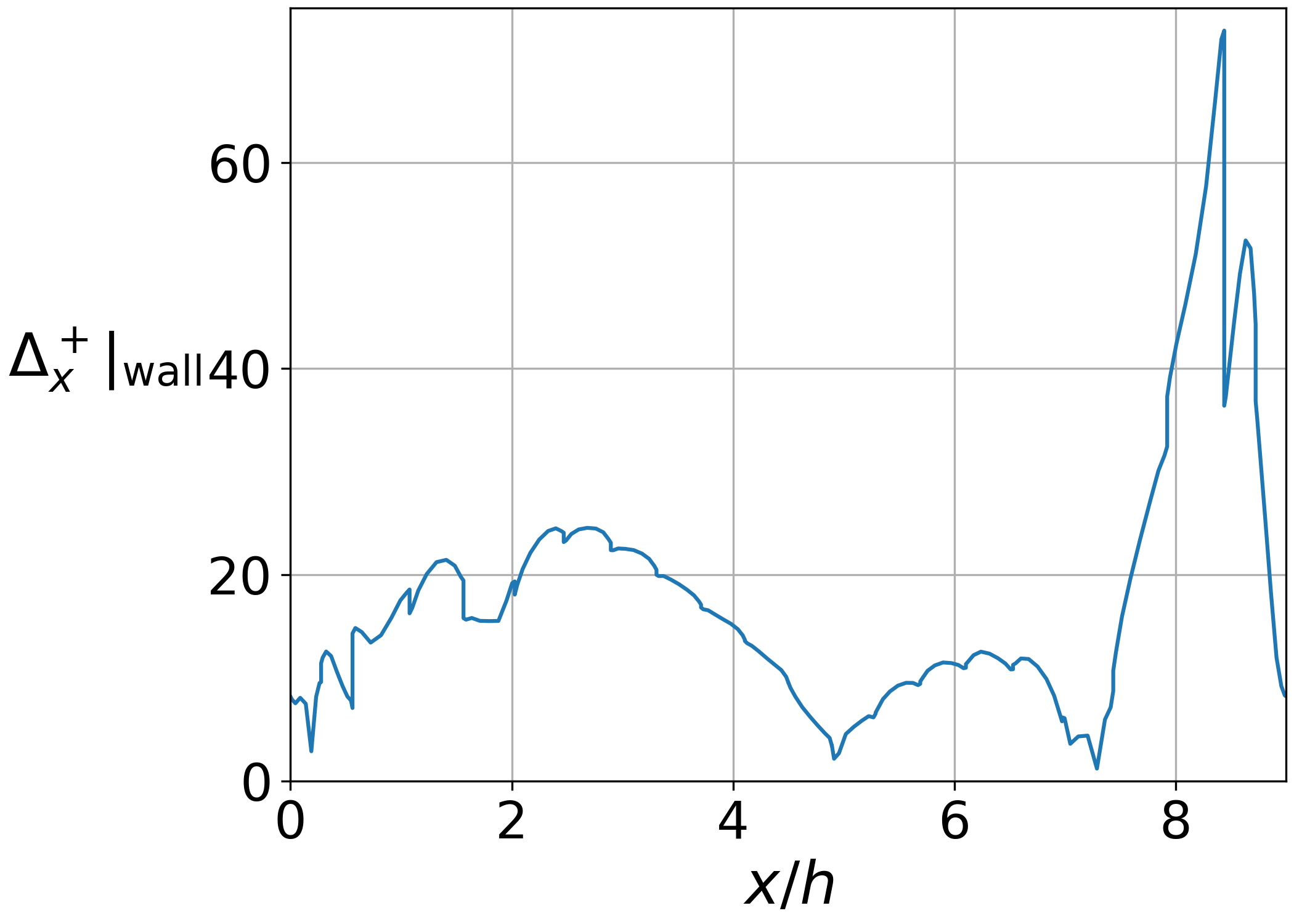}
            \caption{$\Delta x ^+\vert_{\rm wall}$, $Re_b=10595$}
            \label{fig:phill_xplus_10595_Vreman}
        \end{subfigure}
        \begin{subfigure}{0.32\linewidth}
            \centering
            \includegraphics[width=\linewidth]{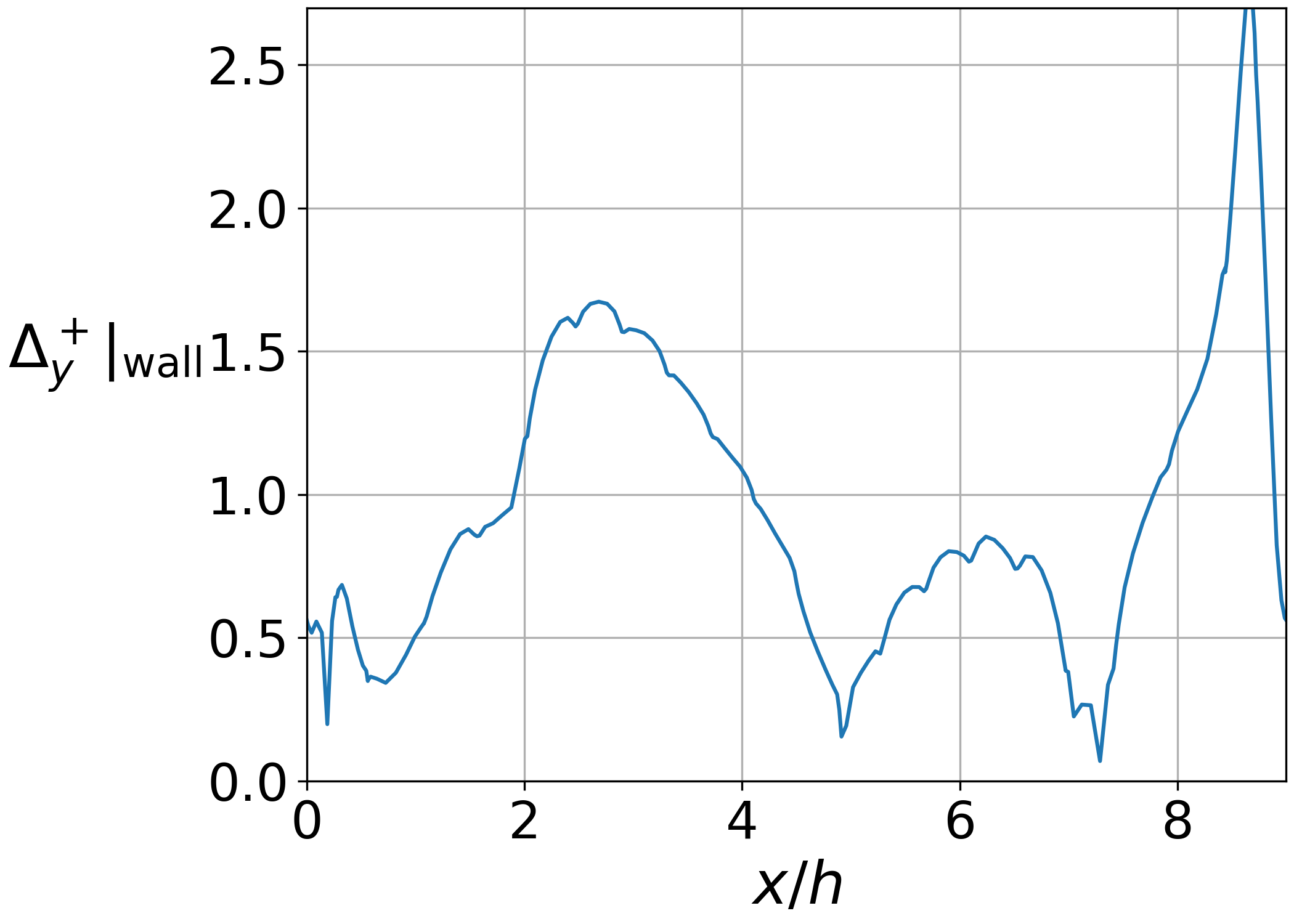}
            \caption{$\Delta y ^+\vert_{\rm wall}$, $Re_b=10595$}
            \label{fig:phill_yplus_10595_Vreman}
        \end{subfigure}
        \begin{subfigure}{0.32\linewidth}
            \centering
            \includegraphics[width=\linewidth]{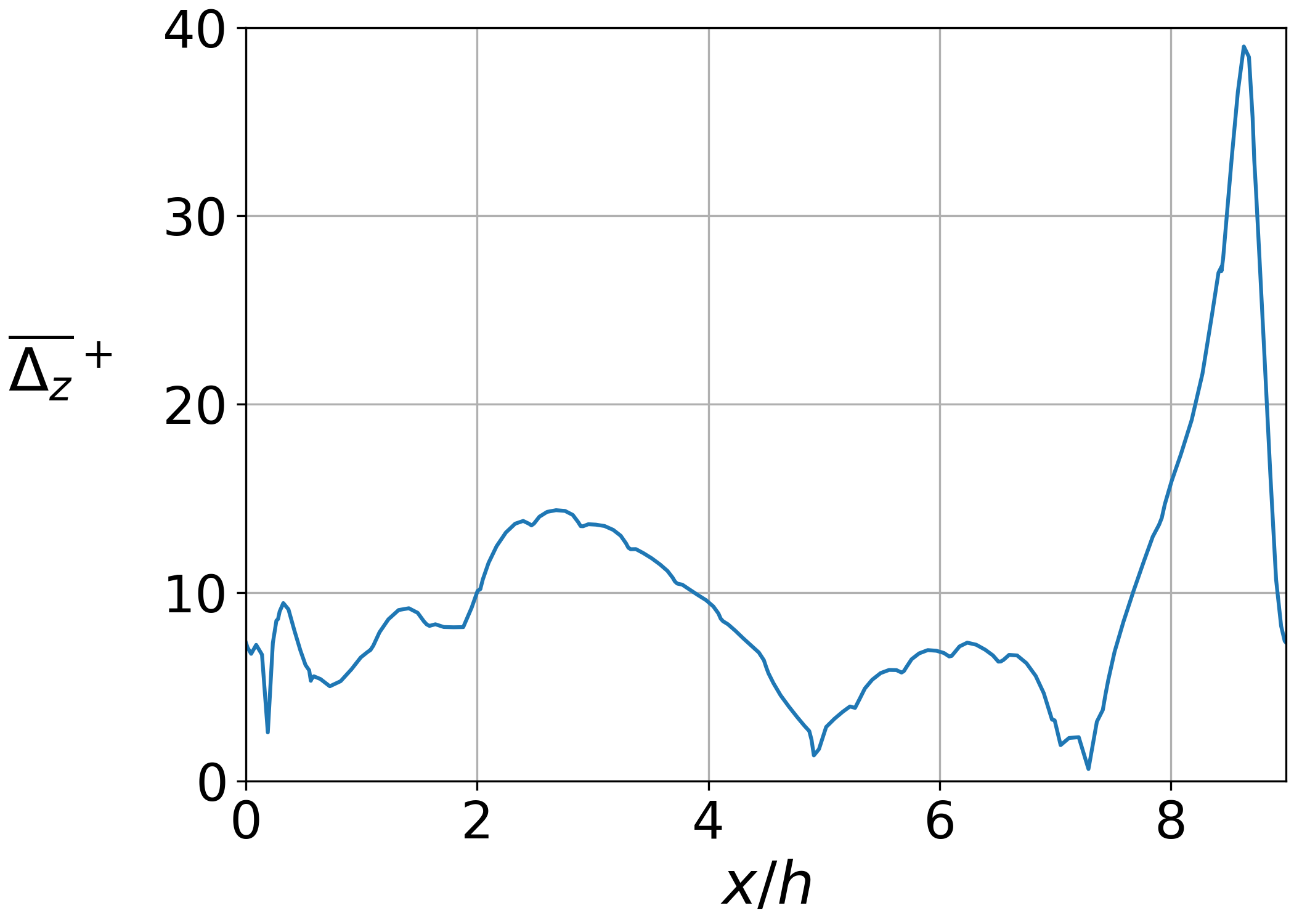}
            \caption{$\overline{\Delta z} ^+$, $Re_b=10595$}
            \label{fig:phill_zplus_10595_Vreman}
        \end{subfigure}
        \begin{subfigure}{0.32\linewidth}
            \centering
            \includegraphics[width=\linewidth]{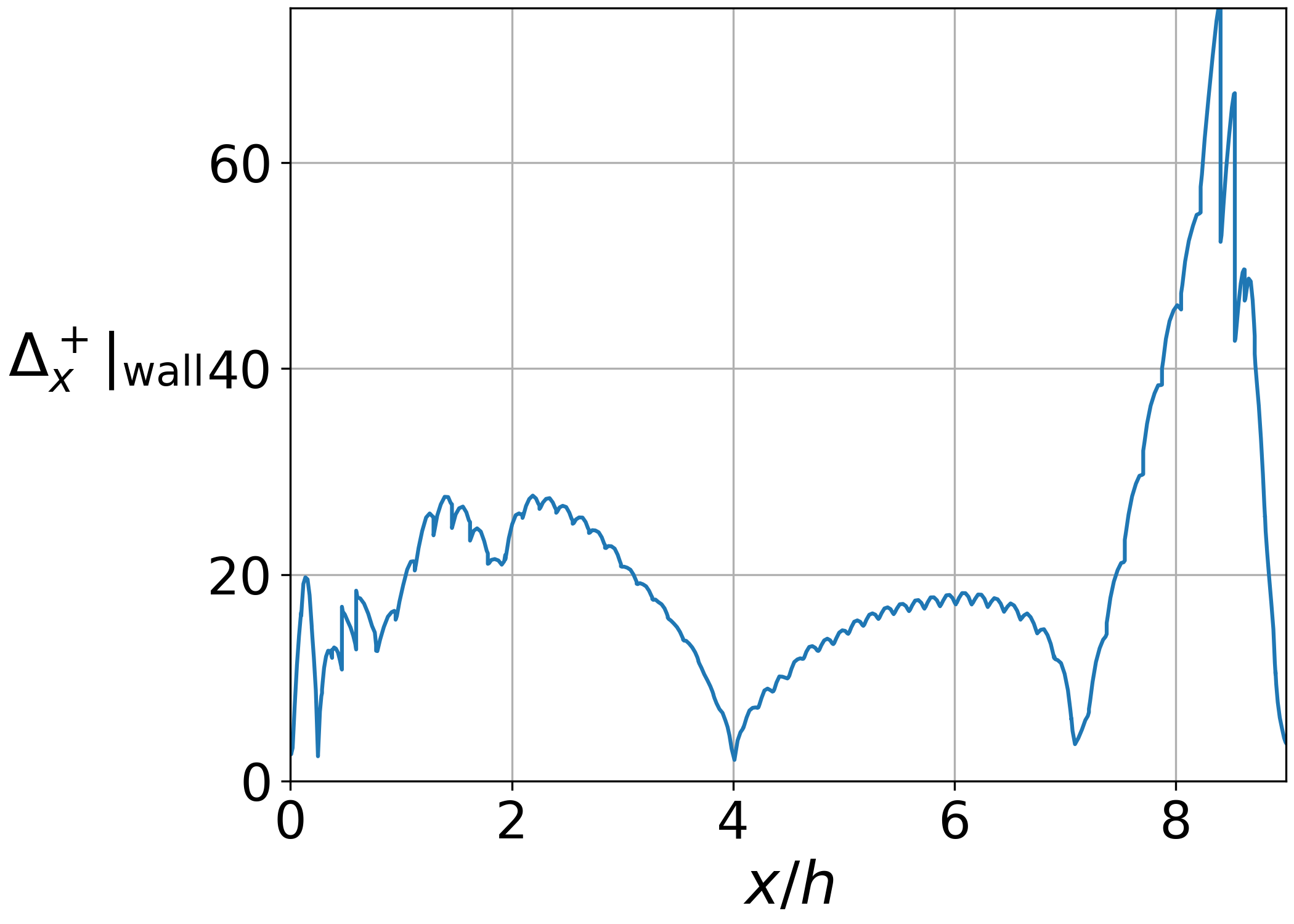}
            \caption{$\Delta x ^+\vert_{\rm wall}$, $Re_b=37000$}
            \label{fig:phill_xplus_37000_Vreman}
        \end{subfigure}
        \begin{subfigure}{0.32\linewidth}
            \centering
            \includegraphics[width=\linewidth]{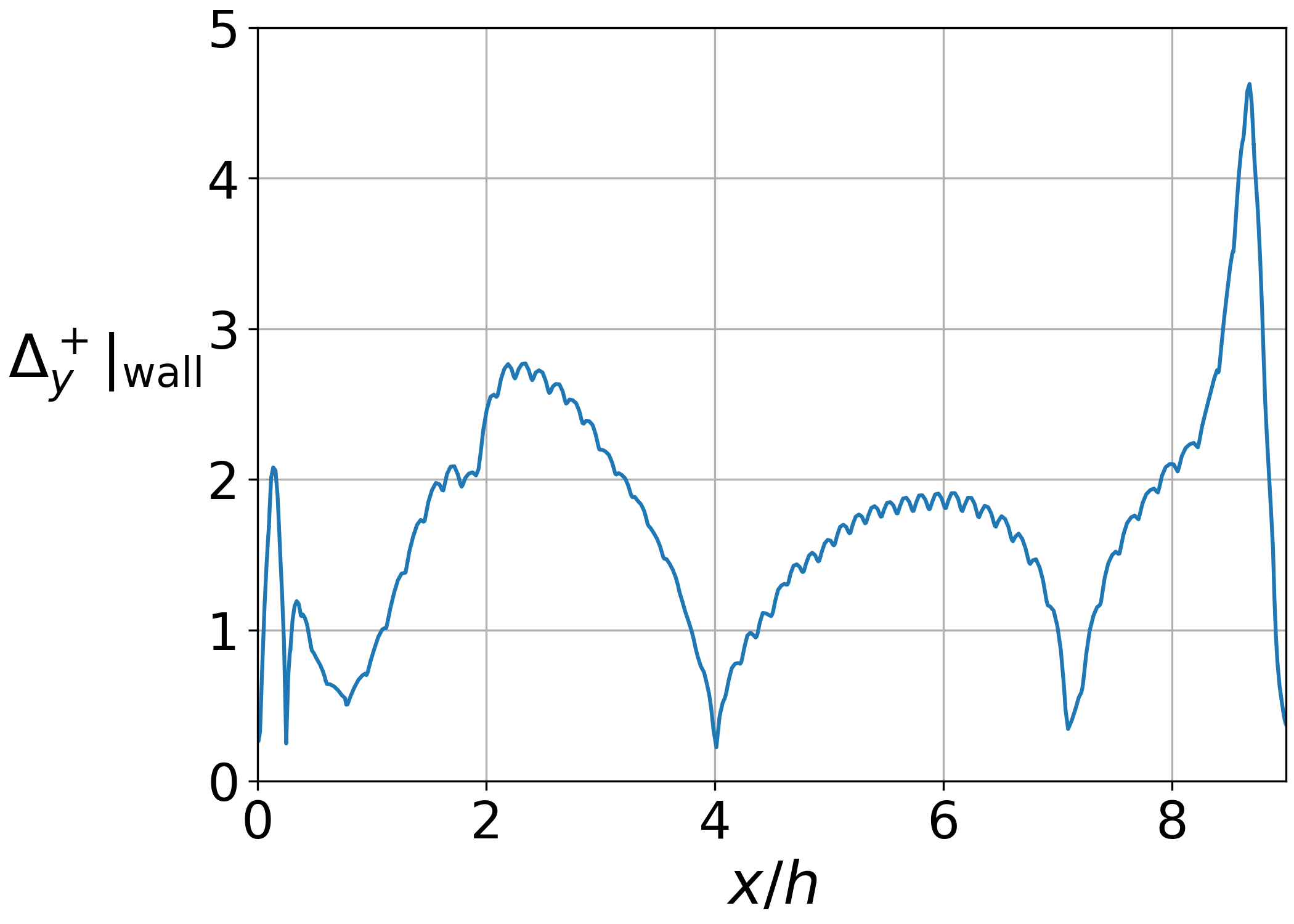}
            \caption{$\Delta y ^+\vert_{\rm wall}$, $Re_b=37000$}
            \label{fig:phill_yplus_37000_Vreman}
        \end{subfigure}
        \begin{subfigure}{0.32\linewidth}
            \centering
            \includegraphics[width=\linewidth]{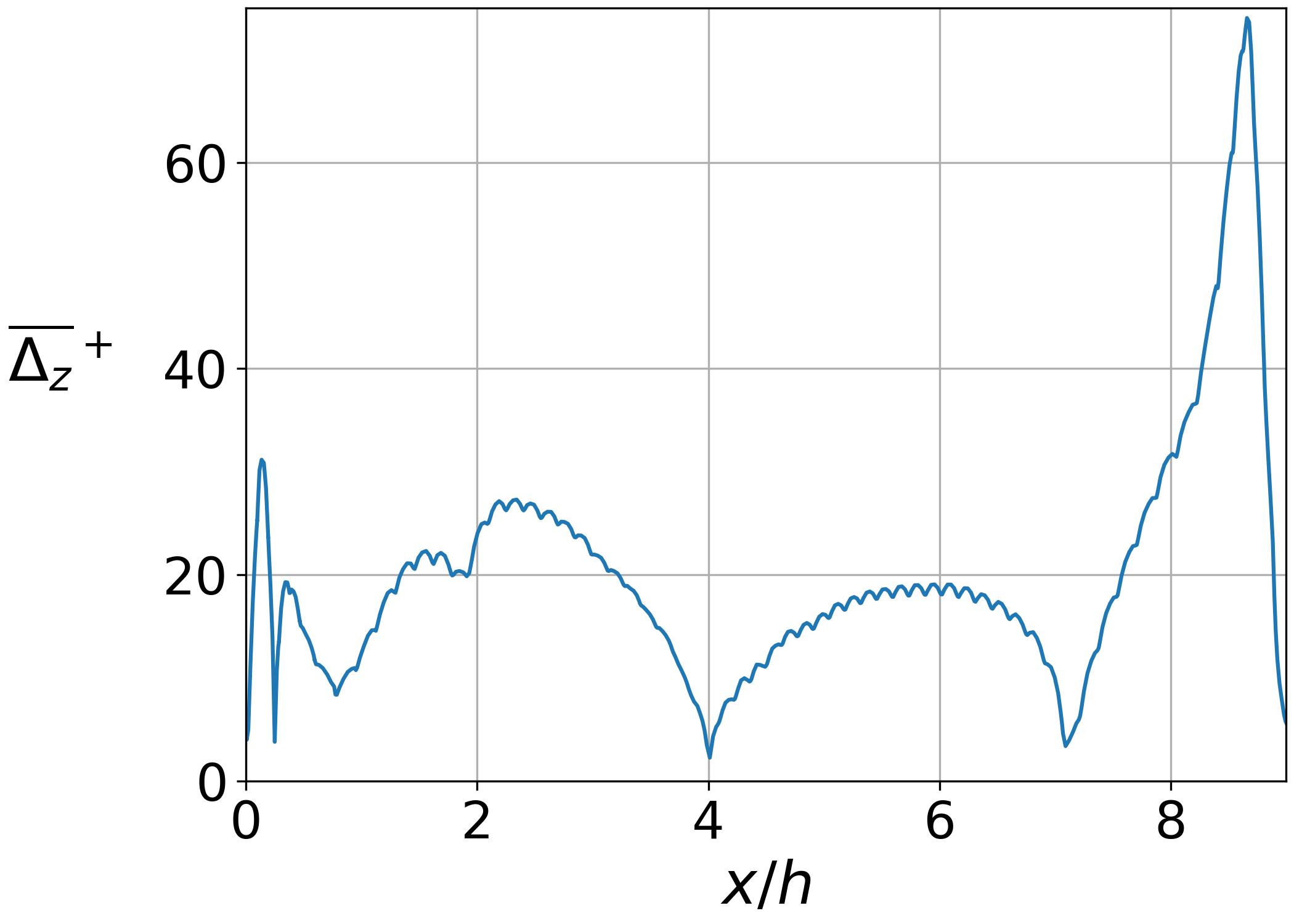}
            \caption{$\overline{\Delta z} ^+$, $Re_b=37000$}
            \label{fig:phill_zplus_37000_Vreman}
        \end{subfigure}
        \caption{The resolution distribution of the wall closest GLL-points above the bottom surface of the periodic hill at $Re_b=10595$ (a-c) and $Re_b=37000$ (b-f) with respect to $x/h$ for the Vreman model. $\Delta x ^+\vert_{\rm wall}$ refers to the streamwise tangential pointwise resolution at the bottom wall; $\Delta y ^+\vert_{\rm wall}$ refers to the $y^+$ of the closest GLL-points to the bottom wall; and $\overline{\Delta z} ^+$ refers to the spanwise average GLL-spacing.}
        \label{fig:phill_resolution}
    \end{figure}

For both of the Reynolds numbers an individual mesh is considered accordingly. Thus, a mesh of 22×16×19 elements and a polynomial order of 8 is chosen for the $Re_b=10595$ case. The corresponding resolution in terms of $\Delta x^+\vert_{\rm wall}$, $\Delta y ^+\vert_{\rm wall} $ and $\overline{\Delta z} ^+$ is shown in the upper row of Figure \ref{fig:phill_resolution} indicating a very good resolution in terms of LES requirements beside the upwind section of the second crest. 

For the $Re_b=37000$ case, a mesh resolution of 512×256×256 grid points as applied in \cite{Gloerfelt2019} was considered as a reference. Thus, basing on a polynomial order of 7, a resolution of 64×32×32 elements is used in this case, as shown in Figure \ref{fig:phill_mesh}. The resolution in terms of $\Delta x^+\vert_{\rm wall}$, $\Delta y ^+\vert_{\rm wall} $ and $\overline{\Delta z} ^+$ is shown in the bottom row of Figure \ref{fig:phill_resolution}. The $\Delta y ^+\vert_{\rm wall} $ values obtained for the first  GLL point of the wall range from  0.25 on top of the first crest to 2.8 on the leeward bottom of the first hill up to 4.6 upwind of the second crest. Obviously, these values do not match the ideal requirements of a wall-resolved LES where everywhere $y^+ \approx 1$ for the closest point is required. However, this somewhat under-resolved setting is considered to be an adequate scenario to test GJP.


Within the investigations a time integration of 3rd order is applied. The flow prediction is advanced in time such that the turbulence is fully developed. Starting from this point, averaging in time is performed until stationary statistics are achieved.  The LES filter width for both of the sub-grid scale models is based on the pointwise distance between the GLL-points. Gradient jump penalisation is applied with $a=0.8$ for all the cases. Table \ref{tab:phill_cases} provides an overview of the various test cases.

\begin{table}[ht]
        \centering
        \begin{tabular}{ccccccccc}
        \hline\hline
             Case   & $Re_b$ & SGS & $N_x$ & $N_y$ & $N_z$ & $P$ & GJP: $a$ \\ \hline
             LES    & 10595 & Sigma & 22 & 16 & 19 & 8 & 0   \\
             LES    & 10595 & Sigma & 22 & 16 & 19 & 8 & 0.8  \\
             LES    & 10595 & Vreman & 22 & 16 & 19 & 8 & 0   \\
             LES    & 10595 & Vreman & 22 & 16 & 19 & 8 & 0.8  \\ \midrule
             LES    & 37000 & Sigma & 64 & 32 & 32 & 7 & 0   \\
             LES    & 37000 & Sigma & 64 & 32 & 32 & 7 & 0.8  \\
             LES    & 37000 & Vreman & 64 & 32 & 32 & 7 & 0   \\
             LES    & 37000 & Vreman & 64 & 32 & 32 & 7 & 0.8  \\\midrule
             LES \cite{Gloerfelt2019}    & 37000 & - & 512 & 256 & 256 & - & -  \\
        \hline\hline
        \end{tabular}
        \caption{Overview over the periodic hill cases considered within this contribution.}
        \label{tab:phill_cases}
    \end{table}

\subsection{Instantaneous flow fields}

\begin{figure}[ht]
        \centering
        \begin{subfigure}{0.49\linewidth}
            \centering
            \includegraphics[width=\linewidth]{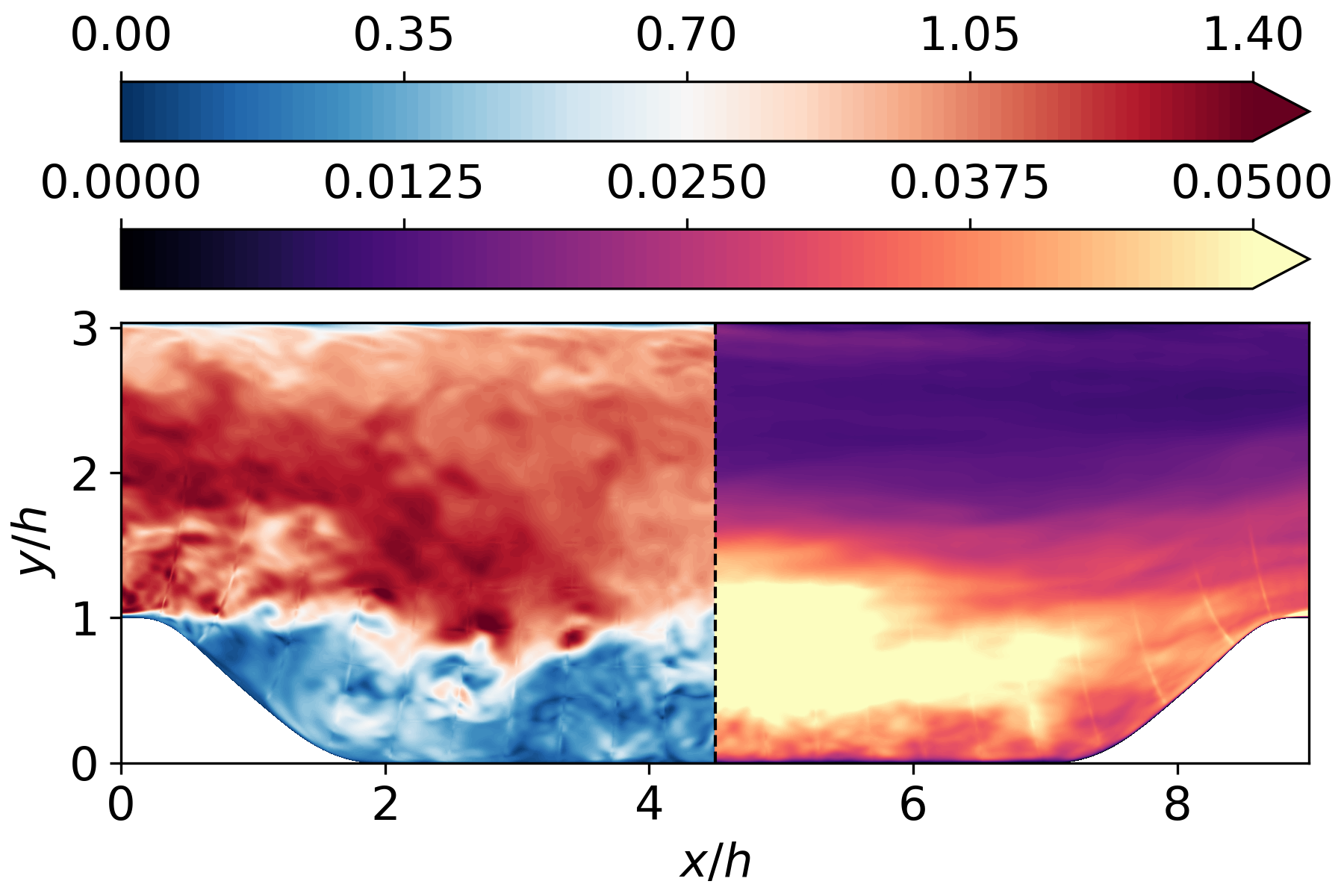}
            \caption{Sigma, $a=0$}
            \label{fig:phill_10595_Sigma_velmag_noGJP}
        \end{subfigure}
        \begin{subfigure}{0.49\linewidth}
            \centering
            \includegraphics[width=\linewidth]{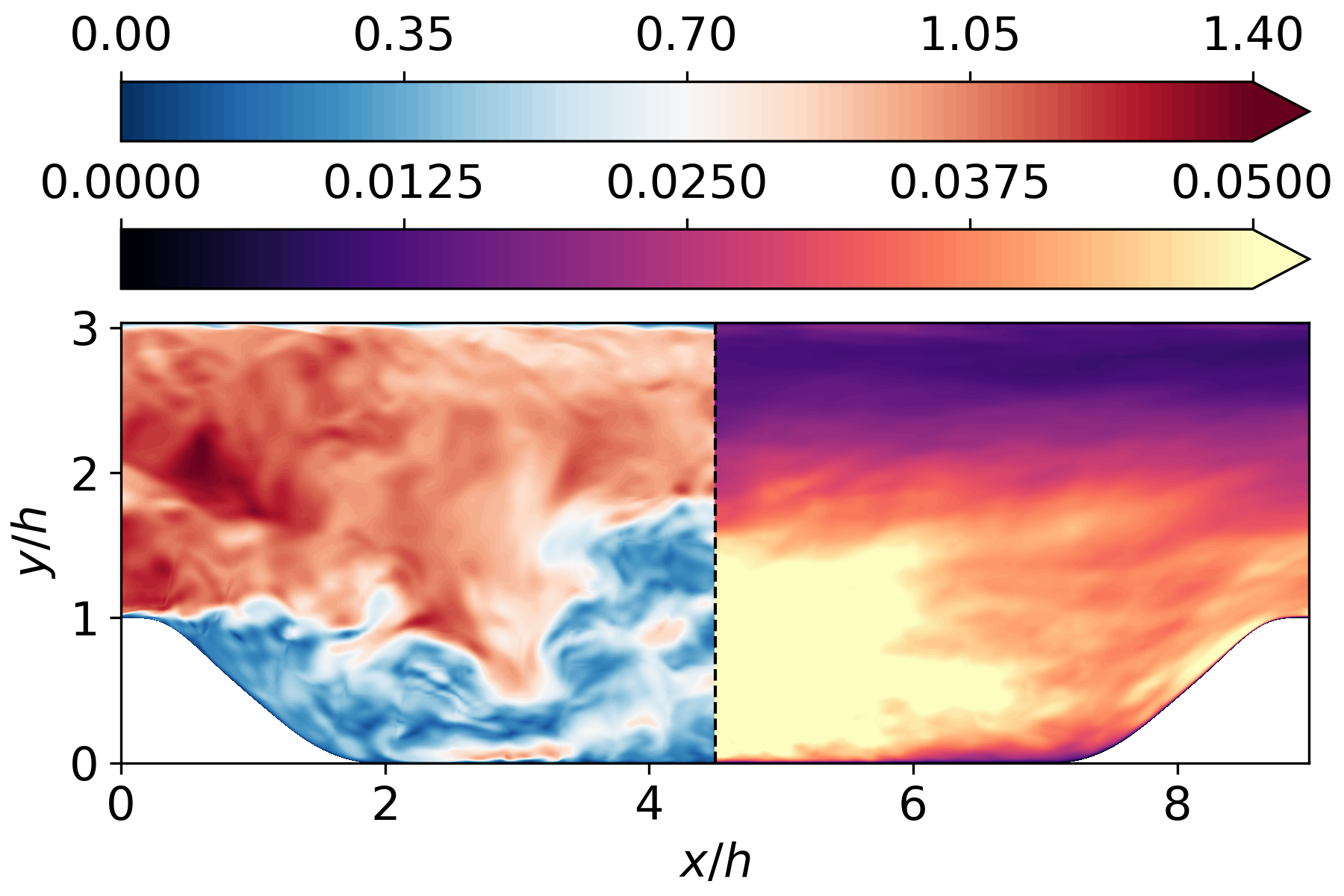}
            \caption{Sigma, $a=0.8$}
            \label{fig:phill_10595_Sigma_velmag_GJP}
        \end{subfigure}
        \begin{subfigure}{0.49\linewidth}
            \centering
            \includegraphics[width=\linewidth]{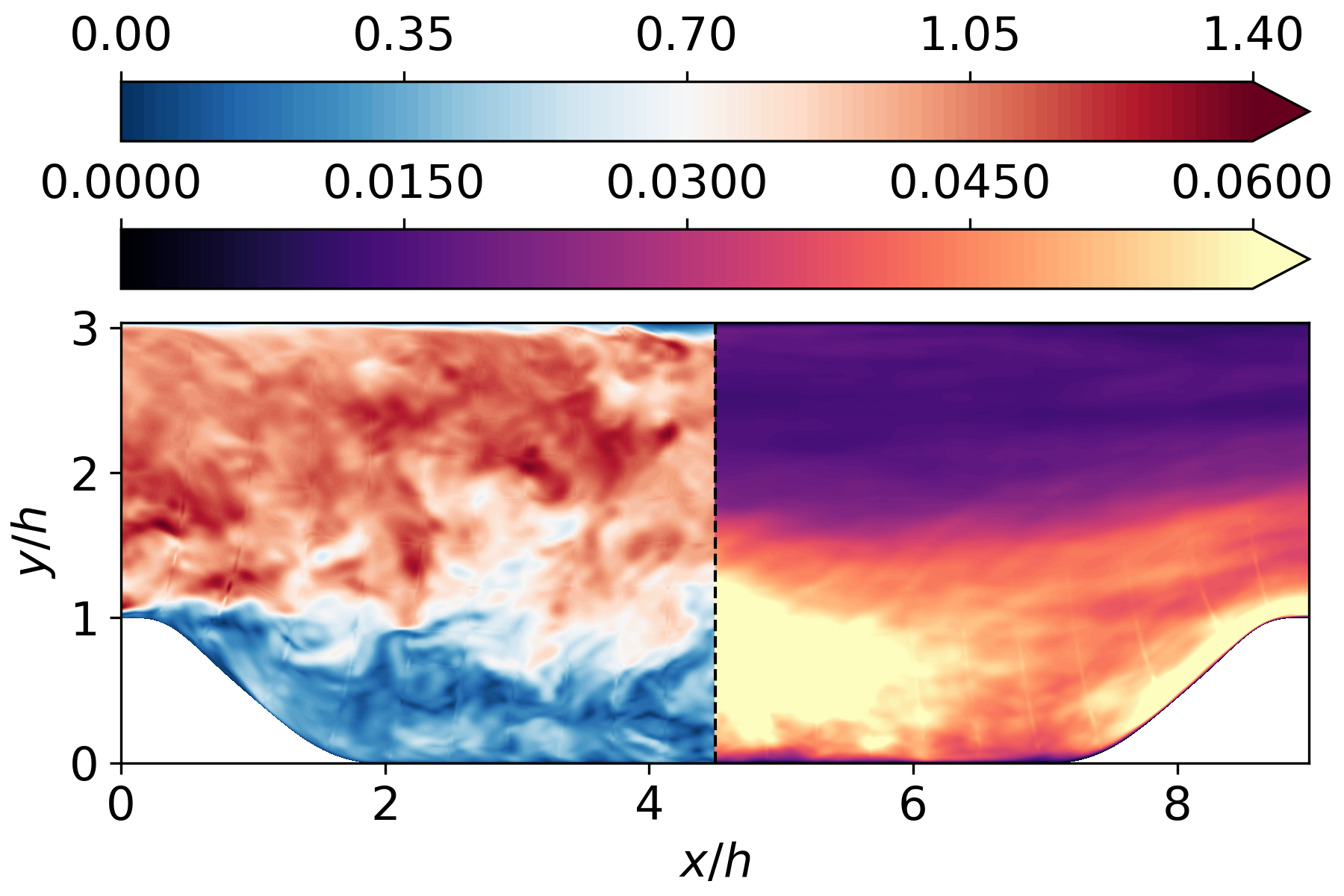}
            \caption{Vreman, $a=0$}
            \label{fig:phill_10595_Vreman_velmag_noGJP}
        \end{subfigure}
        \begin{subfigure}{0.49\linewidth}
            \centering
            \includegraphics[width=\linewidth]{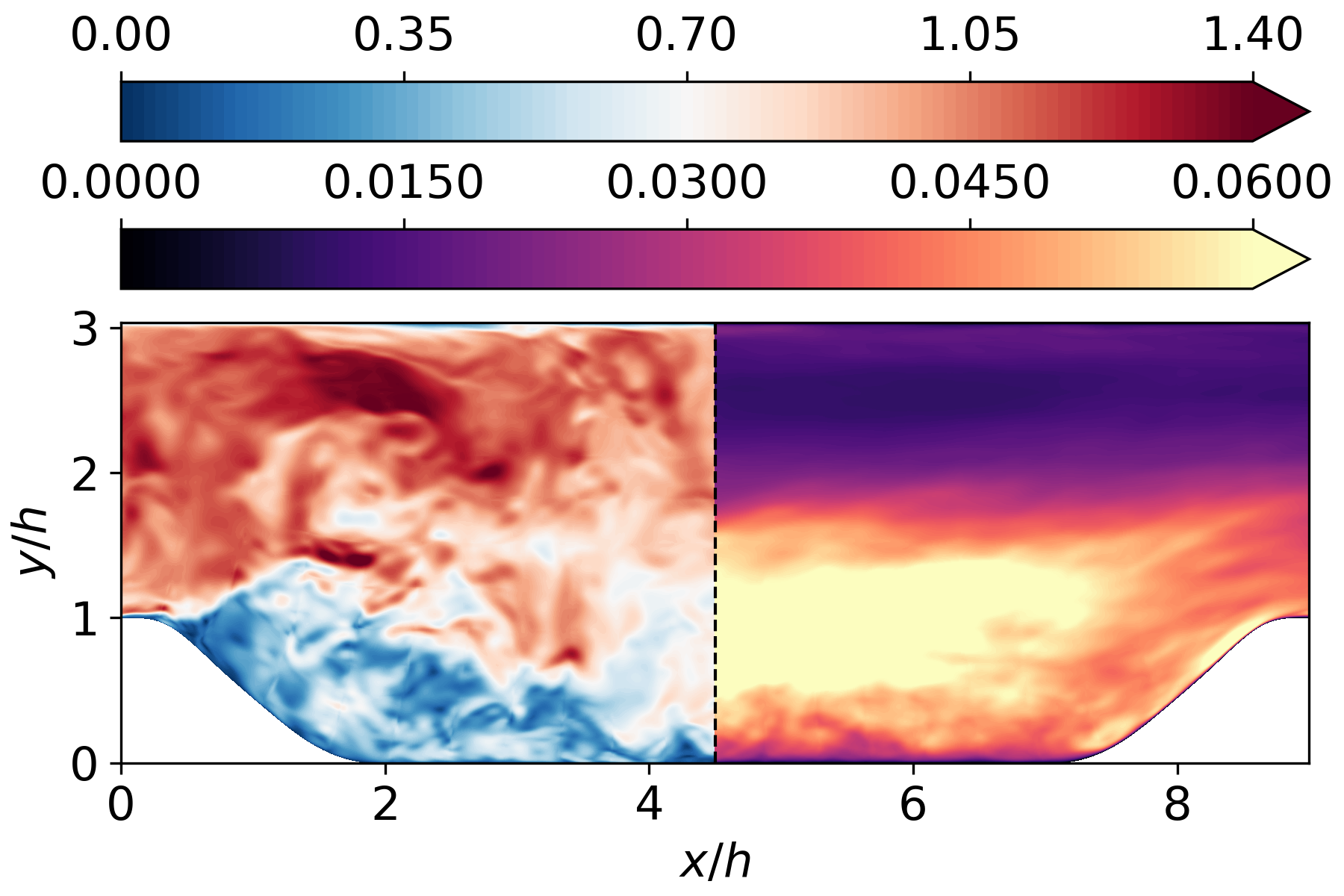}
            \caption{Vreman, $a=0.8$}
            \label{fig:phill_10595_Vreman_velmag_GJP}
        \end{subfigure}
        
        \caption{The velocity magnitude field scaled by $U_b$ (left) and the "short-period" TKE $\langle u^2+v^2+w^2\rangle_{\tau=100h/U_b}/2U_b^2$ (right) of the periodic hill at $Re_b=10595$ in an arbitrary plane for the Sigma and the Vreman model. The GJP is either turned off ($a=0$) or active with the recommended value $a=0.8$.}
        \label{fig:phill_10595_velmag}
    \end{figure}

\begin{figure}[ht]
        \centering
        \begin{subfigure}{0.49\linewidth}
            \centering
            \includegraphics[width=\linewidth]{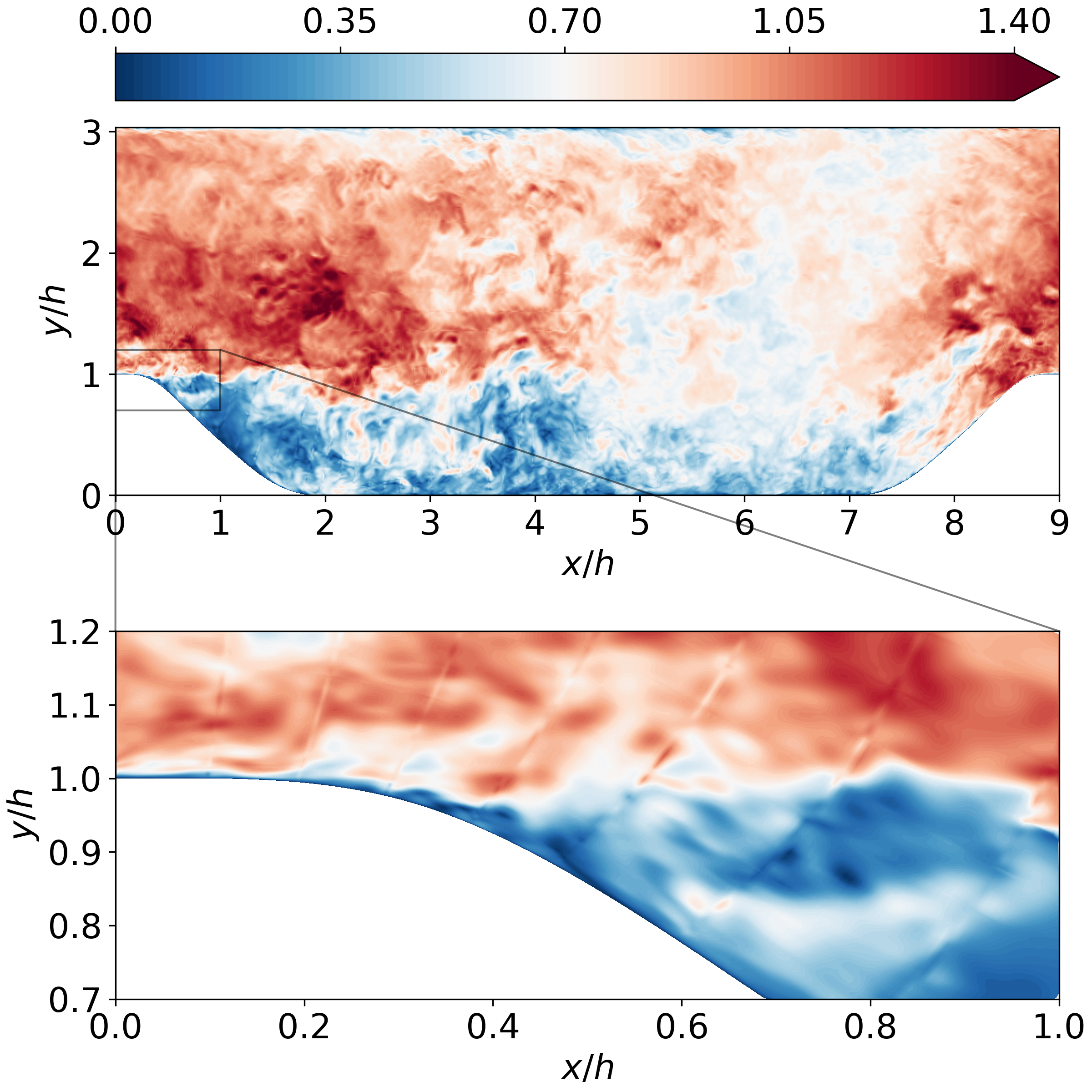}
            \caption{Sigma, $a=0$}
            \label{fig:phill_37000_Sigma_velmag_noGJP}
        \end{subfigure}
        \begin{subfigure}{0.49\linewidth}
            \centering
            \includegraphics[width=\linewidth]{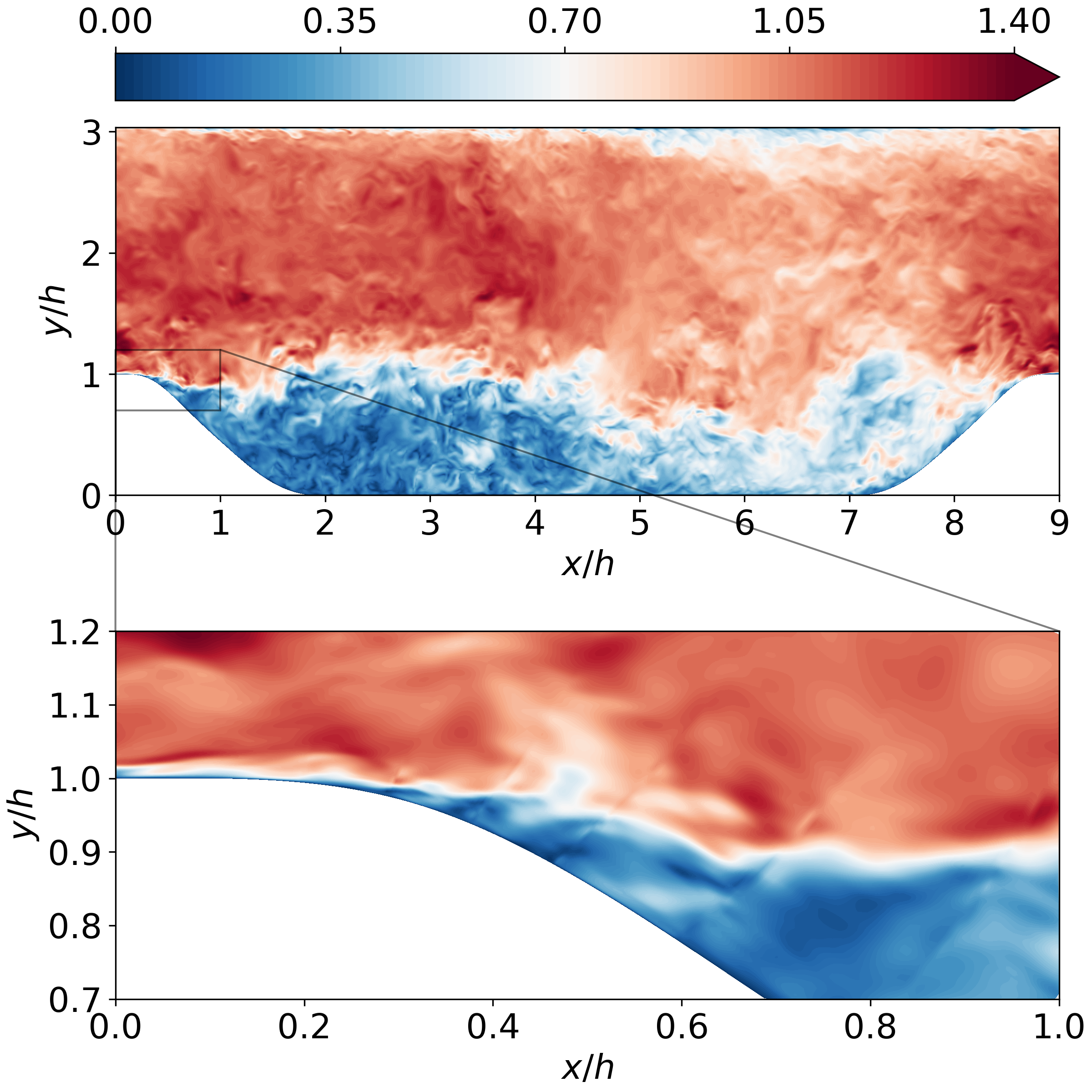}
            \caption{Sigma, $a=0.8$}
            \label{fig:phill_37000_Sigma_velmag_GJP}
        \end{subfigure}
        \begin{subfigure}{0.49\linewidth}
            \centering
            \includegraphics[width=\linewidth]{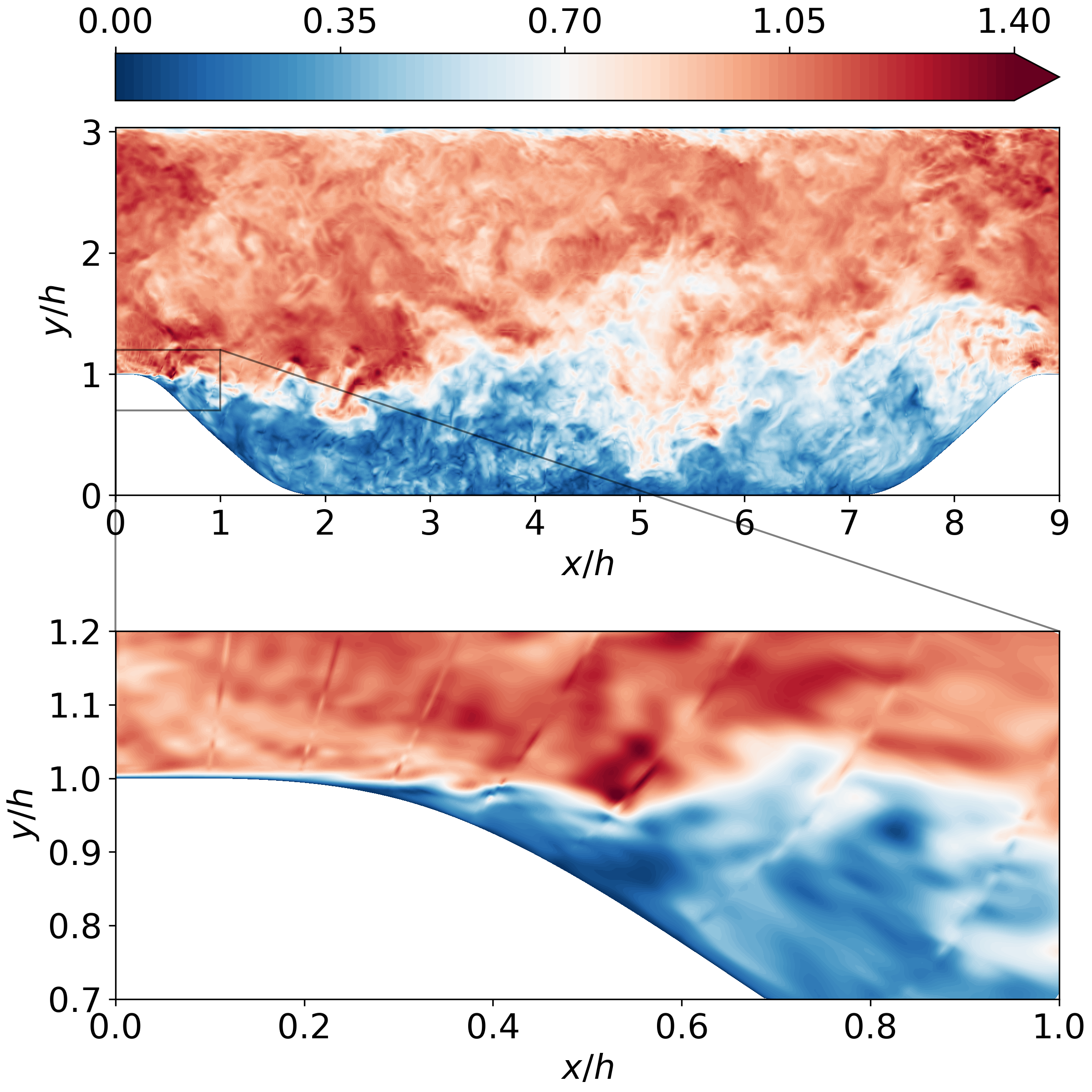}
            \caption{Vreman, $a=0$}
            \label{fig:phill_37000_Vreman_velmag_noGJP}
        \end{subfigure}
        \begin{subfigure}{0.49\linewidth}
            \centering
            \includegraphics[width=\linewidth]{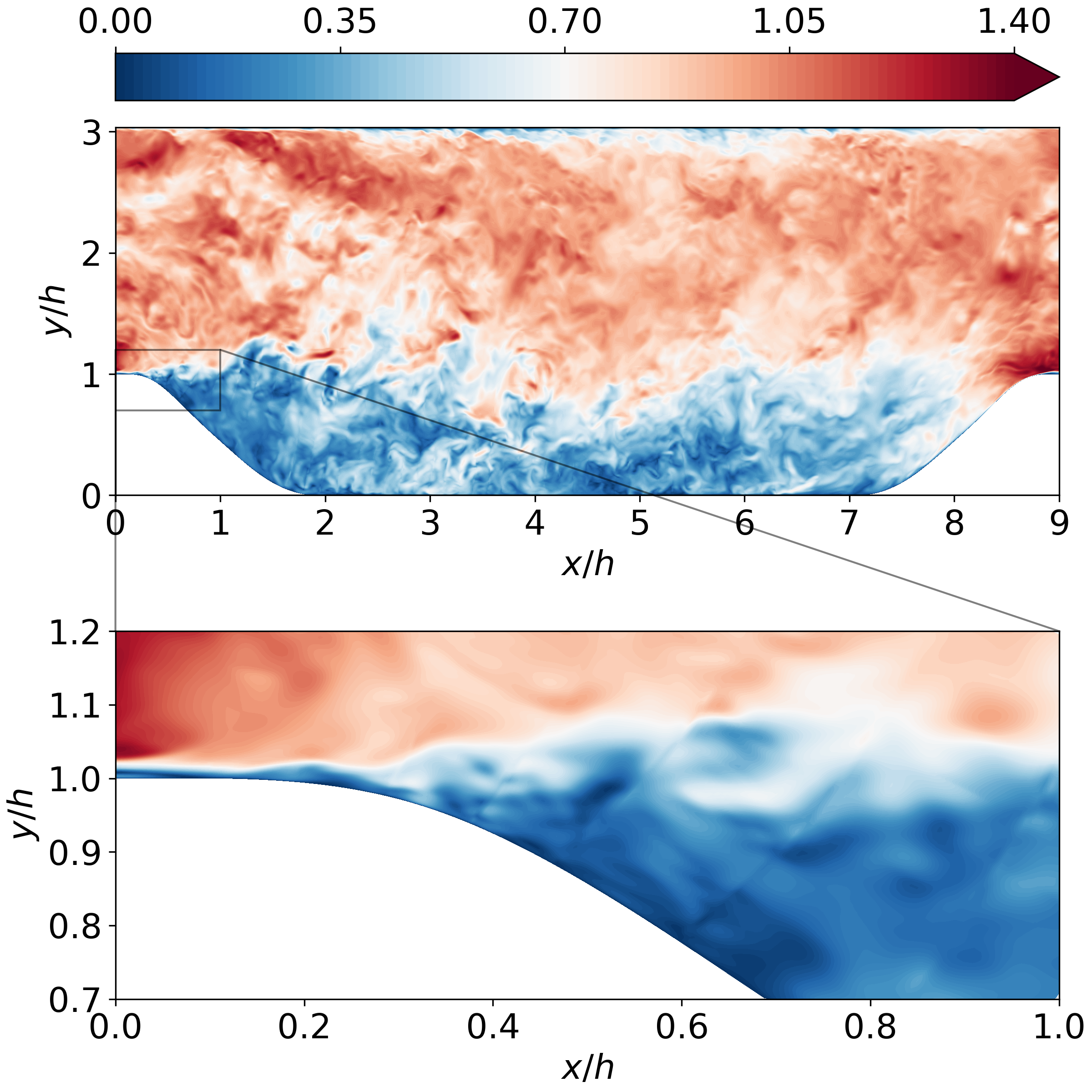}
            \caption{Vreman, $a=0.8$}
            \label{fig:phill_37000_Vreman_velmag_GJP}
        \end{subfigure}
        
        \caption{The velocity magnitude field scaled by $U_b$ of the periodic hill at $Re_b=37000$ in an arbitrary $x-y$ plane for the Sigma and the Vreman model. The upper plot of each panel exhibits the entire plane while the bottom plot zooms in to show the minute details in subset of the domain $x/h\in[0,1], y/h\in[0.7,1.2]$.
        }
        \label{fig:phill_37000_velmag}
    \end{figure}

In Figures \ref{fig:phill_10595_velmag} and \ref{fig:phill_37000_velmag}  instantaneous fully turbulent flow fields (at arbitrary time) in an arbitrary plane of the domain are shown for $Re_b=10595$ and $37000$, for both SGS models, with and without GJP. The flow enters the domain from the left side and separates from the first crest. Large vortical structures, at relatively higher velocities from the upper part of the domain interact with the flow field at lower velocities close to the lower wall, \emph{i.e.,} in between the valley between the two crests.

For a better visualisation of the wiggles, a "short-period" turbulent kinetic energy (TKE) $\langle u^2+v^2+w^2\rangle/2U_b^2$ is calculated by taking one half of the average of the squared velocity fluctuation field over a time period $\tau=100h/U_b$ without averaging over the spatial homogeneous direction. Since the purpose is just to visualise the temporal fluctuation of the wiggles, it needs not to be long enough to reach statistical stationarity. Given the knowledge of the strong vortical motion on the leeward side which might mask the wiggles out visually, we plot such field only at the right half of the plotted subdomain in Figure \ref{fig:phill_10595_velmag} and observe some indications. In Figure \ref{fig:phill_10595_Sigma_velmag_noGJP} for example, high values of the "short-period" TKE  emerge at the element interfaces, showing a rib-like pattern. This indicates the wiggles oscillate at a comparable level with the flow structures during this time period. Such pattern of element-related high values also emerges when applying Vreman model without GJP in Figure \ref{fig:phill_10595_Vreman_velmag_noGJP}.
Here, comparing the individual cases at $Re_b=10595$ in Figure \ref{fig:phill_10595_velmag}, the wiggles observed for both SGS models in Figure \ref{fig:phill_10595_Sigma_velmag_noGJP} and \ref{fig:phill_10595_Vreman_velmag_noGJP} are clearly less pronounced when applying GJP, as in Figure \ref{fig:phill_10595_Sigma_velmag_GJP} and \ref{fig:phill_10595_Vreman_velmag_GJP}.

    \newpage
    
    For the higher $Re_b=37000$, Figure \ref{fig:phill_37000_velmag} demonstrates that the GJP does not change the physics significantly when proceeding to a higher $Re_b$. At the same time, the wiggles are reduced as for the lower Reynolds number, but not completely suppressed either. Note that the wiggles are more difficult to see  due to the higher mesh density.
    To see the wiggles in detail, the bottom plots in each panel of Figure \ref{fig:phill_37000_velmag} show a detailed view around the starting point of the separation zone $x/h\in[0,1]$, $y/h\in[0.7,1.2]$, showcasing the mentioned effect of the application of the GJP on the dissipation of the wiggles, especially in the section $x/h\in[0,0.75]$ where the separation bubble starts to emerge. In particular, the details of the solution appears to be smoother inside the elements when applying GJP, especially for the Vreman model. It could be related to a  loss of physical information or the advected structures from the wiggles do not exist in GJP cases. This phenomenon will be further discussed in Section \ref{sec:turb_channel_550_PSD} on WRLES of the turbulent channel flow. 

    \clearpage
    \subsection{Turbulent Kinetic Energy}
    The following Figures \ref{fig:phill_10595_TKE} and \ref{fig:phill_37000_TKE} show the distribution of the turbulent kinetic energy scaled by $U^2_b$. In all cases where GJP is not applied ($a=0$), there is some non-physical pattern caused by wiggles visible that aligns with the orientation of the elements of the computational mesh. This is reflected as local peaks of TKE since the wiggles produce additional local fluctuations. With GJP in operation ($a=0.8)$, as expected, these wiggles are damped with most of the effect in combination with the Vreman model as seen in Figure \ref{fig:phill_37000_Vreman_TKE_GJP}. Note however that the wiggles have not completely disappeared, but their amplitude is significantly reduced, which will demonstrated via the following Figure \ref{fig:phill_TKE_y1} in detail.
    
    \begin{figure}[ht]
        \centering
        \begin{subfigure}{0.49\linewidth}
            \centering
            \includegraphics[width=\linewidth]{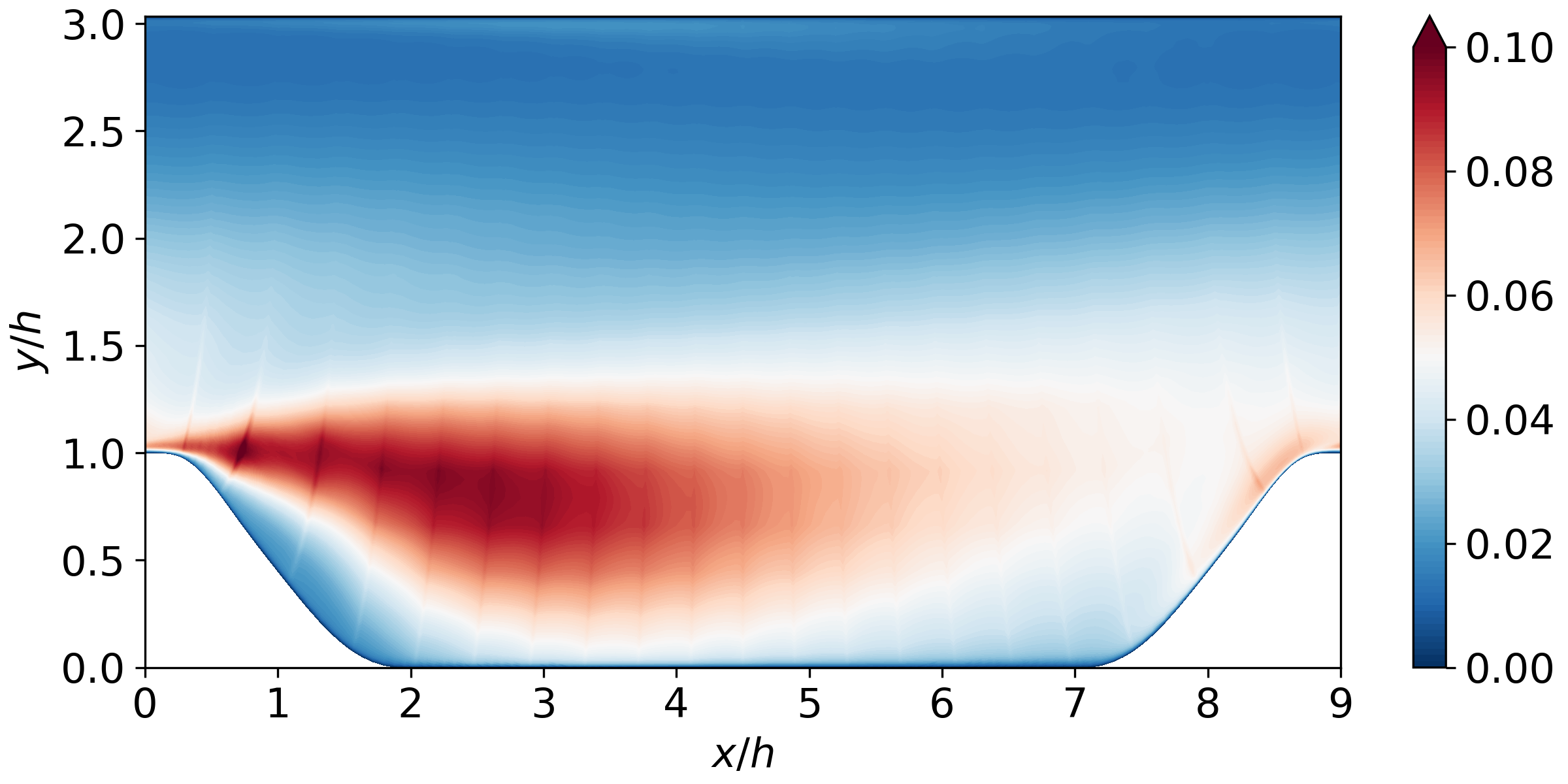}
            \caption{Sigma, $a=0$}
            \label{fig:phill_10595_Sigma_TKE_noGJP}
        \end{subfigure}
        \begin{subfigure}{0.49\linewidth}
            \centering
            \includegraphics[width=\linewidth]{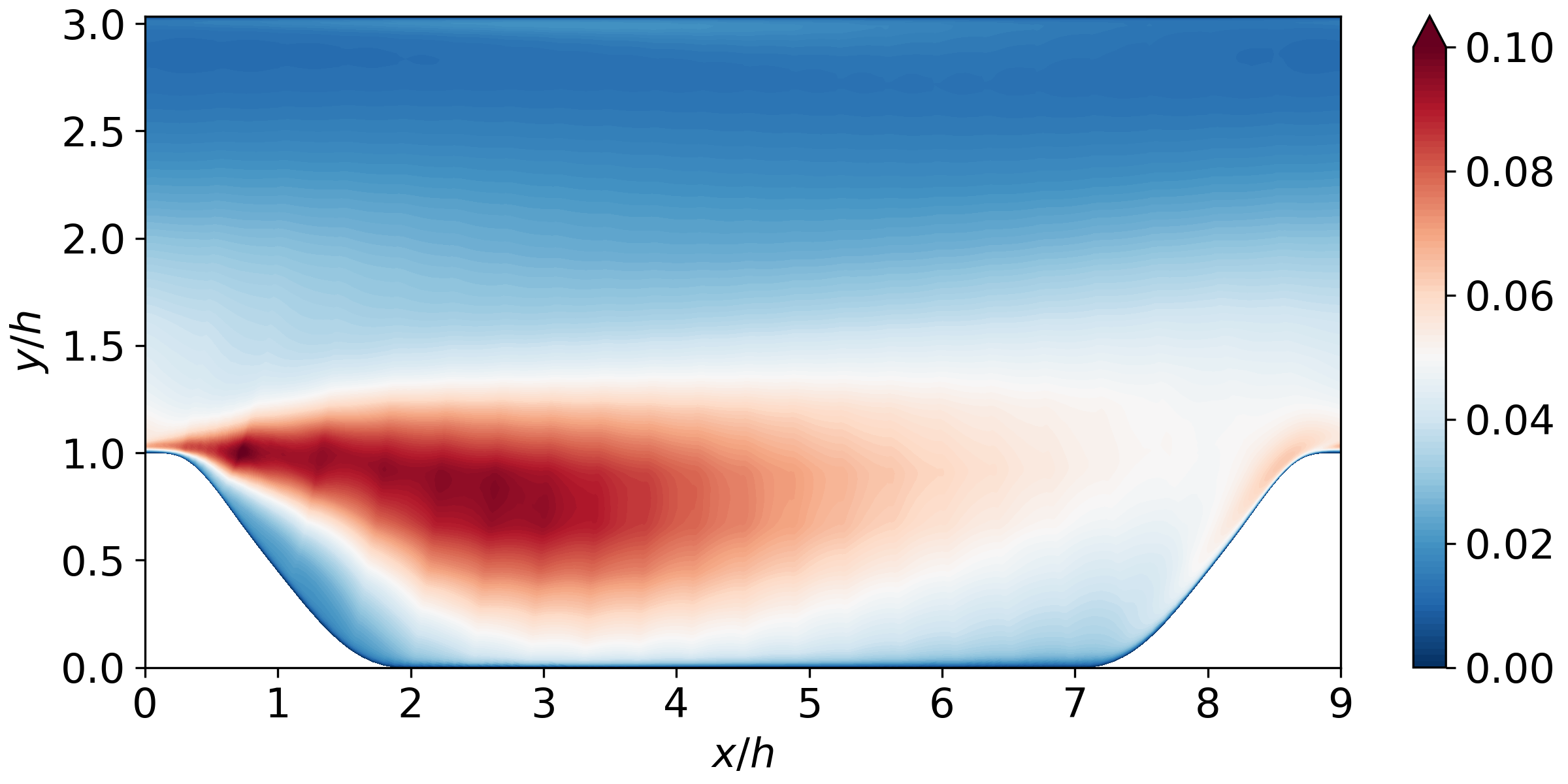}
            \caption{Sigma, $a=0.8$}
            \label{fig:phill_10595_Sigma_TKE_GJP}
        \end{subfigure}
        \begin{subfigure}{0.49\linewidth}
            \centering
            \includegraphics[width=\linewidth]{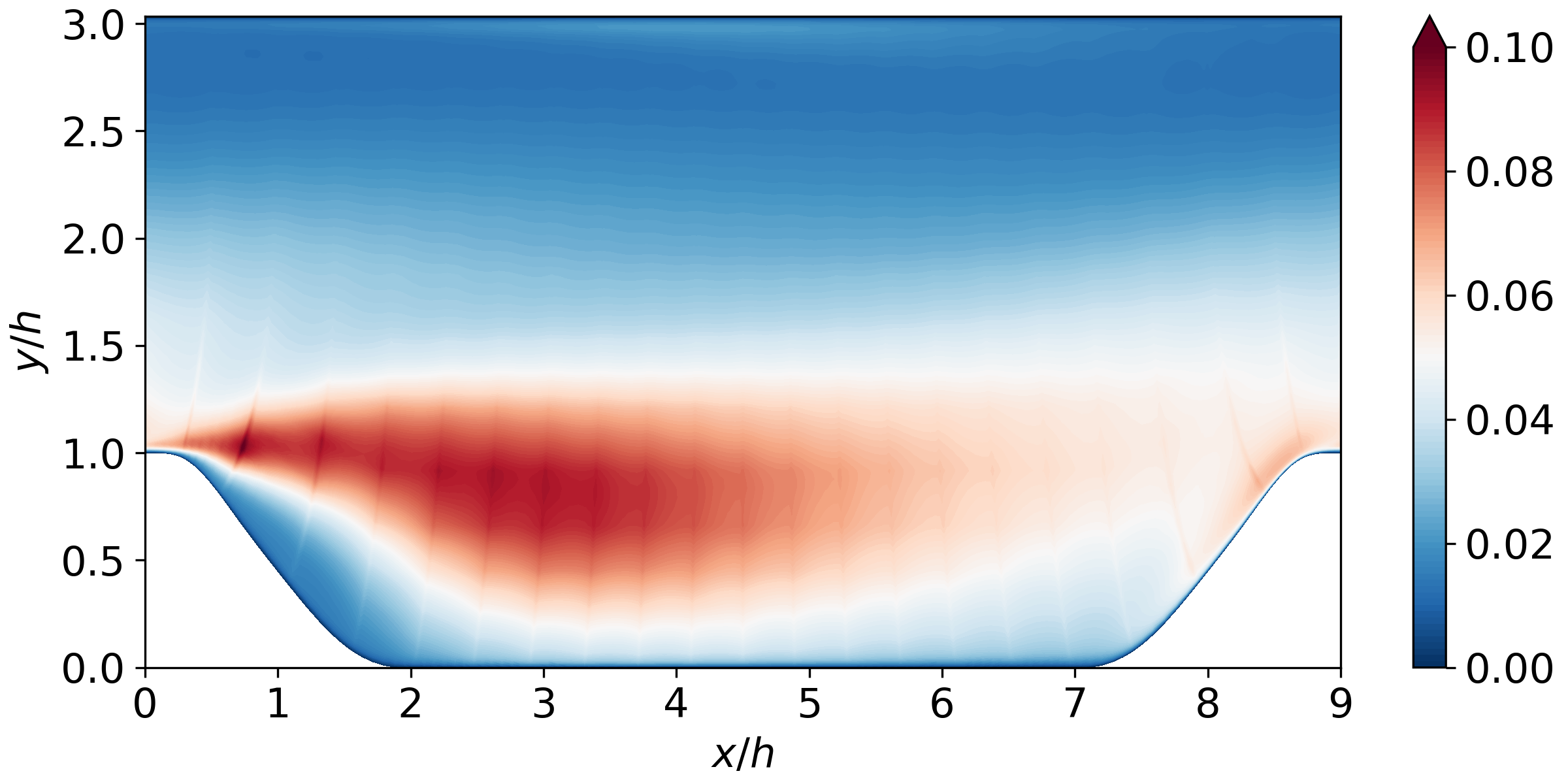}
            \caption{Vreman, $a=0$}
            \label{fig:phill_10595_Vreman_TKE_noGJP}
        \end{subfigure} 
        \begin{subfigure}{0.49\linewidth}
            \centering
            \includegraphics[width=\linewidth]{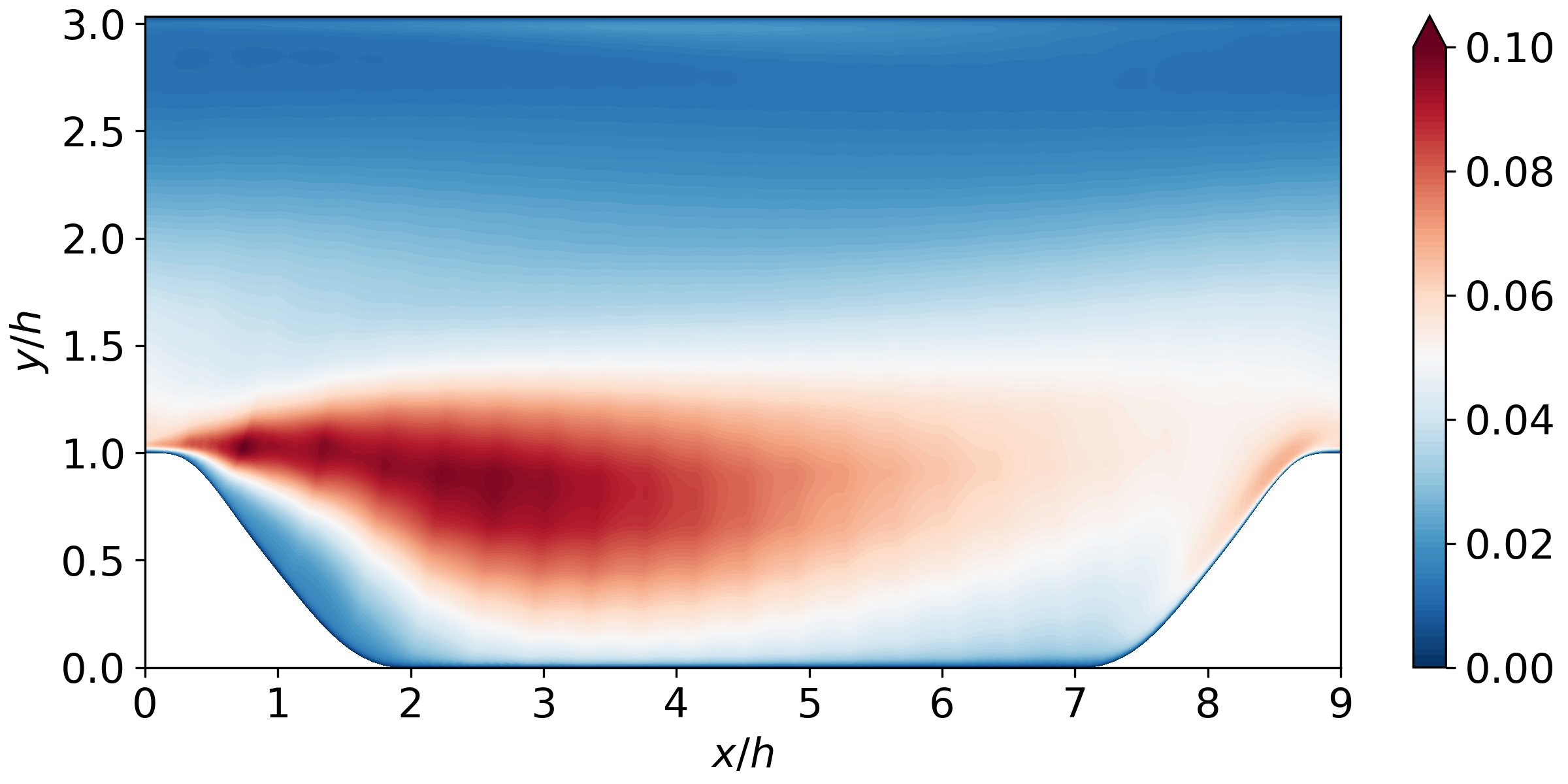}
            \caption{Vreman, $a=0.8$}
            \label{fig:phill_10595_Vreman_TKE_GJP}
        \end{subfigure}
        
        \caption{TKE field scaled by $U_b^2$ of the periodic hill at $Re_b=10595$ for the Sigma and the Vreman model.}
        \label{fig:phill_10595_TKE}
    \end{figure}

    \begin{figure}[ht]
        \centering
        \begin{subfigure}{0.49\linewidth}
            \centering
            \includegraphics[width=\linewidth]{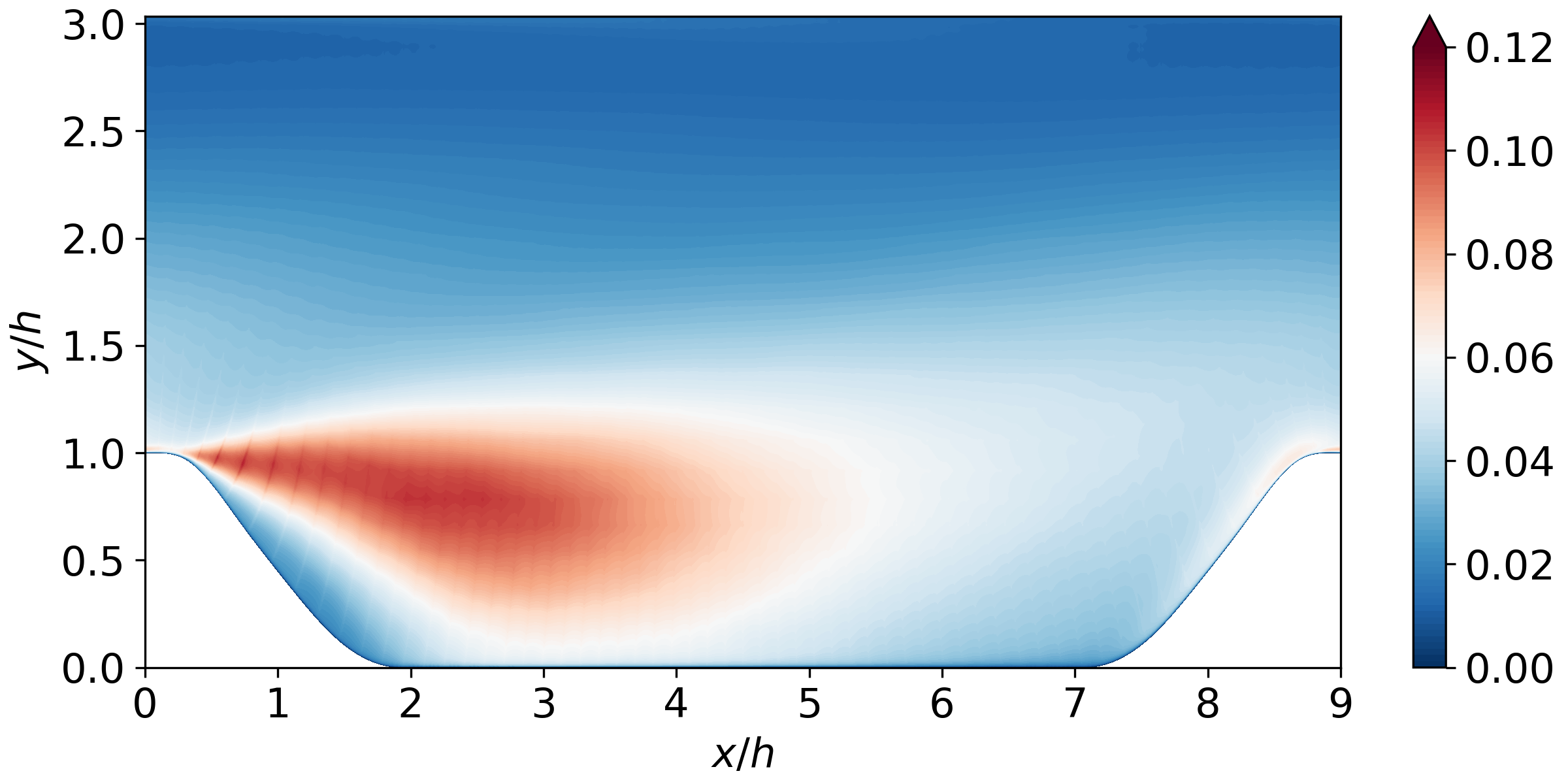}
            \caption{Sigma, $a=0$}
            \label{fig:phill_37000_Sigma_TKE_noGJP}
        \end{subfigure}
        \begin{subfigure}{0.49\linewidth}
            \centering
            \includegraphics[width=\linewidth]{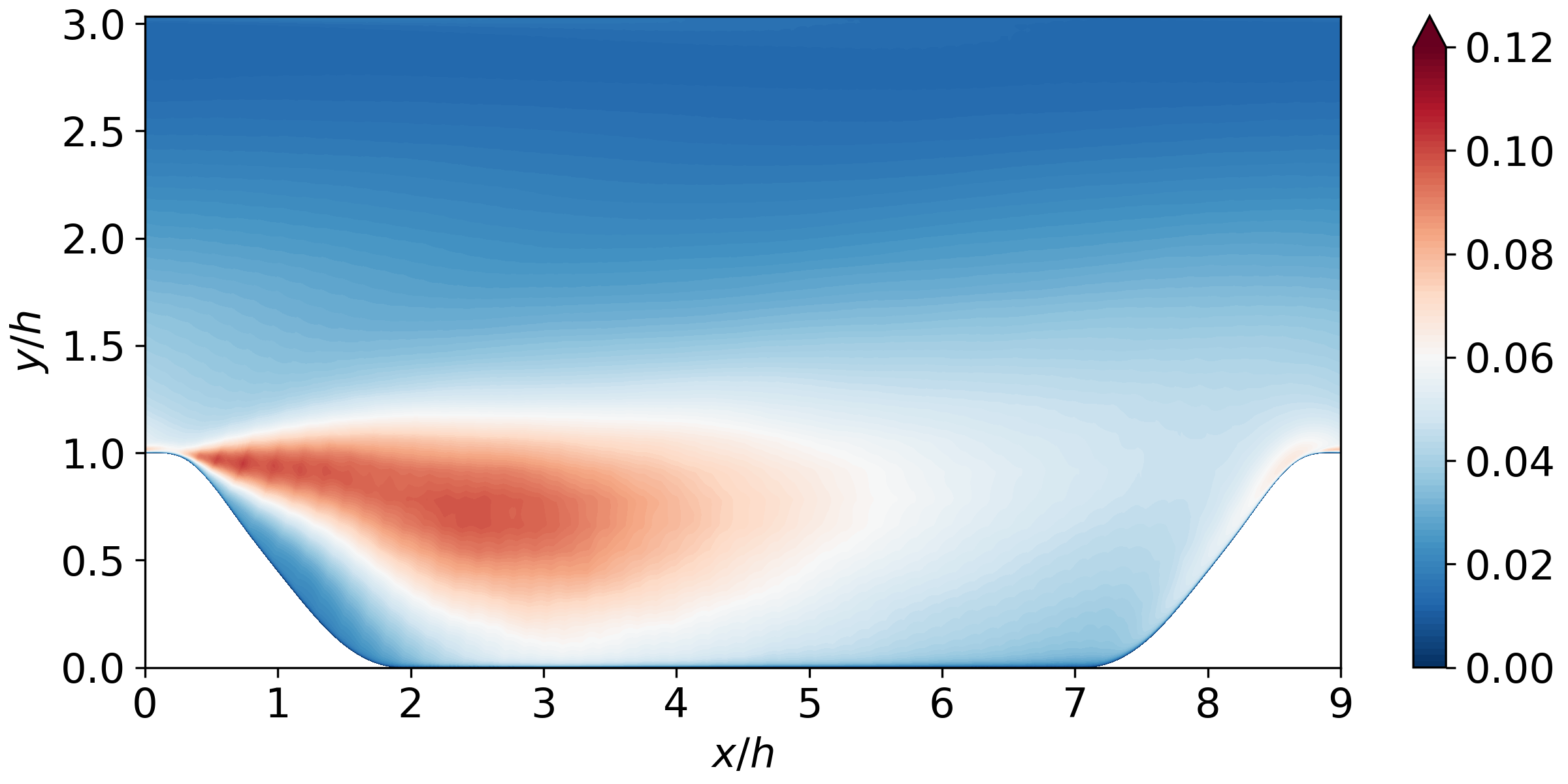}
            \caption{Sigma, $a=0.8$}
            \label{fig:phill_37000_Sigma_TKE_GJP}
        \end{subfigure}
        \begin{subfigure}{0.49\linewidth}
            \centering
            \includegraphics[width=\linewidth]{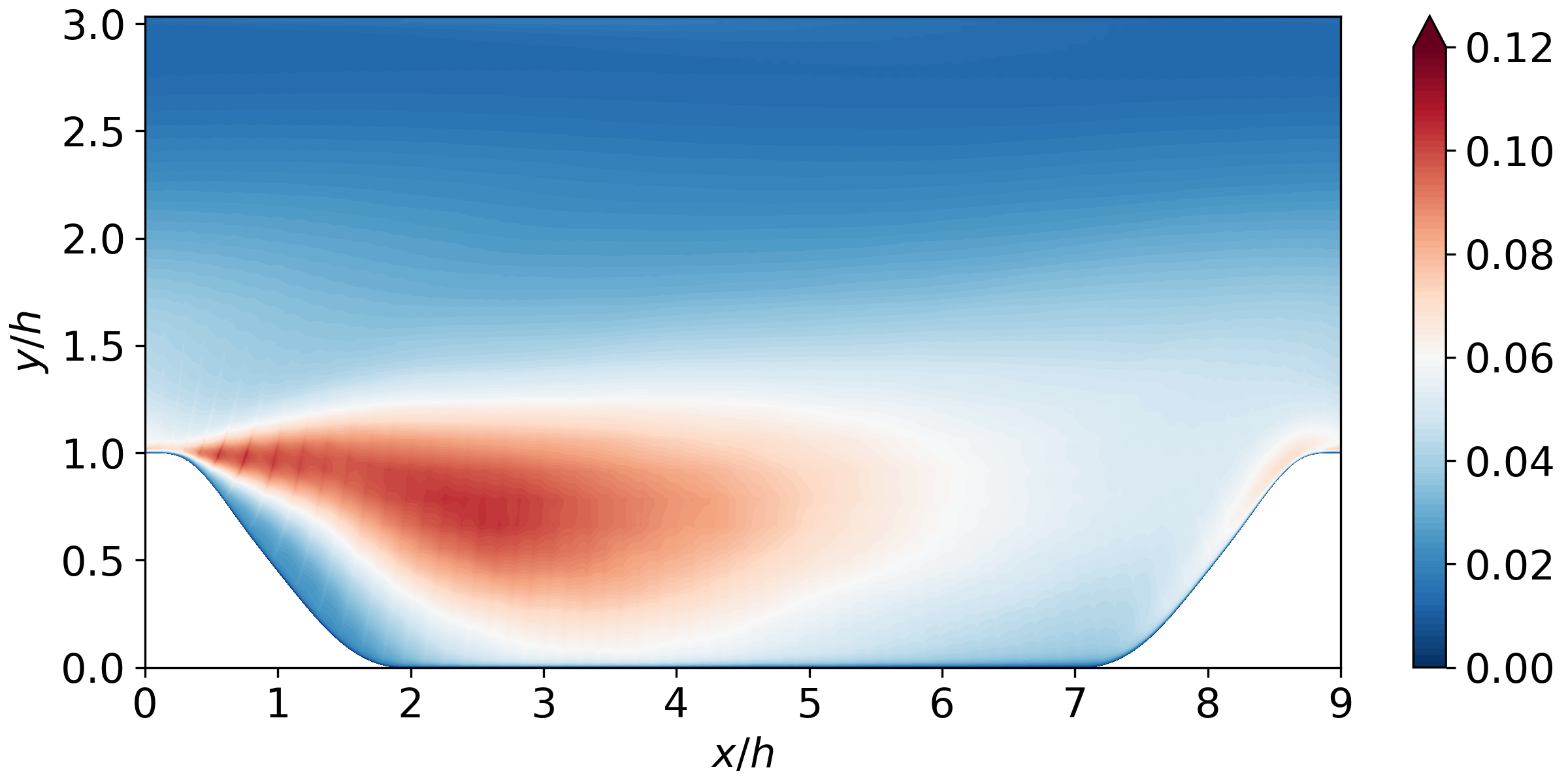}
            \caption{Vreman, $a=0$}
            \label{fig:phill_37000_Vreman_TKE_noGJP}
        \end{subfigure}
        \begin{subfigure}{0.49\linewidth}
            \centering
            \includegraphics[width=\linewidth]{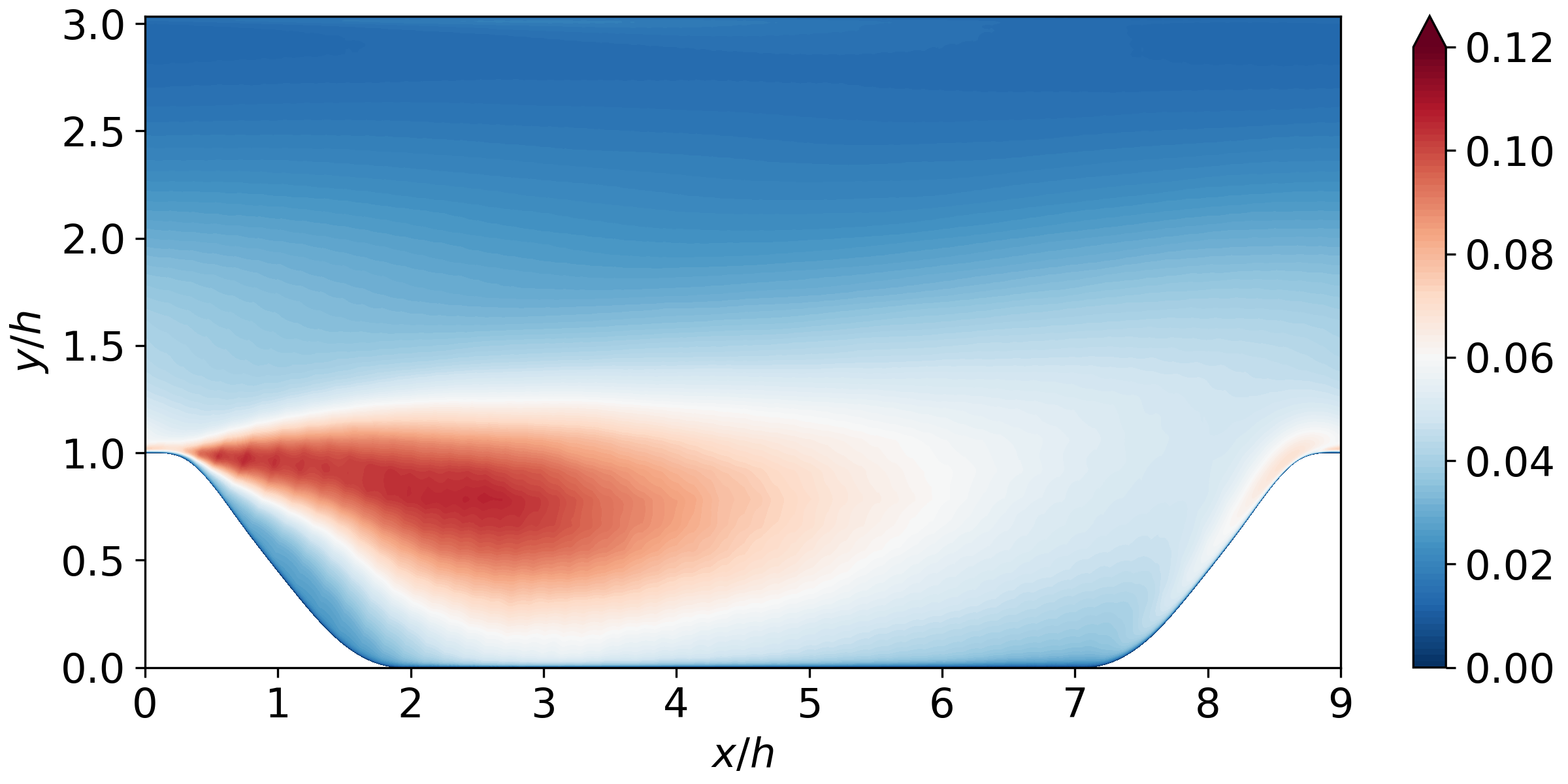}
            \caption{Vreman, $a=0.8$}
            \label{fig:phill_37000_Vreman_TKE_GJP}
        \end{subfigure}
        
        \caption{TKE field scaled by $U_b^2$ of the periodic hill at $Re_b=37000$ for the Sigma and the Vreman model.}
        \label{fig:phill_37000_TKE}
    \end{figure} 

        

    Figure \ref{fig:phill_TKE_y1} shows line profiles at $y/h=1$ extracted from Figures \ref{fig:phill_10595_TKE} and \ref{fig:phill_37000_TKE}, corresponding to the crest-to-crest region where pronounced oscillations are observed. While the application of GJP introduces damping across all cases, the effect varies with both the SGS model and Reynolds number. For the Sigma model at $Re_b=10595$, the TKE profile with GJP ($a=0.8$) closely follows the case without GJP ($a=0$), apart from a reduction in TKE at the oscillatory locations. At $Re_b=37000$, the Sigma model with GJP exhibits a consistently lower TKE in the upstream portion of the line compared to the case without GJP, while also reducing the wiggle amplitude. In contrast, the Vreman model demonstrates a more similar behaviour for both Reynolds numbers: GJP tends to preserve or even increase the TKE at the oscillatory locations while enhancing TKE within the elements, resulting in a smoother distribution.
    
    \begin{figure}[ht]
        \centering
        \begin{subfigure}{0.49\linewidth}
            \centering
            \includegraphics[width=\linewidth]{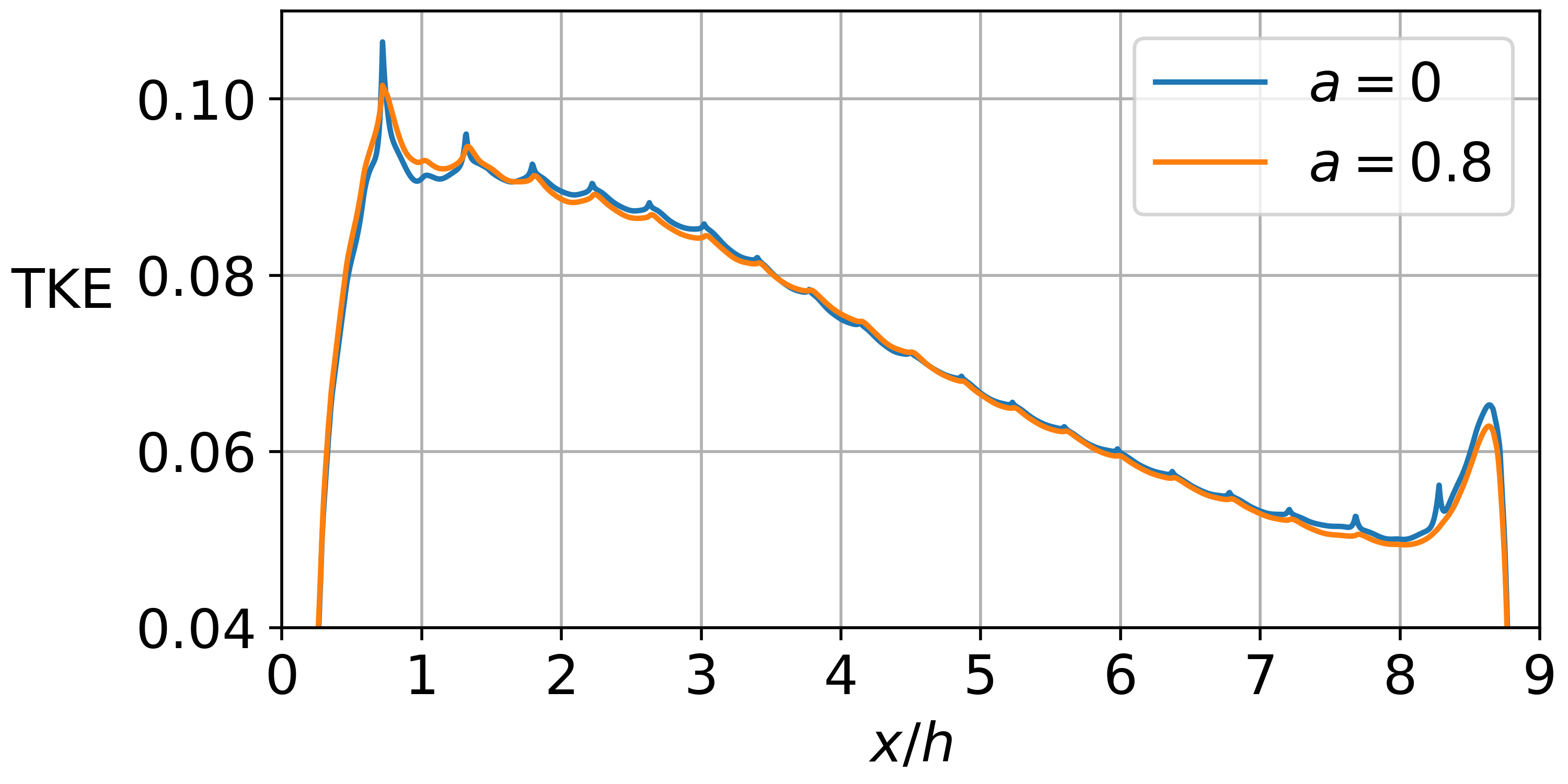}
            \caption{$Re_b=10595$, Sigma}
            \label{fig:phill_10595_Sigma_TKE_y1}
        \end{subfigure}
        \begin{subfigure}{0.49\linewidth}
            \centering
            \includegraphics[width=\linewidth]{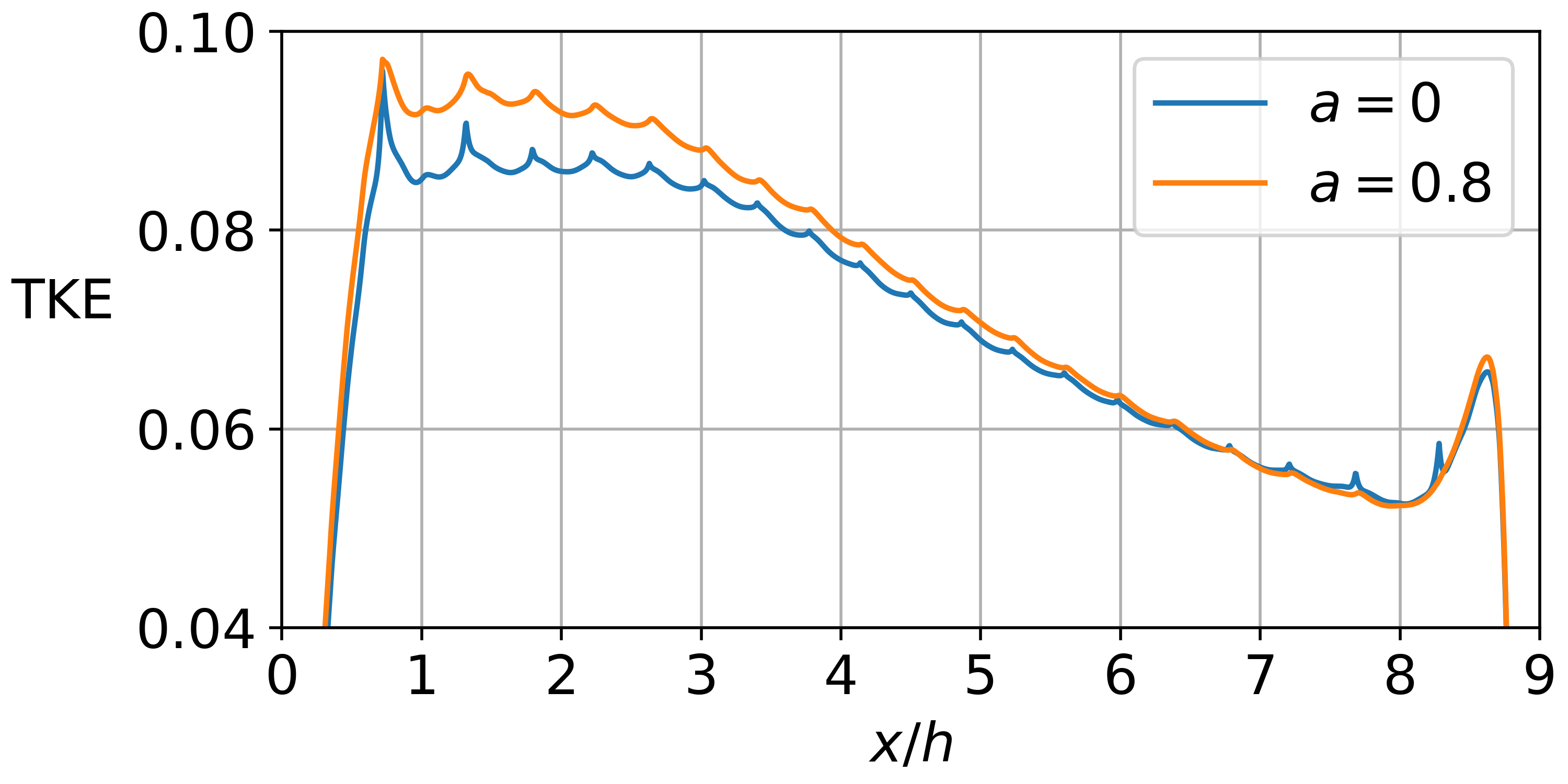}
            \caption{$Re_b=10595$, Vreman}\label{fig:phill_10595_Vreman_TKE_y1}
        \end{subfigure}
        \begin{subfigure}{0.49\linewidth}
            \centering
            \includegraphics[width=\linewidth]{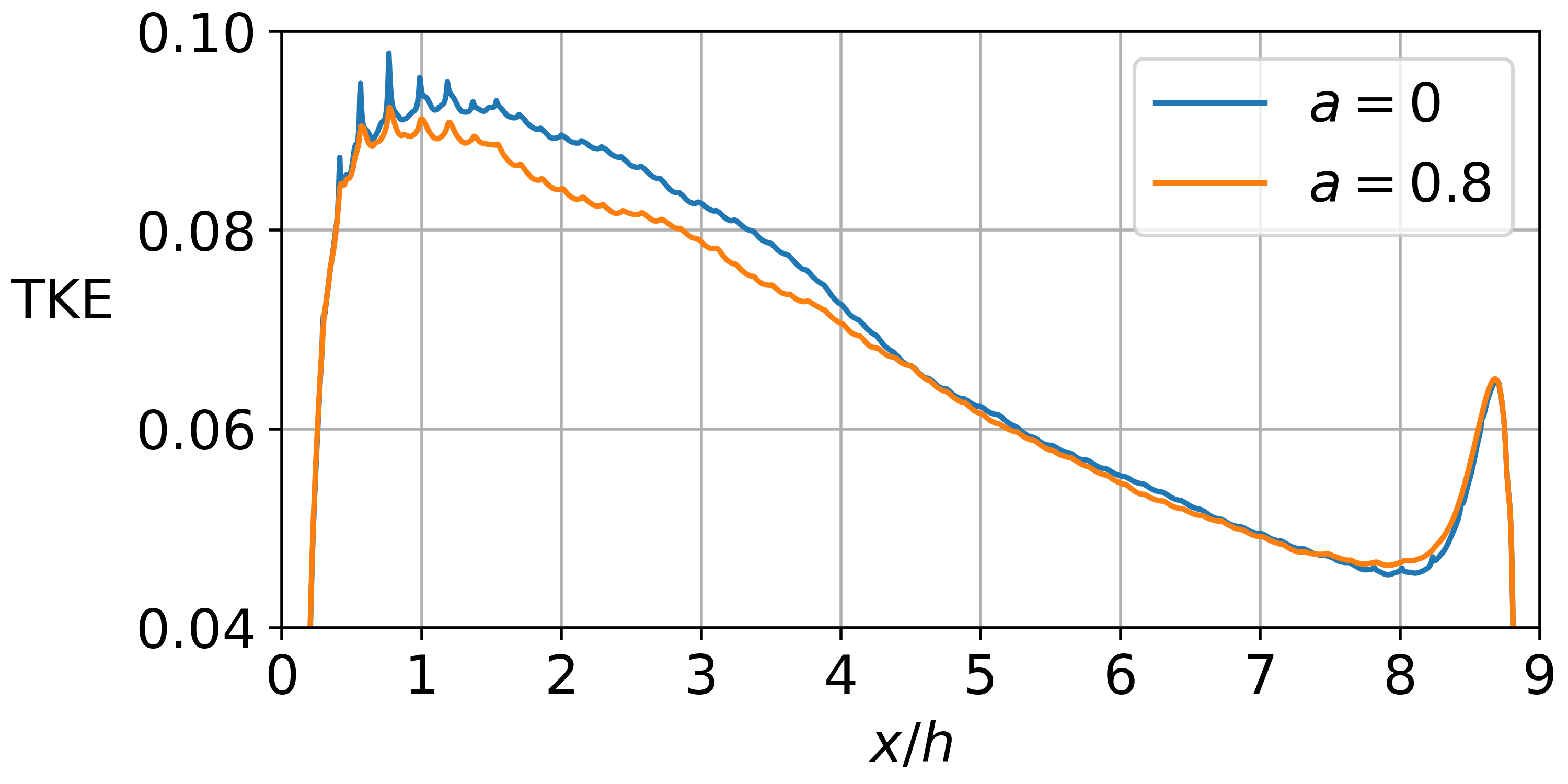}
            \caption{$Re_b=37000$, Sigma}
            \label{fig:phill_37000_Sigma_TKE_y1}
        \end{subfigure}
        \begin{subfigure}{0.49\linewidth}
            \centering
            \includegraphics[width=\linewidth]{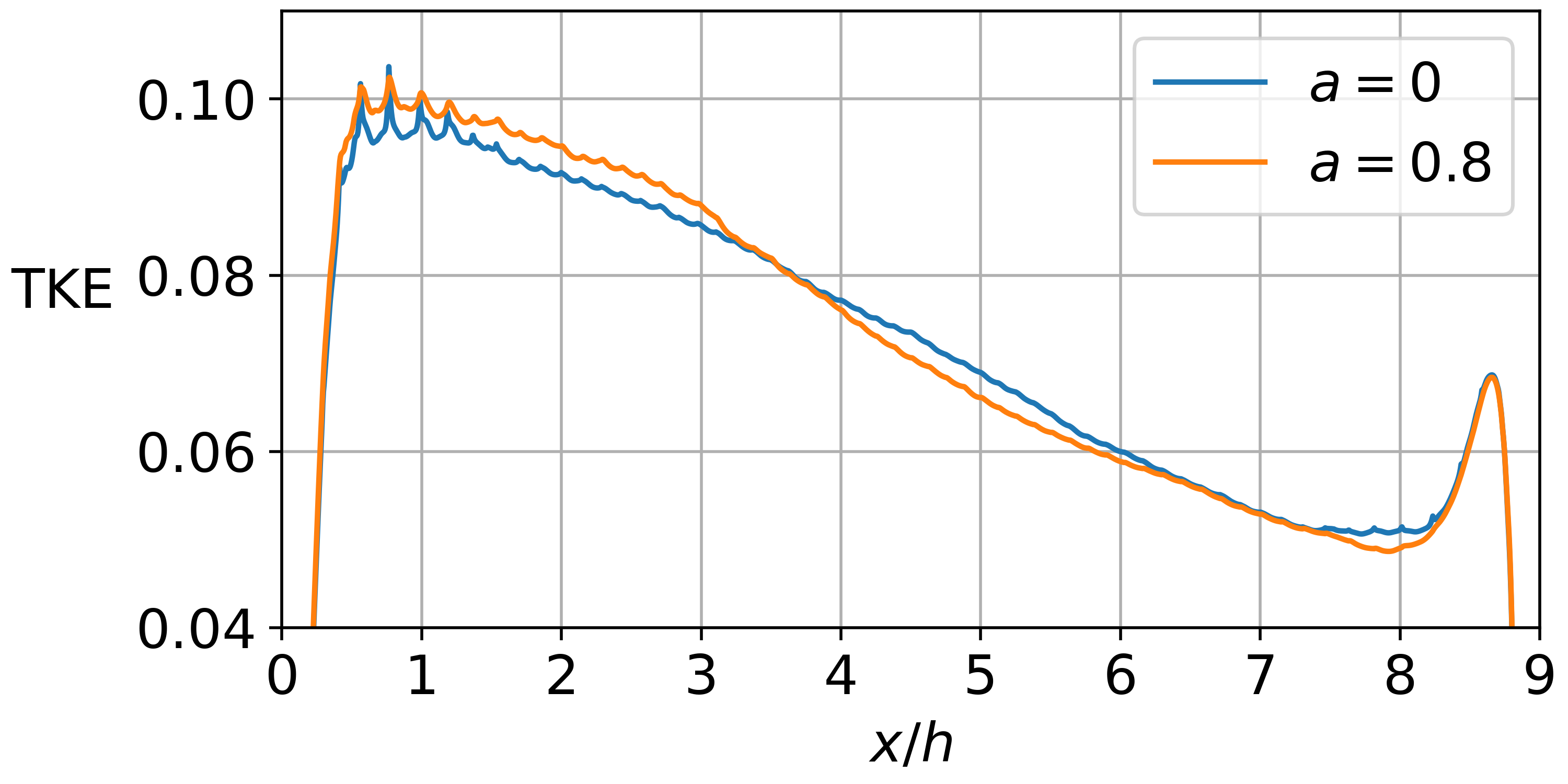}
            \caption{$Re_b=37000$, Vreman}\label{fig:phill_37000_Vreman_TKE_y1}
        \end{subfigure}
        
        \caption{Profiles of the turbulent kinetic energy (TKE) scaled by $U_b^2$ pertaining to the  periodic hill at $Re_b=10595$ and $37000$ for the Sigma and the Vreman model at $y/h=1.0$.}
        \label{fig:phill_TKE_y1}
    \end{figure}
    \clearpage

To summarise the results obtained for the periodic-hill case we can clearly see a beneficial effect of GJP, in reducing the wiggles around element boundaries, and at the same time reducing the fluctuation level inside elements as well. However, for relatively coarse LES, the wiggles in the turbulent kinetic energy do not fully disappear. In addition, there is a noticeable effect on the general fluctuation level. Therefore, we analyse in more detail the effect of GJP on a canonical turbulent case, in an effort to identify the characteristics of the damping introduced by GJP.

\section{Fully developed turbulent channel flow}\label{sec:turb_channel_550}

In this section, we study canonical turbulent channel flow, with particular emphasis on the impact of GJP on the spectral energy distribution. 

    \subsection{Case setup}
    While GJP improves the smoothness of the WRLES of periodic hill flows, the insight of GJP's effect on a spectral space still remains unknown. For this purpose, we further vary the amplitude of the GJP and consider a canonical fully developed channel flow at $Re_\tau \approx 550$, by setting the bulk Reynolds number to $Re_b=10000$. The size of the computational domain is $L_x \times L_y \times L_z = 8\pi\delta \times 2\delta \times 3\pi\delta$ to capture the largest flow scales as was adopted in~\cite{Hoyas2006}. The information of the meshes is shown in Table \ref{tab:case_all}. The spatial resolution of LES is rather coarse in the horizontal $x$ and $z$ directions while keeps the same as the DNS case in the wall normal $y$ direction, thus being in the realm of a  WRLES. The resolution is modulated by the averaging Gauss--Lobatto--Legendre (GLL) spacing $\Delta_x := L_x/N_x/P$, $\Delta_z := L_z/N_z/P$ and $\Delta_{y,{\rm wall}} := L_{y,{\rm wall}}/P$, where $N_x, N_z$ are number of elements in $x,y,z$ directions, $P$ is the polynomial order, and $L_{y,{\rm wall}}$ is the height of the element adjacent to the wall. To show how well the flow resolved, all length quantities are scaled in inner units by the length scale   $\delta_\nu := \sqrt{\nu/\frac{d <U>}{d y}|_{\rm wall}}$, where $\nu$ is the kinematic viscosity and $\frac{d <U>}{d y}|_{\rm wall}$ is the mean $x-$velocity gradient at the wall obtained from DNS \emph{a priori}. Thus, $\Delta_x^+ = \Delta_x / \delta_\nu$, and likewise for $\Delta_{y,{\rm wall}}^+$ and $\Delta_z^+$. The filter width $\Delta$ used to evaluate the eddy viscosity is estimated in an averaged way: for all GLL-points within the same element, they share a same $\Delta = (V_{\Omega_e}/(P+1)^3)^{1/3}$. The purpose is mainly to avoid artifacts on the eddy viscosity from the variation of $\Delta$ in the two homogeneous directions.
    \begin{table}[ht]
        \centering
                \caption{Computational meshes of channel flow simulations.}

        \begin{tabular}{ccccccccc}
        \hline\hline
             Case   & $N_x$ & $N_y$ & $N_z$ & $P$ & $\Delta_x^+$ & $\Delta_{y,{\rm wall}}^+$ & $\Delta_z^+$ \\
        \hline
             DNS    & $144$ & $36$  & $108$ & $7$ & 9.1  & 1.6 & 4.5   \\
             LES    &  $40$  & $36$  & $30$  & $7$ & 48.9 & 1.6 & 24.5  \\
        \hline\hline
        \end{tabular}
        \label{tab:case_all}
    \end{table}
    
    \subsection{Instantaneous fields}
\begin{figure}[ht]
        \centering
        \begin{subfigure}{0.49\linewidth}
            \centering
            \includegraphics[width=\linewidth]{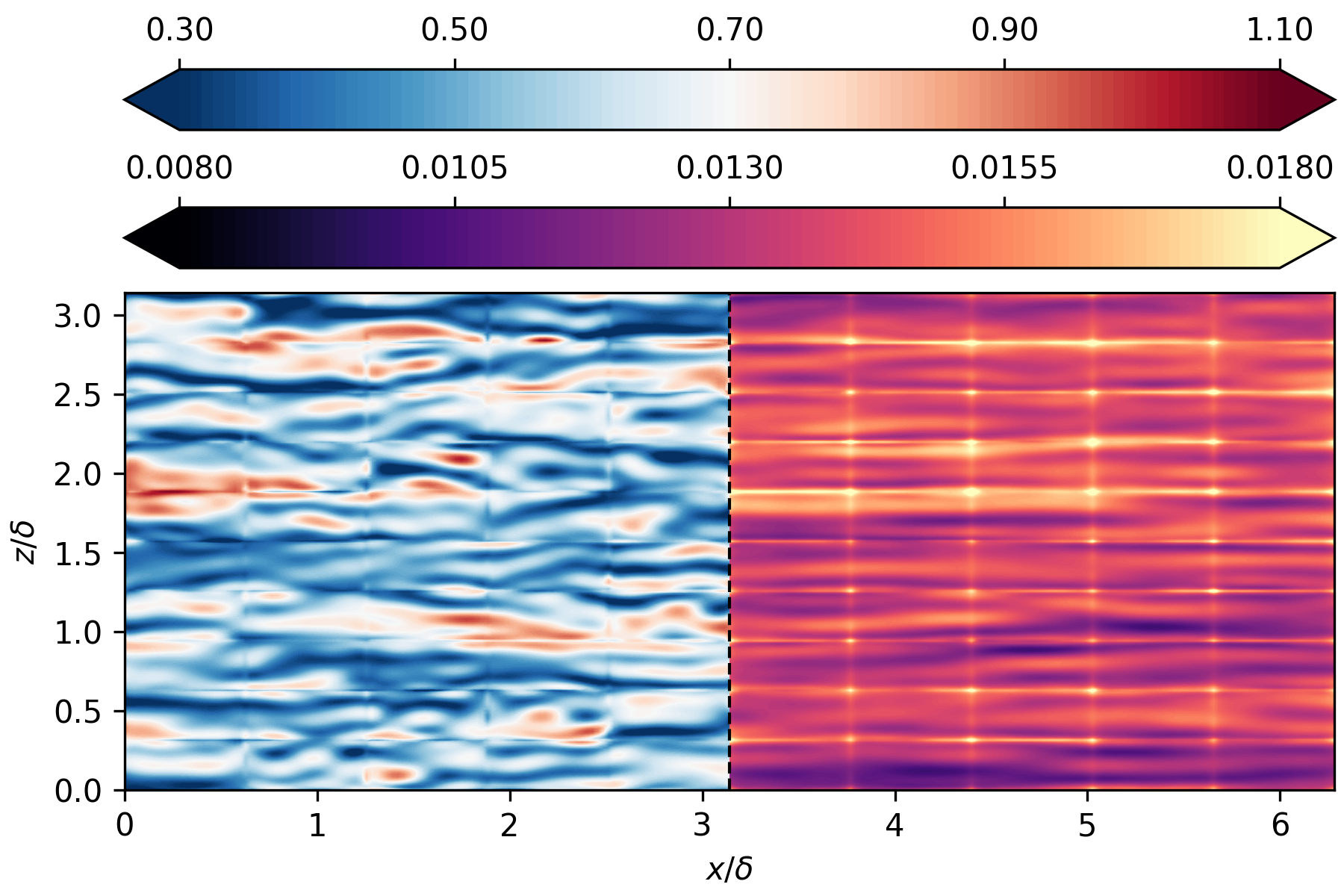}
            \caption{Sigma model, $a=0.0$}
            \label{fig:channel_Sigma_a0_field_yplus15}
        \end{subfigure}
        \begin{subfigure}{0.49\linewidth}
            \centering
            \includegraphics[width=\linewidth]{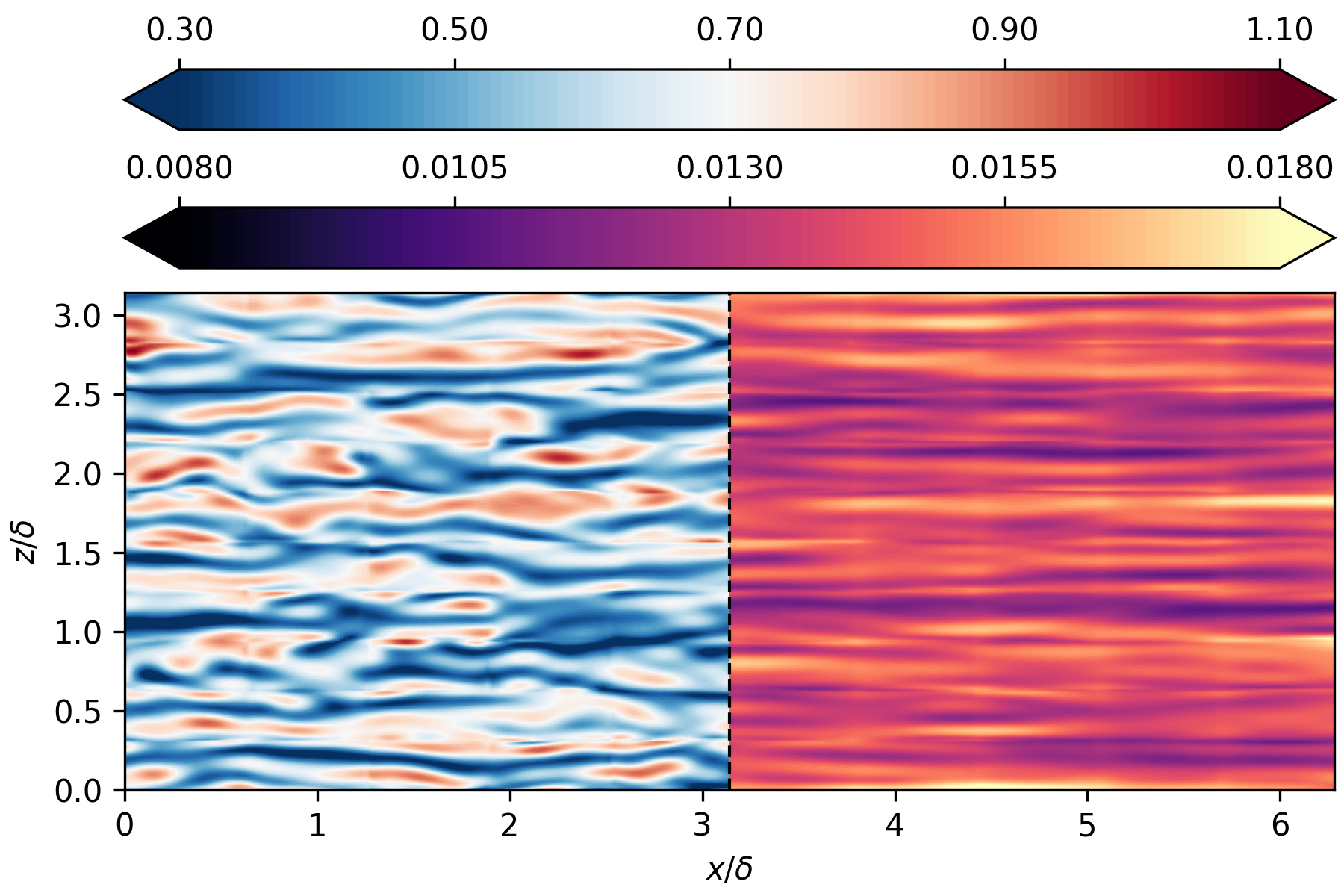}
            \caption{Sigma model, $a=0.8$}
            \label{fig:channel_Sigma_a08_field_yplus15}
        \end{subfigure}
        \centering
        \begin{subfigure}{0.49\linewidth}
            \centering
            \includegraphics[width=\linewidth]{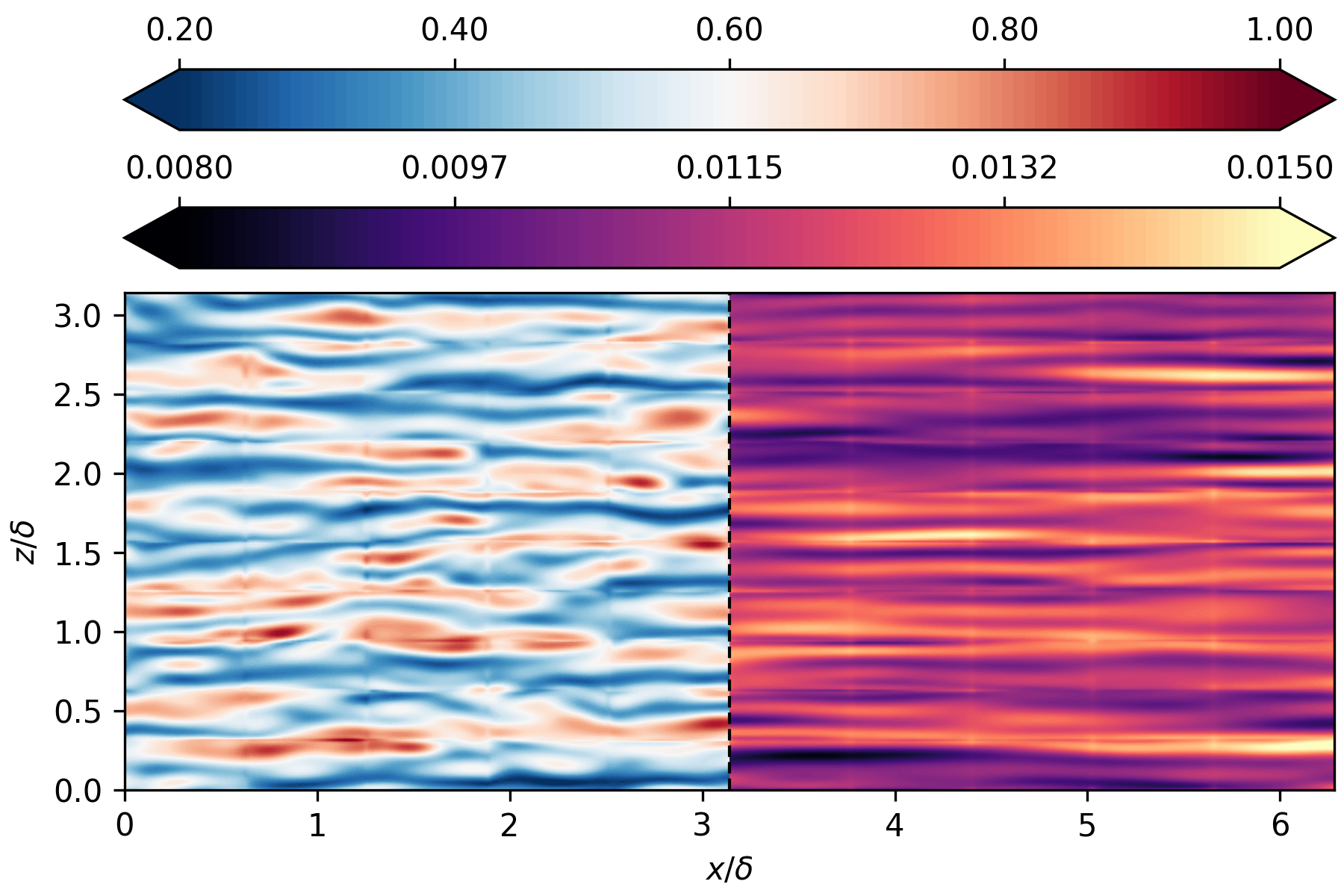}
            \caption{Vreman model, $a=0.0$}
            \label{fig:channel_Vreman_a0_field_yplus15}
        \end{subfigure}
        \begin{subfigure}{0.49\linewidth}
            \centering
            \includegraphics[width=\linewidth]{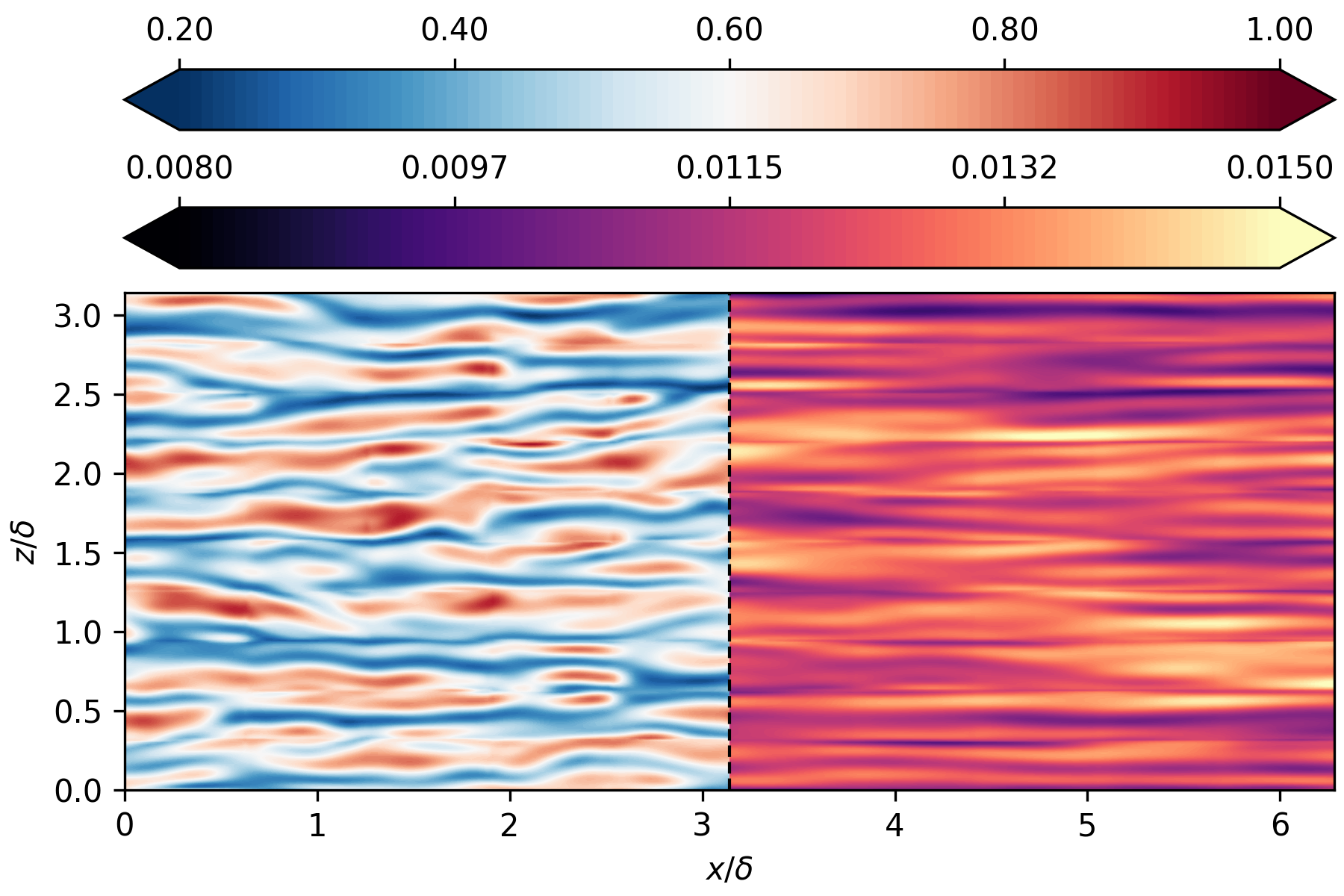}
            \caption{Vreman model, $a=0.8$}
            \label{fig:channel_Vreman_a08_field_yplus15}
        \end{subfigure}
        \caption{Velocity magnitude field scaled by the bulk velocity $\sqrt{U^2+V^2+W^2}/U_b$ (left) and the "short-period" TKE $\langle u^2+v^2+w^2\rangle_{\tau=100\delta/U_b}/2U_b^2$ (right) at $y^+\approx15$ of LES of turbulent channel flow at $Re_\tau=550$ using different SGS model with different GJP parameters. The range of the figures is a subset $x\in[0,2\pi],z\in[0,\pi]$ of the full domain.}
        \label{fig:channel_field_yplus15}
    \end{figure}

    In order to have a direct visual demonstration of the effect of GJP, we start with the instantaneous velocity magnitude field at a near wall height $y^+\approx15$, as shown in Figure \ref{fig:channel_field_yplus15}. Although the data obtained from the Sigma and Vreman models result in visually different fields when comparing the panels \ref{fig:channel_Sigma_a0_field_yplus15} and \ref{fig:channel_Vreman_a0_field_yplus15}, both models suffer from non-physical oscillations appearing at the element interfaces, with sizes close to the finest resolved scale. Similar to the periodic hill cases, a "short-period" TKE is calculated over a time period $\tau=100\delta/U_b$ and not averaging over spatial homogeneous directions for a better visualisation of the wiggles. We plot such field at the right half of the plotted subdomain in Figure \ref{fig:channel_field_yplus15} and observe some indications. In Figure \ref{fig:channel_Sigma_a0_field_yplus15} for example, high values of the "short-period" TKE emerge at the element interfaces, which maximises at the interfaces which connects $4$ elements, showing a sparkling pattern. This indicates the wiggles oscillates dramatically even beyond the level of the flow structures during this time period. Such pattern of element-related high values also emerges when applying Vreman model without GJP in Figure \ref{fig:channel_Vreman_a0_field_yplus15}, albeit with a weaker intensity and without an obvious sparkling pattern.
    
    We also notice that even adopting an averaged approach to estimate the filter size $\Delta$, the wiggles do not vanish, which means they are not solely the results of the artifacts of the non-uniform $\Delta$ distribution. Since in our CG-SEM formulation, the communication across elements merely aims to ensure the $C^0$ continuity of the field, the non-physical oscillations are included in the evolution of the turbulent flow due to the lack of a dissipation mechanism. By applying GJP with the proposed parameter $a=0.8$, the mesh-dependent wiggles are visually damped significantly, see the comparison of Figure \ref{fig:channel_Sigma_a08_field_yplus15} with \ref{fig:channel_Sigma_a0_field_yplus15} and Figure \ref{fig:channel_Vreman_a08_field_yplus15} with \ref{fig:channel_Vreman_a0_field_yplus15}. Particularly, for the Vreman model, the flow field using GJP is rather visually smooth which could be attributed to the strong dissipation from the SGS model itself. However, for the Sigma model, the wiggles are not fully dissipated by GJP using the proposed parameter $a=0.8$ as a weak element-related pattern still exits on the right half of Figure \ref{fig:channel_Sigma_a08_field_yplus15}.
 
    \subsection{Reynolds stresses}
\begin{figure}[ht]
        \centering
        \begin{subfigure}{0.4\linewidth}
            \centering
            \includegraphics[width=\linewidth]{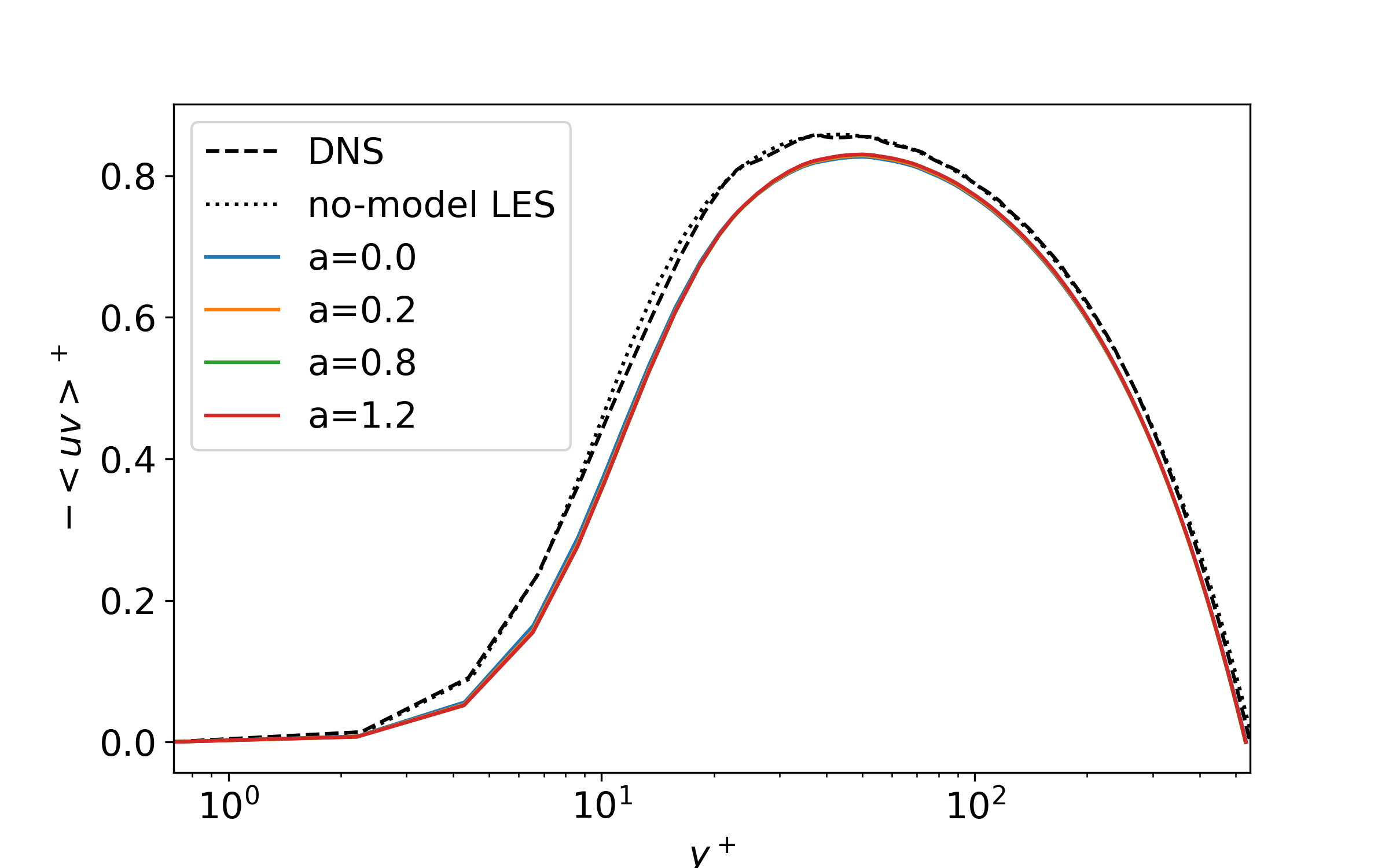}
            \caption{Sigma model}
            \label{fig:channel_Sigma_stats_uv}
        \end{subfigure}
        \begin{subfigure}{0.4\linewidth}
            \centering
            \includegraphics[width=\linewidth]{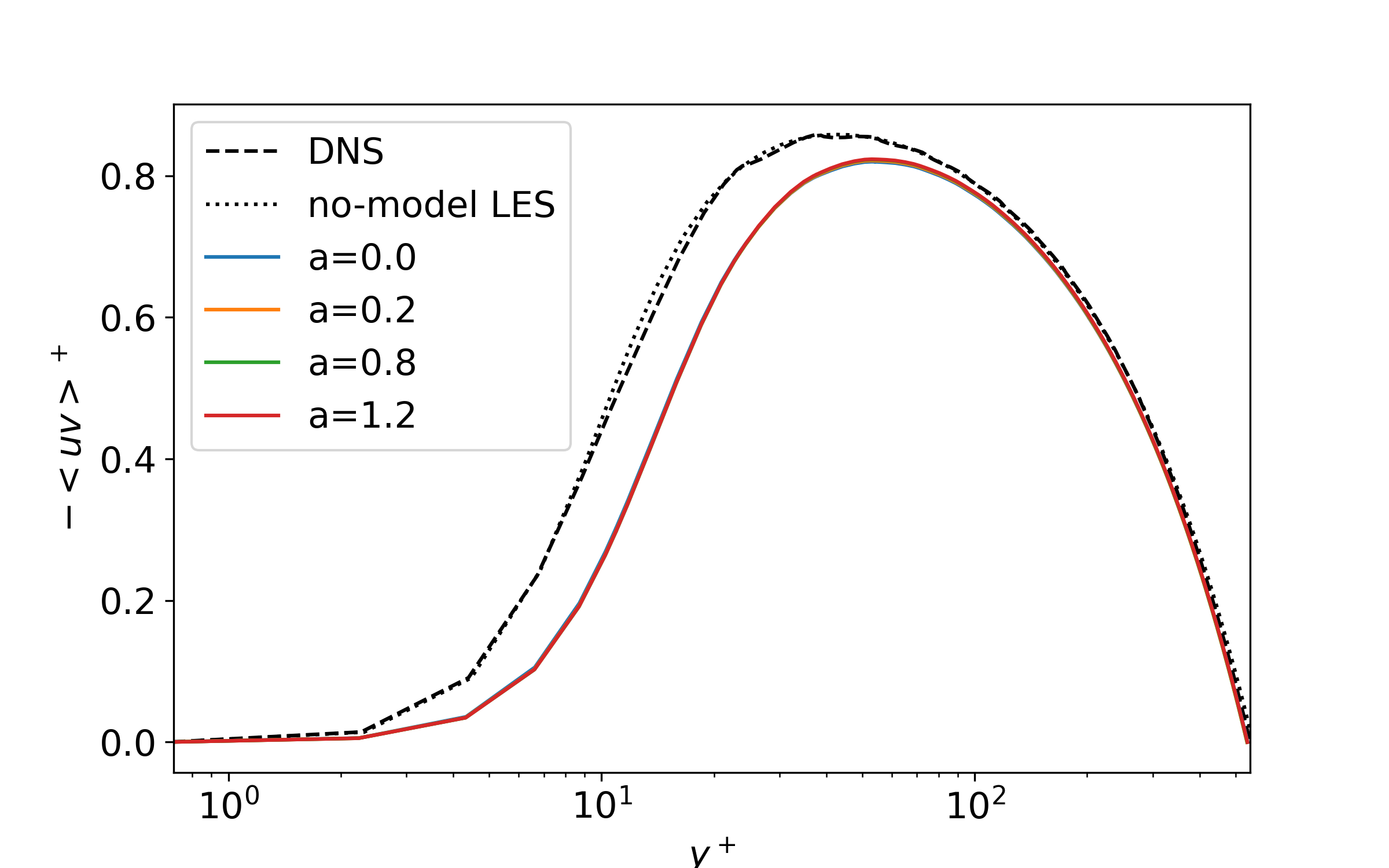}
            \caption{Vreman model}
            \label{fig:channel_Vreman_stats_uv}
        \end{subfigure}
        \caption{$\langle uv\rangle^+$ as a function of wall distance $y^+$ for LES of turbulent channel flow at $Re_\tau\approx550$, with $a \in [0.0, 0.4, 0.8, 1.2]$. No model LES pertains to a case with LES resolution but without active SGS model or GJP.}
        \label{fig:channel_stats_uv}
    \end{figure}

\begin{figure}[ht]
        \centering
        \begin{subfigure}{0.4\linewidth}
            \centering
            \includegraphics[width=\linewidth]{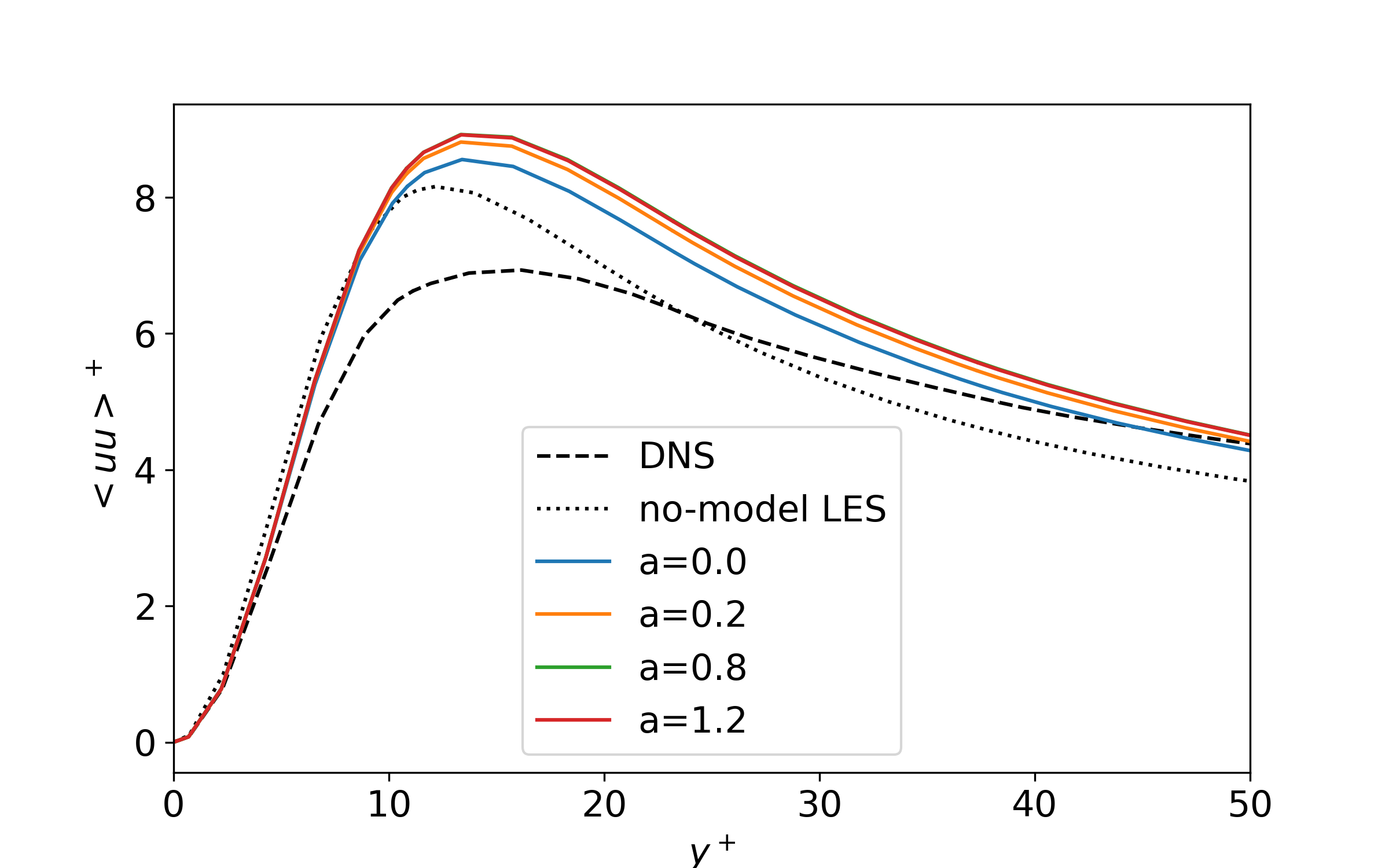}
            \caption{Sigma model, $y^+<50$}
            \label{fig:channel_Sigma_stats_uu_nw}
        \end{subfigure}
        \begin{subfigure}{0.4\linewidth}
            \centering
            \includegraphics[width=\linewidth]{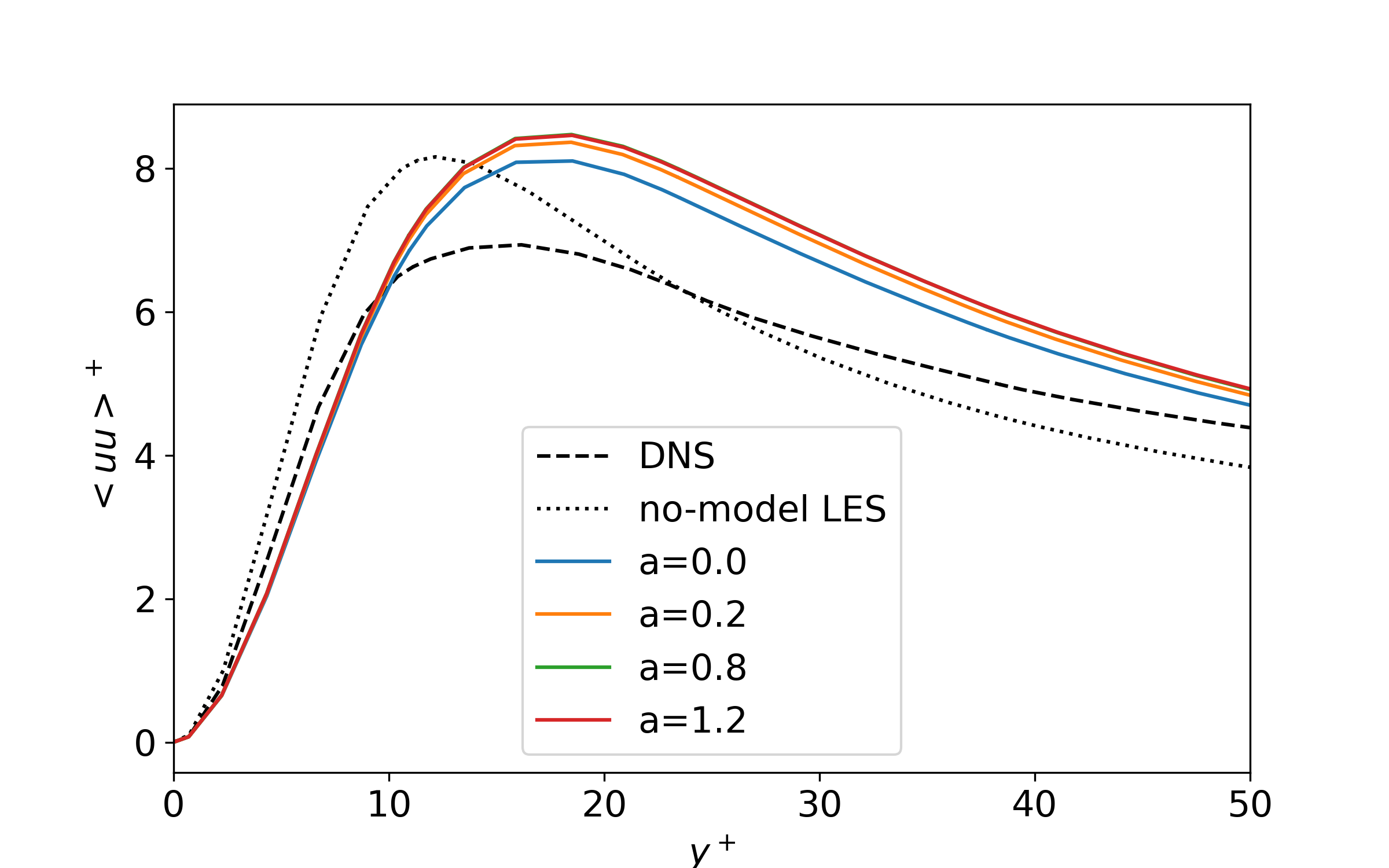}
            \caption{Vreman model, $y^+<50$}
            \label{fig:channel_Vreman_stats_uu_nw}
        \end{subfigure}
        \begin{subfigure}{0.4\linewidth}
            \centering
            \includegraphics[width=\linewidth]{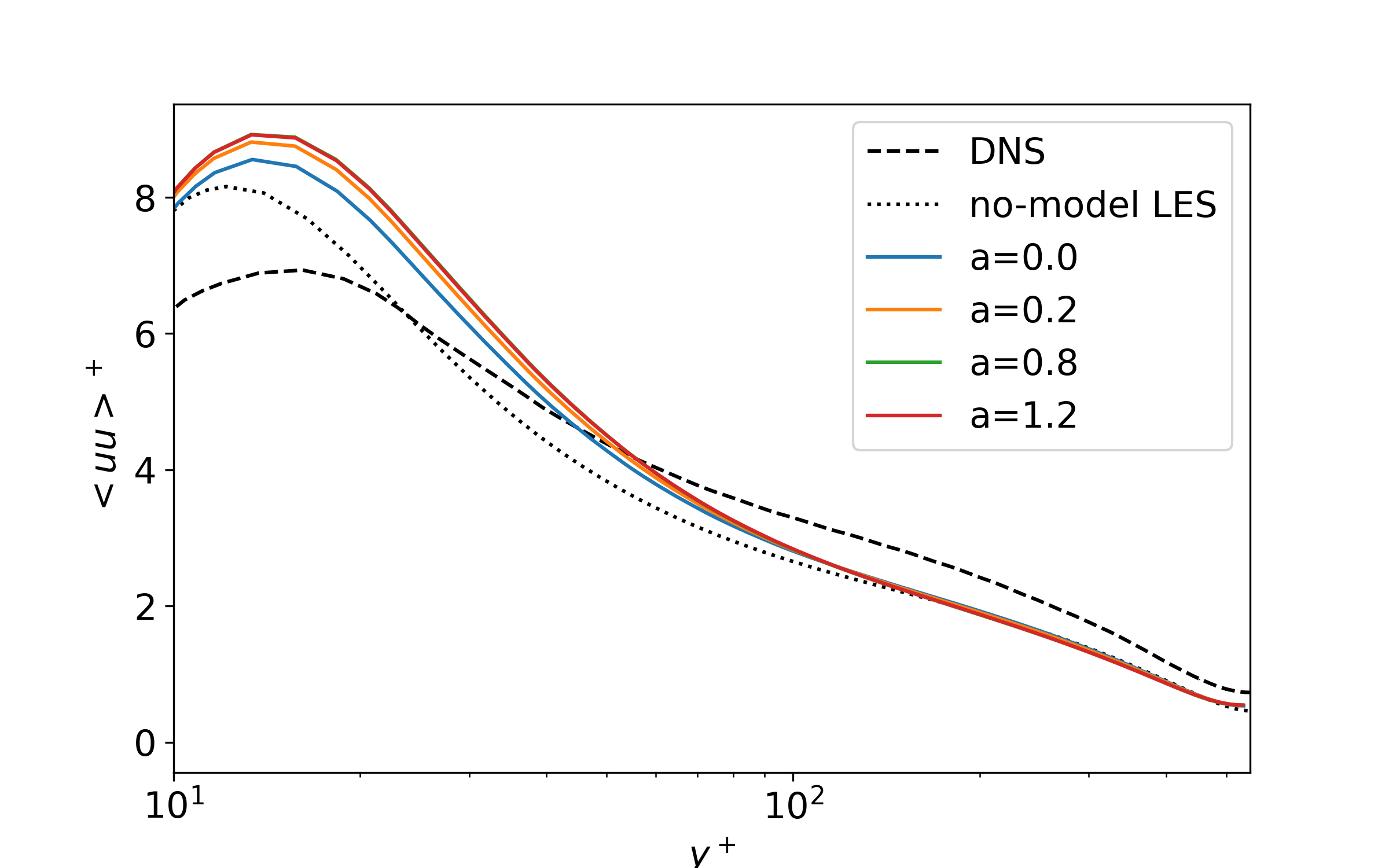}
            \caption{Sigma model, $y^+>10$}
            \label{fig:channel_Sigma_stats_uu_ct}
        \end{subfigure}
        \begin{subfigure}{0.4\linewidth}
            \centering
            \includegraphics[width=\linewidth]{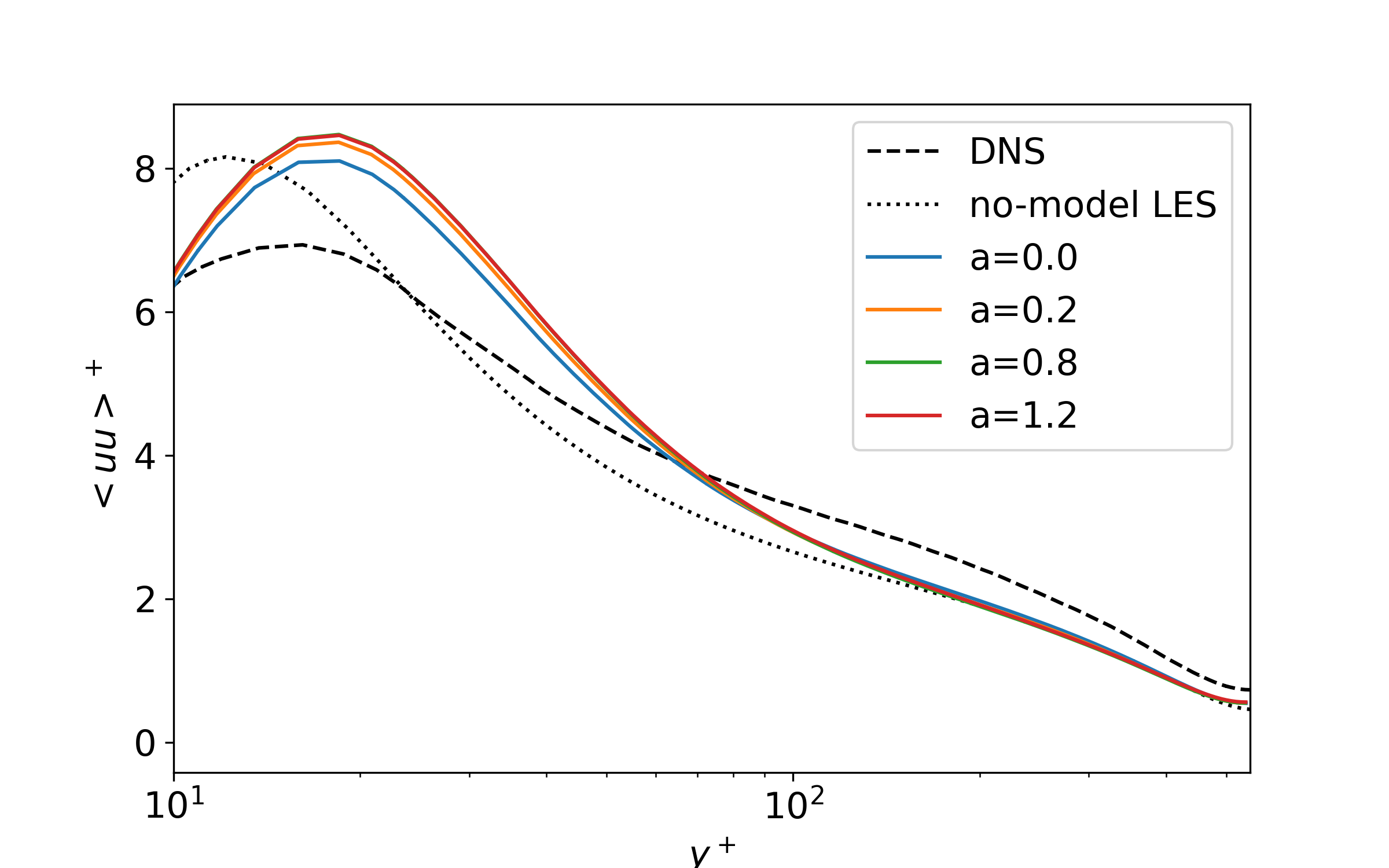}
            \caption{Vreman model, $y^+>10$}
            \label{fig:channel_Vreman_stats_uu_ct}
        \end{subfigure}
        \caption{$\langle uu\rangle^+$ v.s. $y^+$of LES of turbulent channel flow at $Re_\tau\approx550$, with $a \in [0.0, 0.2, 0.8, 1.2]$}
        \label{fig:channel_stats_uu}
    \end{figure}

    To investigate the influence of the SGS models and the GJP to the flow statistics, Reynolds stress components $\langle uv\rangle^+$ and $\langle uu\rangle^+$ are plotted against $y^+$ in Figure \ref{fig:channel_stats_uv} and \ref{fig:channel_stats_uu}, respectively. The former reflects the intensity of turbulent transport of momentum in the wall normal direction while the latter reflects the intensity of spatial fluctuation. Note that the plus unit scaling is based on $\delta_\nu$ and $u_\tau$ obtained from the respective LES. From Figure \ref{fig:channel_stats_uv}, both SGS models gives weaker vertical turbulent momentum transport comparing with DNS and the no-model LES case, which could be mainly attributed to the SGS motions being modelled as an additional viscous term and the corresponding fluctuating transport thus being reduced. The GJP does not affect the vertical turbulent momentum transport. 

    On the other hand, for $<uu>^+$, the discrepancy due to GJP is more obvious, see Figure \ref{fig:channel_stats_uu}. In the near wall region, see Figure \ref{fig:channel_Sigma_stats_uu_nw} and \ref{fig:channel_Vreman_stats_uu_nw}, all results obtained on the LES grid have stronger fluctuations compared with DNS. It is also observed in~\cite{Fröhlich_Rodi_2002}. They attributed it to the lack of adequate resolution in all three directions near the wall hence a wall model is needed. In addition, all explicit SGS models give a stronger inner peak than the no-model LES at $y^+ \approx 10-20$, which is not expected. By applying GJP in addition to the SGS models, the fluctuation amplitudes get even stronger. However, in the region farther away from the wall, $\langle uu\rangle^+$ from all LES collapse onto the same magnitude. This could be attributed to the smoothness of the flow close to the center of the channel, which gives weaker strain rate and thus gradient jump across the element interfaces. Therefore the SGS models and the GJP does not play a big role; however the discrepancy to the DNS is still noteworthy.
    
    \subsection{Power spectral density} \label{sec:turb_channel_550_PSD}
\begin{figure}[ht]
        \centering
        \begin{subfigure}{0.32\linewidth}
            \centering
            \includegraphics[width=\linewidth]{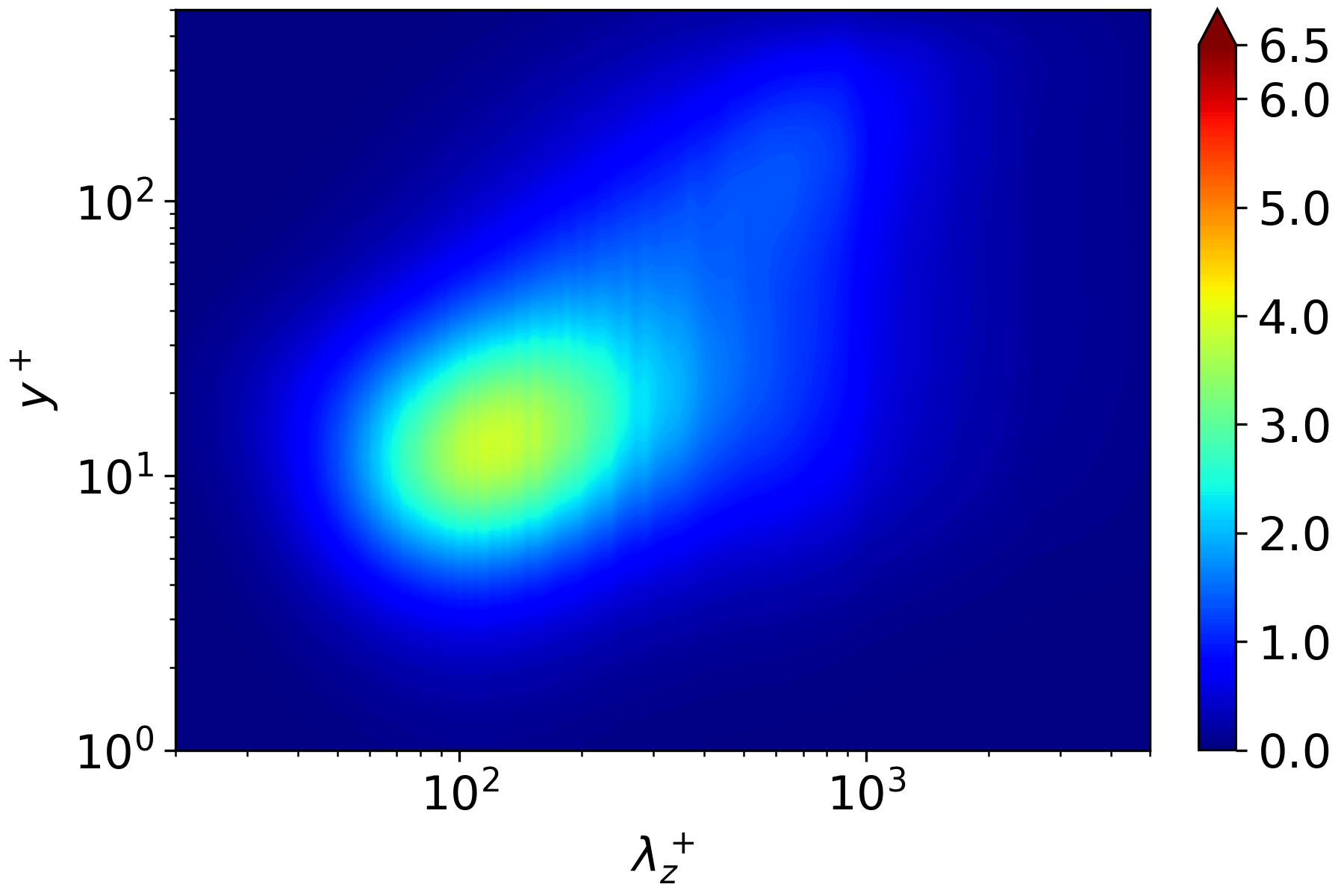}
            \caption{DNS}
            \label{fig:psd_DNS}
        \end{subfigure}
        \begin{subfigure}{0.32\linewidth}
            \centering
            \includegraphics[width=\linewidth]{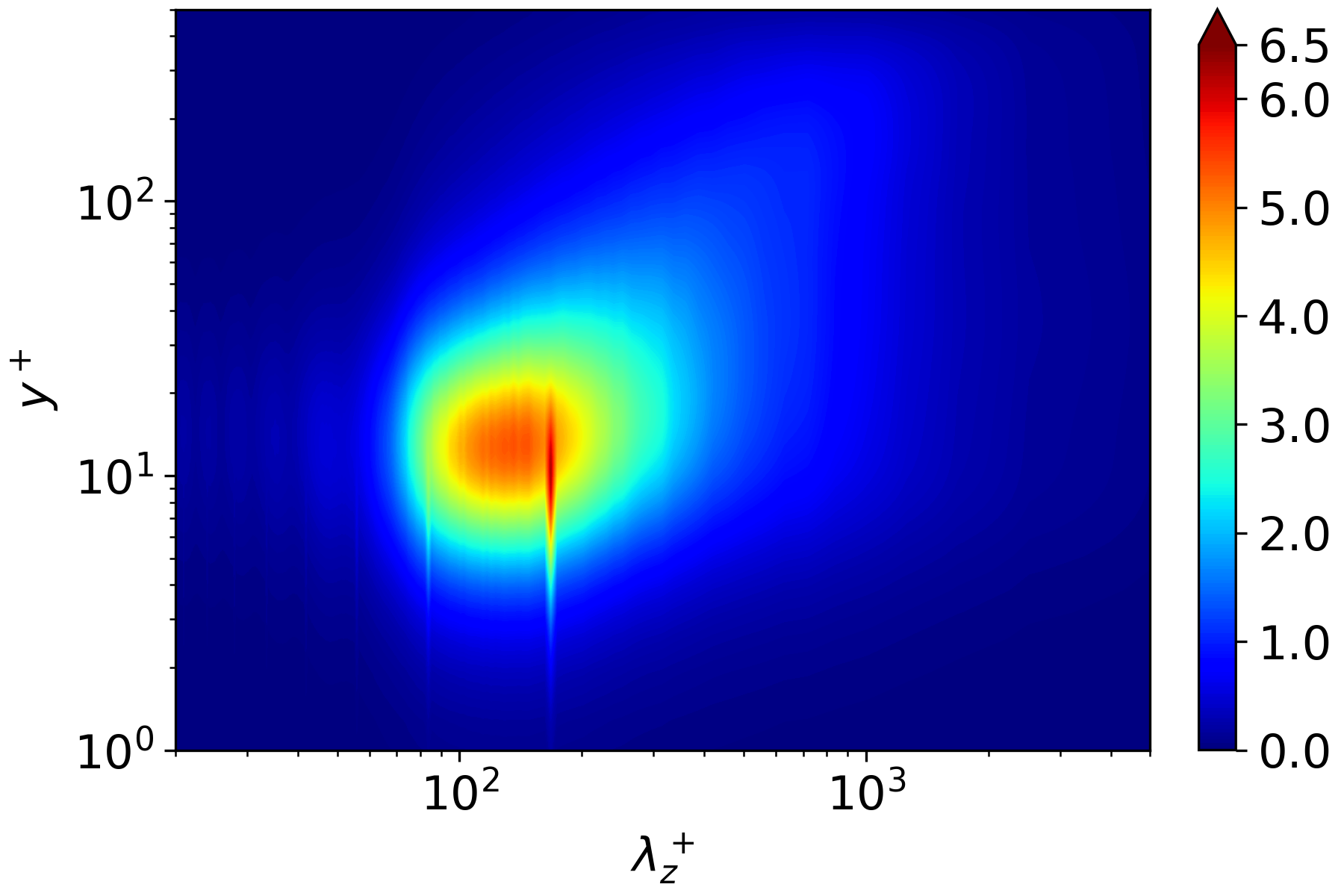}
            \caption{Sigma model, $a=0.0$}
            \label{fig:psd_Sigma_a0}
        \end{subfigure}
        \begin{subfigure}{0.32\linewidth}
            \centering
            \includegraphics[width=\linewidth]{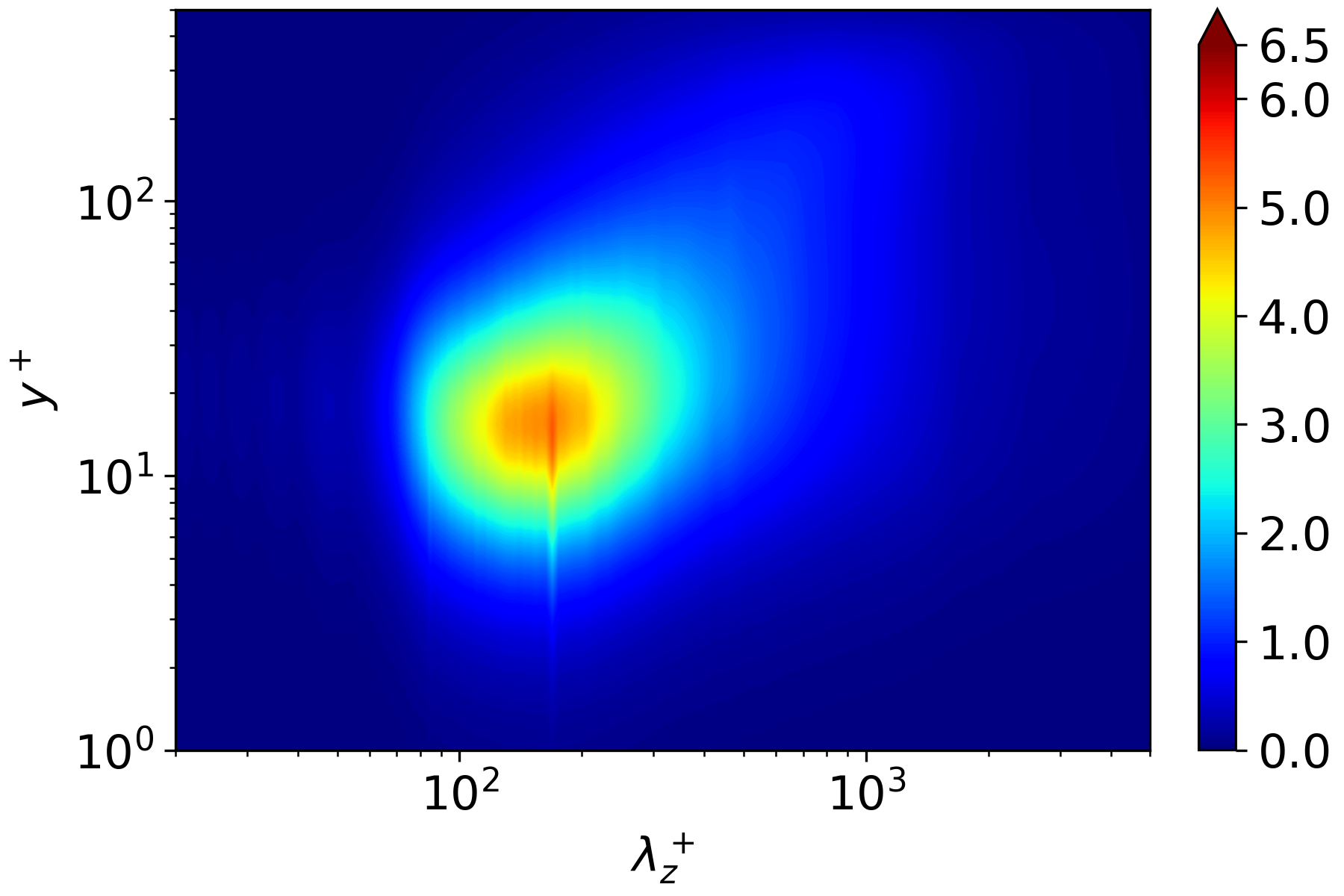}
            \caption{Vreman model, $a=0.0$}
            \label{fig:psd_Vreman_a0}
        \end{subfigure}
        \begin{subfigure}{0.32\linewidth}
            \centering
            \includegraphics[width=\linewidth]{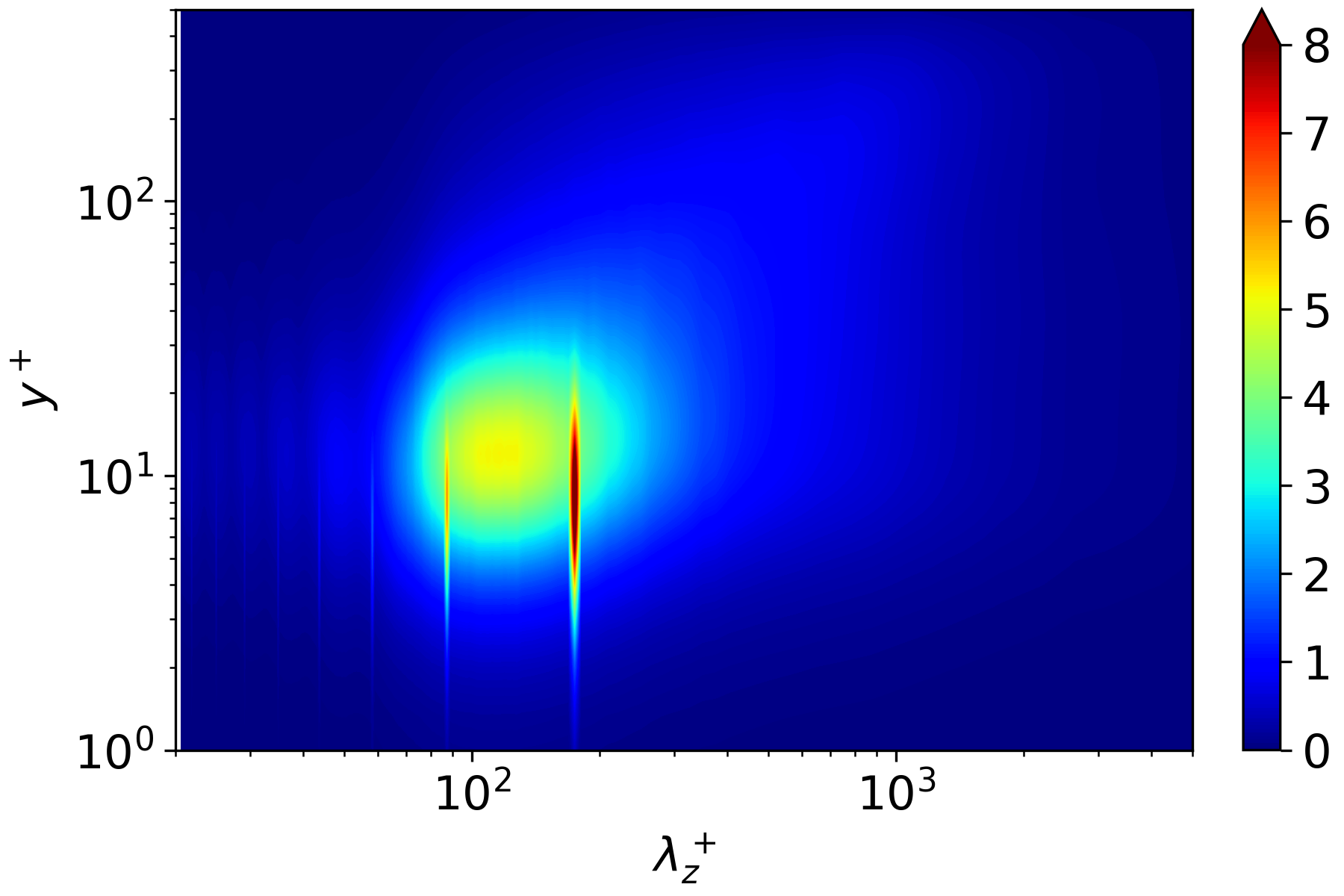}
            \caption{no-model LES}
            \label{fig:psd_nmLES}
        \end{subfigure}
        \begin{subfigure}{0.32\linewidth}
            \centering
            \includegraphics[width=\linewidth]{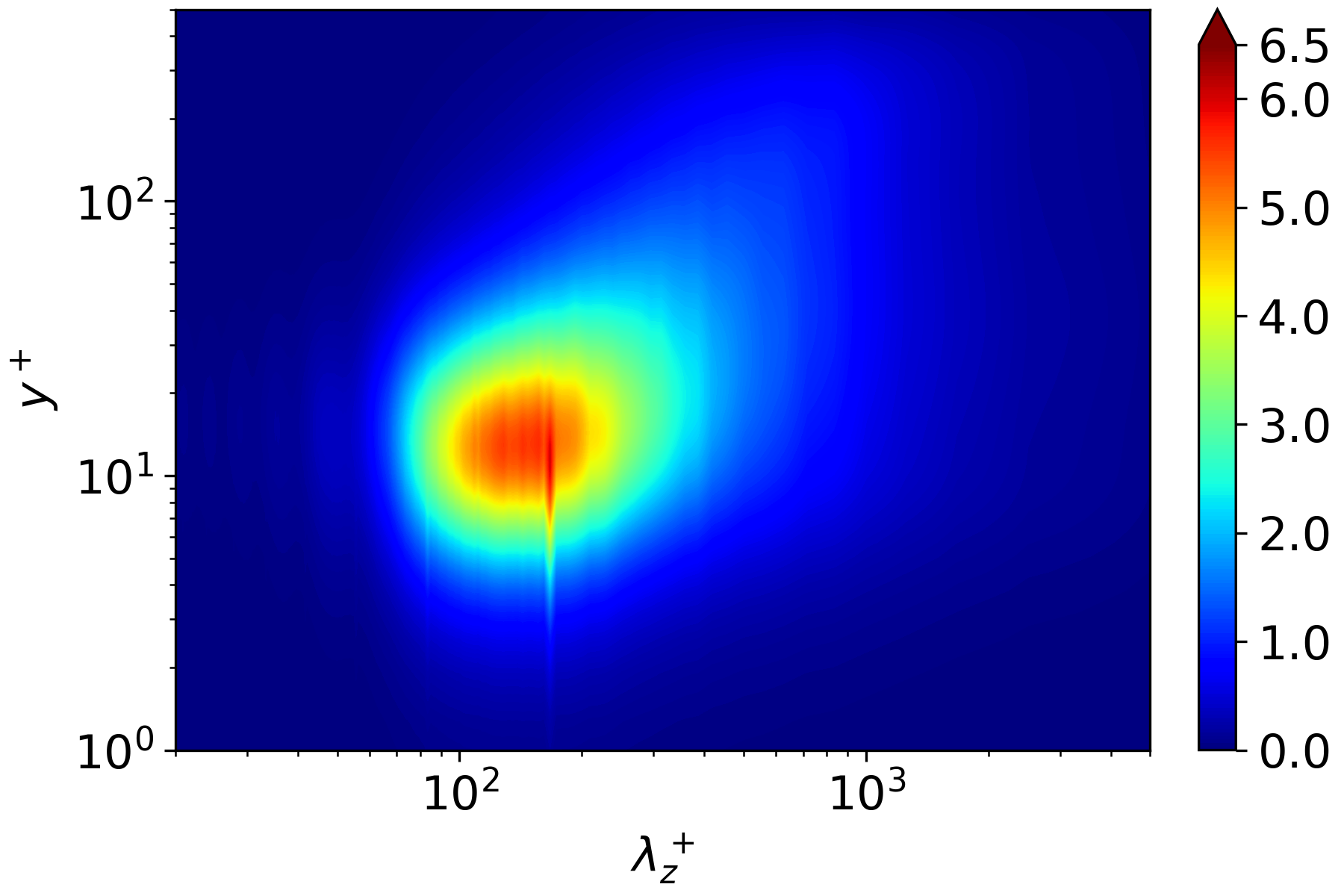}
            \caption{Sigma model, $a=0.8$}
            \label{fig:psd_Sigma_a08}
        \end{subfigure}
        \begin{subfigure}{0.32\linewidth}
            \centering
            \includegraphics[width=\linewidth]{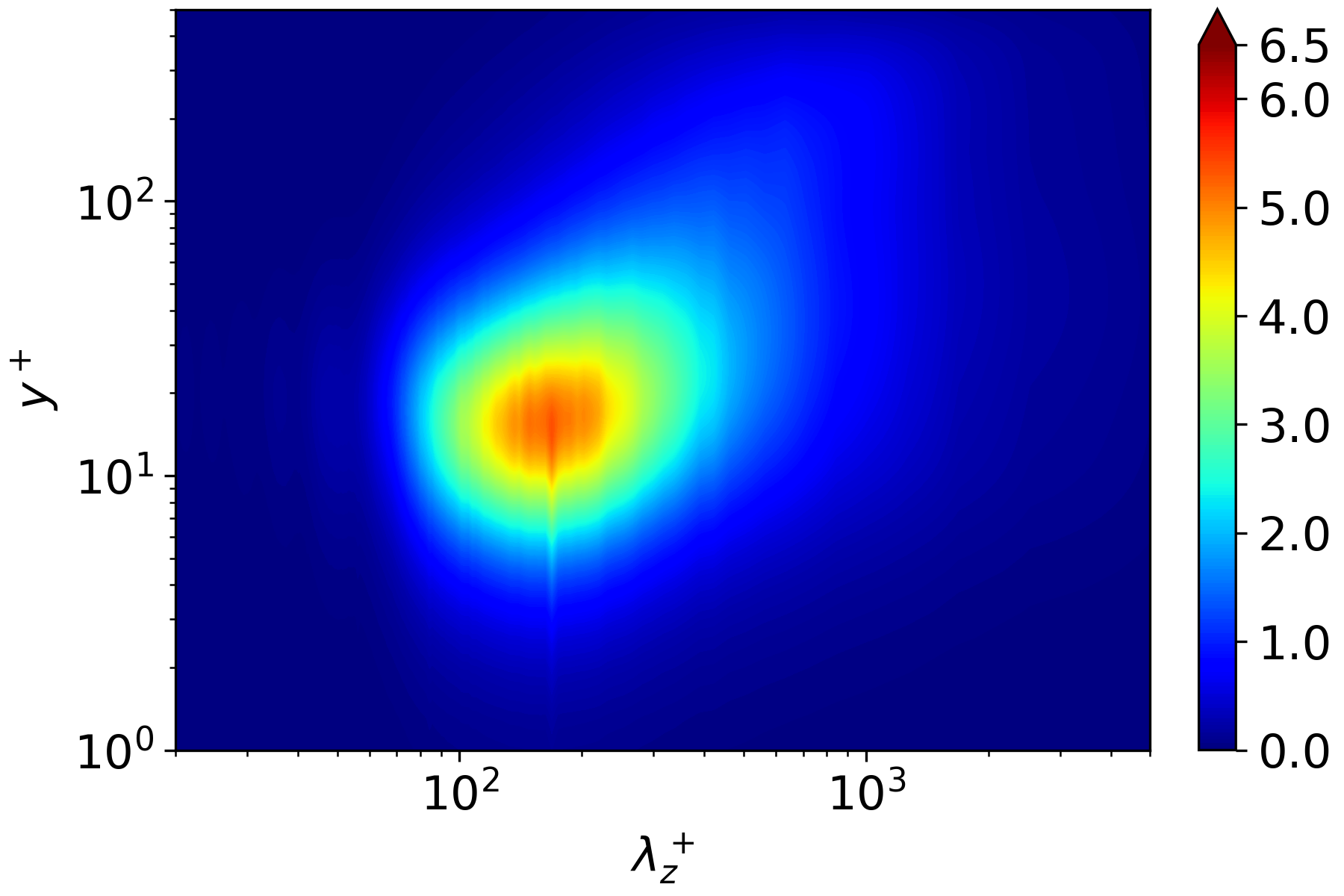}
            \caption{Vreman model, $a=0.8$}
            \label{fig:psd_Vreman_a08}
        \end{subfigure}
        \caption{Pre-multiplied PSD of $u$ in terms of the spanwise wave length $\lambda_z^+$ and the wall distance in plus unit $y^+$, \emph{i.e.,} $E_{uu}(k_z, y^+)\cdot k_z /u_\tau^2$ of turbulent channel flow at $Re_\tau=550$}
        \label{fig:psd}
    \end{figure}

    The above Reynolds stress profiles show no wiggles because the flow is well resolved to a DNS level in the wall-normal direction for all LES cases; the averaging performed over the homogeneous wall-parallel directions will smoothen any oscillations in those directions. In order to show the wiggles quantitatively, we present the premultiplied power spectral density (PSD) in the spanwise direction $E_{uu}(k_z, y^+)\cdot k_z /u_\tau^2$ in Figure \ref{fig:psd}. As commonly done, we show the spectra as a function of  wavelength $\lambda_z^+$  instead  of wavenumber $k_z^+$.  %
    Panel \ref{fig:psd_DNS} shows the well-known so-called inner peak at $\lambda_z^+ \approx 100-200$ and $y^+ \approx 15$. In panels \ref{fig:psd_Sigma_a0} to \ref{fig:psd_Vreman_a08}, the inner peak contains more power for all LES cases. This result is consistent with the stronger peak of $<uu>^+$ in Figure \ref{fig:channel_stats_uu}. 
    
\begin{figure}[ht]
        \centering
        \begin{subfigure}{0.4\linewidth}
            \centering
            \includegraphics[width=\linewidth]{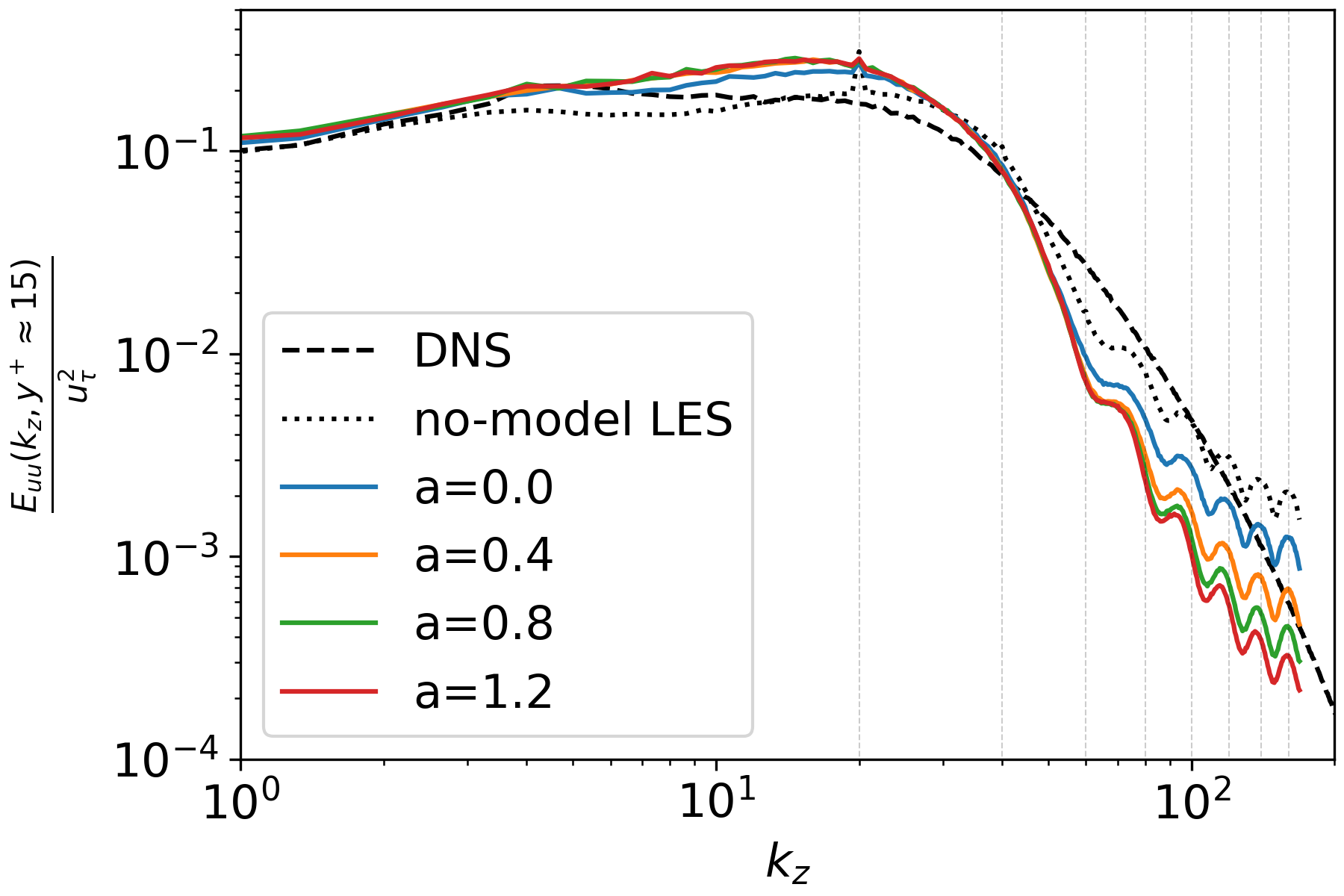}
            \caption{Sigma model}
            \label{fig:psd_1d_kz_Sigma}
        \end{subfigure}
        \begin{subfigure}{0.4\linewidth}
            \centering
            \includegraphics[width=\linewidth]{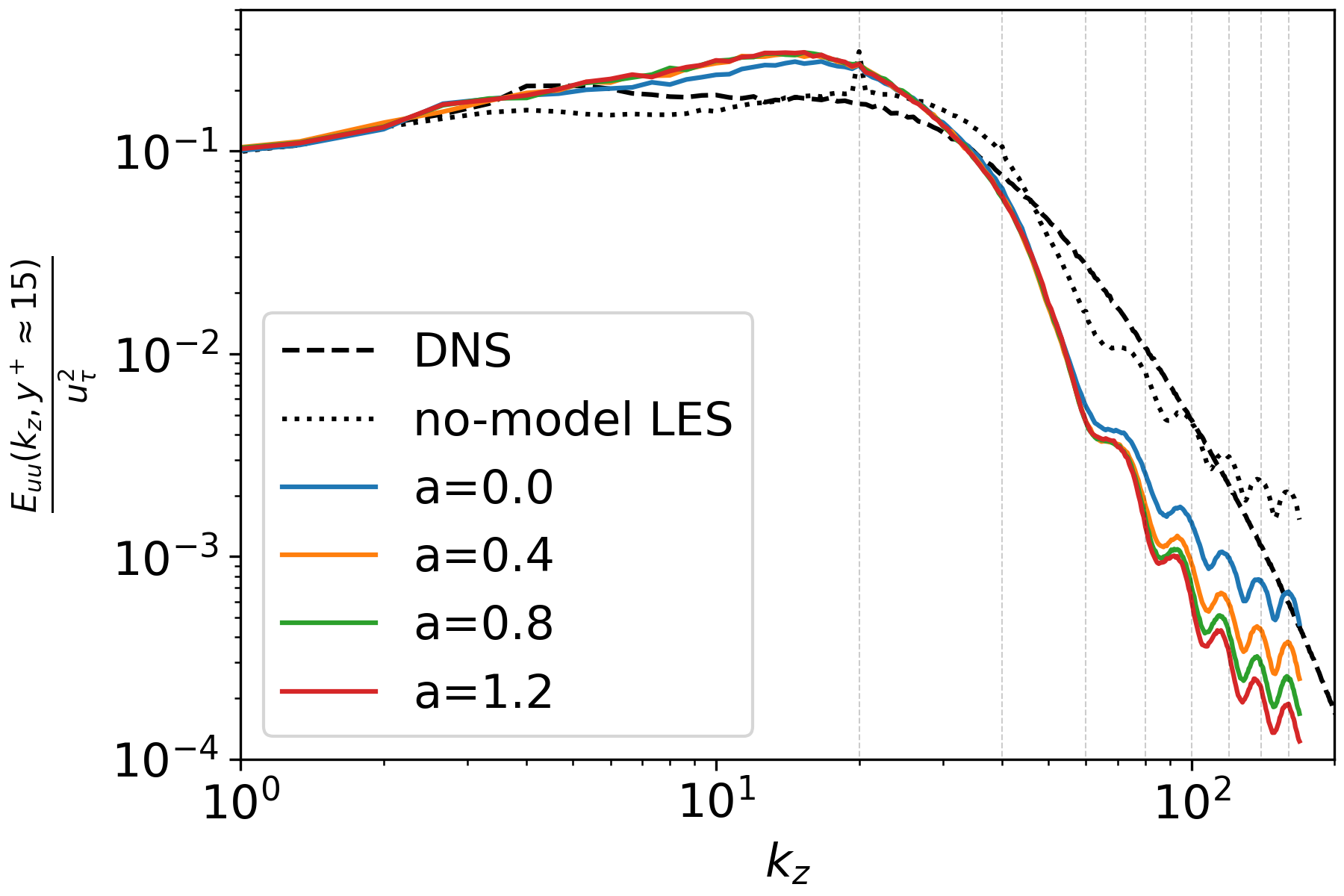}
            \caption{Vreman model}
            \label{fig:psd_1d_kz_Vreman}
        \end{subfigure}
        \caption{PSD of $u$ in terms of the spanwise wave number $k_z$ at the inner peak $y^+ \approx 15$, \emph{i.e.,} $E_{uu}(k_z, y^+\approx15) /u_\tau^2$ of turbulent channel flow at $Re_\tau=550$, with $a \in [0.0, 0.4, 0.8, 1.2]$.}
        \label{fig:psd_1d_kz}
    \end{figure}

    Apart from the general magnitude, one should also expect to see the evidence for the wiggles from the pre-multiplied PSD, recalling the Fourier transform of a pulsating signal with a constant frequency is a spike corresponding to its frequency, and other spikes corresponding to the harmonics as shown in Figure \ref{fig:tgv_spectrum}. The observed wiggles in the flow field are presented in the same way in the pre-multiplied PSD as the spikes in Figure \ref{fig:psd_nmLES} at $\lambda_z^+\approx180, 90, 45, ...$ with a decreasing trend in magnitude with lower wavelengths $\lambda_z^+$. Note that panel \ref{fig:psd_nmLES} uses a different colorbar compared to the rest, showing the strongest spike having much higher amplitude than the neighbouring wavelengths.     
    In panels \ref{fig:psd_Sigma_a0}, \ref{fig:psd_Vreman_a0}, \ref{fig:psd_Sigma_a08} and \ref{fig:psd_Vreman_a08}, the spikes in the pre-multiplied PSD are not fully dissipated by GJP and the quantitative differences are hard to tell by a contour plot. Therefore, the PSD $E_{uu}(k_z, y^+)/u_\tau^2$ is plotted in Figure \ref{fig:psd_1d_kz} as profiles at $y^+ \approx 15$. As one can see from the LES results with SGS models ($a=0.0$), the high wavenumbers are dampened as expected but for the most energetic part $k_z \approx 7-40$, the SGS models increase the energy if comparing with no-model LES solution. For both no-model LES and LES with SGS model (but no GJP cases), the energy on the finest scale exceeds the one from DNS, which is not expected since the solution from LES should reflect the filtered flow field. By applying GJP with increasing magnitude, the high-wavenumber motions are damped more and to a level lower than DNS. However, for low-wavenumber motions, the GJP cases show increased energy. Effectively, GJP and SGS models affect the flow in the same way in a simulation using CG-SEM: dissipate more energy at small scales but raise the energy at large scales.
    
\begin{figure}[ht]
        \centering
        \begin{subfigure}{0.4\linewidth}
            \centering
            \includegraphics[width=\linewidth]{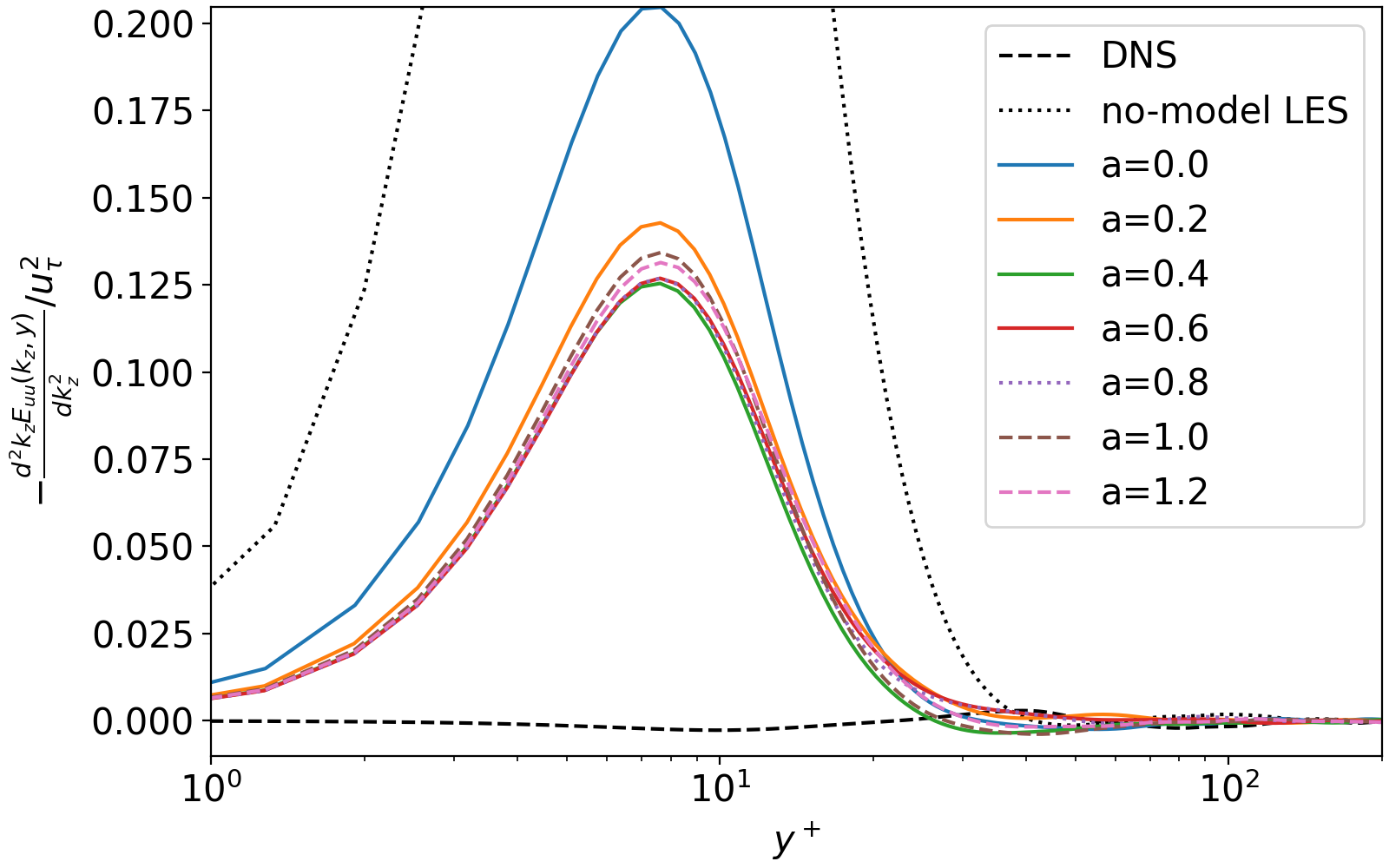}
            \caption{Sigma model}
            \label{fig:psd_spike_jump_Sigma}
        \end{subfigure}
        \begin{subfigure}{0.4\linewidth}
            \centering
            \includegraphics[width=\linewidth]{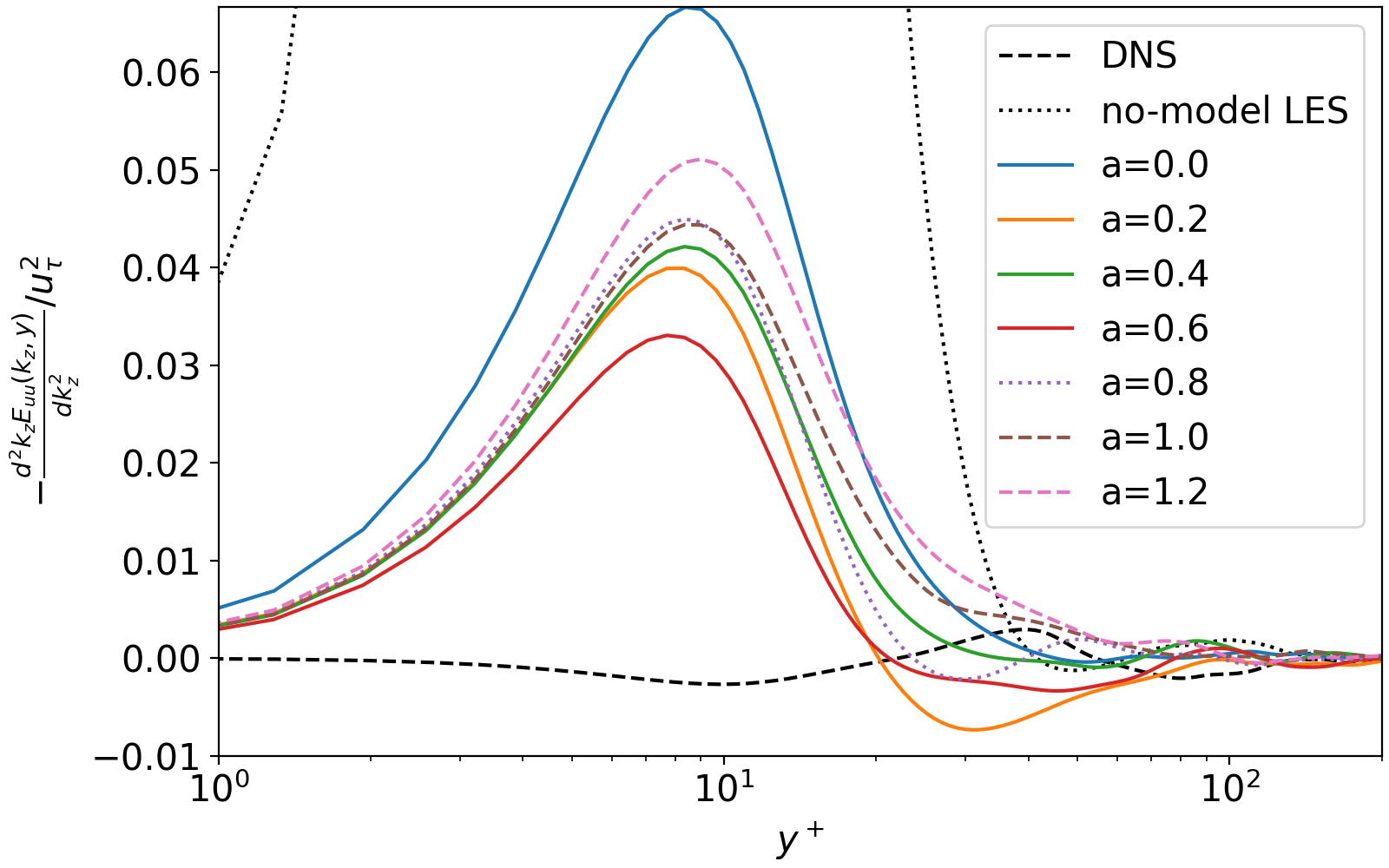}
            \caption{Vreman model}
            \label{fig:psd_spike_jump_Vreman}
        \end{subfigure}
        \caption{The negative second derivative (curvature) of the PSD of $u$ at the strongest spike in terms of the wall distance in plus unit $y^+$, \emph{i.e.,} $\left\{\frac{\partial^2 E_{uu}(k_z, y^+)}{\partial k_z ^2} /u_\tau^2\right\}|_{k_z = k_{z,\rm spike}}$ of turbulent channel flow at $Re_\tau=550$, with $a \in [0.0, 0.2, 0.4, 0.6, 0.8, 1.0, 1.2]$}
        \label{fig:psd_spike_jump}
    \end{figure}

    Apart from the spikes appearing at the element spacing and its higher harmonics, another key concern is the sharpness of the strongest spike in the PSD, or quantitatively speaking, the negative second derivative of the PSD $\left\{-\frac{\partial^2 E_{uu}(k_z, y^+)}{\partial k_z ^2} /u_\tau^2\right\}|_{k_z = k_{z,\rm spike}}$. By using SGS models, the spike is much smoother if comparing $a=0.0$ with "no-model LES" in Figure \ref{fig:psd_spike_jump}. By further applying a proper GJP, the sharpness of the strongest spike decreases significantly. For the Sigma model, it at most reduces the negative second derivative of the PSD from $0.2$ to around $0.125$ for $a=0.4, 0.6$ and $0.8$ with $a=0.4$ to be the optimal; while for the Vreman model, the negative second derivative of the PSD is decreased from above $0.6$ to around $0.3$ by choosing $a=0.6$. But one should also notice that even the best choise for $a$ does not reduce the sharpness of the peak to the low levels given by the reference DNS.
    
From the above consideration in channel flow, we can conclude that LES in SEM, performed at typical LES resolutions, gives noticeable spectral peaks, in particular spanwise directions, corresponding to the element boundaries. These are not sufficiently damped by simply employing an SGS model. GJP clearly help in reducing the sharpness of these spurious spectral peaks, however, given the present formulation cannot reduce them completely. It is important to observe that GJP is not affecting the overall energy distribution, making it an effective way to reduce the spurious oscillations.

\section{Conclusions and Outlook}\label{sec:conclusion}
    This work discusses the possibility to extend the previous application of the gradient jump penalisation (GJP) from the iLES framework into the wall-resolved LES (WRLES) regime. We proceed by analysing three representative cases: the Taylor--Green vortex (TGV) flow, the periodic hill flow and canonical turbulent channel flow. The first case involves transition to turbulence and illustrates the smoothing effect of GJP in a typical three-dimensional flow with usual LES resolution when combined with an active SGS model. 
    The latter two cases are fully developed turbulent flows simulated in the WRLES regime, where the Sigma and Vreman model are adopted as the active SGS models.
    
    In the TGV case, the GJP is coupled with the classical Smagorinsky model. An obvious smoothening effect of GJP is observed on wiggles appearing close to the element interfaces. Via an analysis of the power spectra, the smoothening mechanism is revealed to be damping the energy on small scales, thus making the wiggles less prominent. For the integral quantities such as kinetic energy and enstrophy, the GJP preserves them to a large extent.

    For the periodic hill flow case, GJP smoothens the apparent wiggles visible in  the instantaneous flow fields significantly. This localised dissipative effect on non-physical oscillations is particularly evident in the TKE contours but behaves differently with different SGS models. With the Sigma model, the effect is primarily associated with a reduction of TKE in the wiggle regions, whereas with the Vreman model, it is linked to an increase of TKE within the element interior. Moreover, from the instantaneous fields, the adoption of GJP leads to a slight loss of flow detail, related to the elimination of the small scales close to the elemental interfaces.
    
    Given this observation, we further analyse a canonical fully developed turbulent channel flow case with GJP of different amplitudes. In the well-resolved wall-normal direction, the smooth Reynolds shear stress is not changed by GJP, which means the GJP does not affect the relevant momentum transport in the well-resolved direction. However, the Reynolds normal stress $\langle uu\rangle^+$ is significantly increased by applying GJP, implying that the GJP does not only dissipate the wiggles locally but also influences the velocity field globally in an undesirable way. To investigate the smoothing mechanism of the GJP in more detail, the power spectral density (PSD) is computed and analysed. From the PSD in terms of $\lambda_z^+$ and $y^+$ of the WRLES of the turbulent channel flow, spikes corresponding to the spanwise element spacing and its higher harmonics are observed. The spike corresponding to the spanwise element interval contains the strongest power, and the harmonics decays quickly by applying an active SGS model, and the decay is even steeper by applying stronger GJP. The sharpness of the first spike, quantified by the second  derivative over spanwise wavenumber $k_z$ of the PSD, is significantly alleviated by applying active SGS model and a GJP with a proper intensity, which is found different from the optimal one in~\cite{Moura2022GJP}. For the other wavenumbers not showing a spike, a stronger GJP provides more dissipation for high wavenumbers but insufficient dissipation for low wavenumbers.

    These conclusions suggest that GJP is indeed an effective way to reduce wiggles in SEM appearing across element interfaces. In particular together with an active SGS model, necessary at typical LES resolutions, GJP provides additional targetted damping of spurious oscillations. However, the present formulation does not provide sufficient damping to remove the element-to-element oscillations completely, as evidenced in particular by the spectral decomposition in homogeneous directions. In addition, the effects of the SGS model are affected in such as way that the spectral energy at lower wavenumbers is generally increased (\emph{i.e.,} the dissipation due to the SGS model is reduced).  Therefore, for successful LES using an SEM-type discretisation, additional measures need to be taken. One potential approach would be to consider explicitly filtered LES using a filter with filter width related to the largest spacing in each element. In this way, the smaller scales close to the element boundaries are likely to be suppressed, and the solution is restrained to the scales that may be resolved everywhere. This approach is however not considered in the present work, and will be explored in future research.

    Further work may also include a dynamic model for the GJP coefficient according to how well the turbulence is resolved. Another improvement could be made on an additional dissipation mechanism on low wavenumbers. Further investigation on GJP under wall-modelled LES (WMLES) is also necessary where typically the resolution of turbulence is even lower. Cases such as atmospheric turbulence could be relevant flow cases to study \cite{Huusko2025}.

\bmhead{Acknowledgements}
SD acknowledges financial support by the Swedish e-Science Research Centre (SeRC) as part of the M3 programme.
This work was  partially funded by the European
Union.  This  work  has  received   funding  from  the  European  High
Performance  Computing Joint  Undertaking  (JU)  and Sweden,  Germany,
Spain, Greece and Denmark under grant agreement No 101093393. The post-processing in this work is based on PySEMTools~\cite{PySEMTools}. The computations were enabled by resources provided by the National Academic Infrastructure for Supercomputing in Sweden (NAISS), partially funded by the Swedish Research Council through grant agreement no. 2022-06725

\appendix

\section{Basics of continuous Galerkin spectral element method}\label{appendix:A}
We base the presentation in this section on the text book by~\cite{Karniadakis2005_hpbook}, and customise the formulation for our particular problem, in a purpose to explain the basics numerical notations in detail. Take again the example of the 1D scalar advection equation defined in domain $x\in\Omega$ with a constant advection speed $u=1$, noting that the diffusion term is not discussed here since it is not relevant to the formulation of GJP although it warrants further discussion if one intends to study the CG-SEM itself in detail:
\begin{equation}\label{eq:1d_adv}
    \frac{\partial T(x,t)}{\partial t} + \frac{\partial T(x,t)}{\partial x} = 0.
\end{equation}
To solve the above equation, we first formulate the weak form by multiplying the \textbf{test function} $v(x)$
\begin{equation}\label{eq:weak_1d_adv}
    \int_{\Omega}v(x)(\frac{\partial T(x,t)}{\partial t} + \frac{\partial T(x,t)}{\partial x})dx = 0.
\end{equation}
We could further subdivide the domain into \textbf{elements} and regulate our test function $v(x)$ to be non-zero only in a particular element $\Omega_e$. By mapping the subdomain into a standard computational domain $\hat{\Omega}_e$, we establish
\begin{equation}\label{eq:weak_1d_adv_elem}
    \int_{\Omega_e}v(x)(\frac{\partial T(x,t)}{\partial t} + \frac{\partial T(x,t)}{\partial x})dx = \int_{\hat{\Omega}_e}v(\xi)(\frac{\partial T(\xi,t)}{\partial t} + \frac{\partial T(\xi,t)}{\partial \xi}) J^{-1} d\xi = 0,
\end{equation}
where $J$ is called Jacobian which is defined as $J(x) := \frac{d\xi}{dx}$. If the solution in the subdomain $T(\xi,t)$ is further decomposed into the weighed summation of $P+1$ polynomials ($P$ is called the \textbf{polynomial order}) to decouple the temporal and spatial dimension
\begin{equation}\label{eq:T_decomp_1d}
    T(\xi,t) = \sum_{i=0}^{P} \hat{T}_i(t)\phi_i(\xi),
\end{equation}
and define the test function $v(\xi)$ to be the same as the basis function of $l$-th order $\phi_l(\xi)$, which is the essence of the Galerkin method, Equation (\ref{eq:weak_1d_adv_elem}) becomes
\begin{equation}\label{eq:weak_1d_adv_sem}
    \sum_{i=0}^{P} \left\{\left(\int_{\hat{\Omega}_e} \phi_l(\xi)\phi_i(\xi) d\xi \right)J^{-1} \frac{d\hat{T}_i}{dt} \right\}+ \sum_{i=0}^{P} \left\{\left(\int_{\hat{\Omega}_e} \phi_l(\xi)\frac{d \phi_i(\xi)}{d \xi}d\xi\right) J^{-1}\hat{T}_i\right\} = 0.
\end{equation}
For each $l$, the assembled solution of equation (\ref{eq:weak_1d_adv_sem}) from all elements is a necessary condition to satisfy equation (\ref{eq:1d_adv}) hence is considered to be the numerical solution of equation (\ref{eq:1d_adv}) if the solution is unique. The parenthesis $(\cdot)$ includes operations inside the reference computational domain while the inverse of the Jacobian $J^{-1}$ maps them to the physical domain. Up to this point, there are still two more undetermined degrees of freedom per element at the starting and ending point of the element $\Omega_e$. In CG-SEM, we guarantee the $C^0$ continuity of the variable $T(x,t)$ itself across elements, which solves the closure problem of the system.

For better computational efficiency and numerical properties, the polynomial $\phi_i$ could be chosen to satisfy certain properties. Nevertheless, we stop here with the 1D formulation since those properties are not of the key interest of this study. 

\section{Gradient jump penalty: from 1D to 3D}\label{appendix:B}
In this section, we discuss how we come up with Equation (\ref{eq:GJP_RHS}) from 1D to 3D. The 1D equation to be solved, with the simplification of unity velocity components, is Equation (\ref{eq:GJP_RHS_1d}) as repeated here combining the reference frame mapping in Equation (\ref{eq:weak_1d_adv_elem})
\begin{align}\label{eq:GJP_RHS_1d_appendix}
    \int_{\hat{\Omega}_e}v(\xi)(\frac{\partial T(\xi,t)}{\partial t} + \frac{\partial T(\xi,t)}{\partial \xi})J^{-1} d\xi  &= -\tau \langle h_{\Omega_e}^2 G(T) \frac{\partial v(\xi)}{\partial n}\rangle.\\
    \langle \rangle &:= \int_{\partial \Omega_e \backslash \partial \Omega} ds.
\end{align}
In the 1D scenario, the integration $\int_{\partial \Omega_e \backslash \partial \Omega} ds$ is simply the value on the element boundary points. In the 3D scenario, accordingly, it becomes the surface integral over the element boundaries. Another obvious relevant difference is the computational domain changing from $\xi\in[-1,1]$ into $(\xi_1,\xi_2,\xi_3)\in[-1,1]^3$, hence the test function $v$ changing from $v(\xi)=\phi_i(\xi)$ to $v(\xi_1, \xi_2, \xi_3)=\phi_i(\xi_1)\phi_j(\xi_2)\phi_k(\xi_3)$. As defined in Equation (\ref{eq:GJP_RHS_1d}), $n$ is the outward-pointing normal direction of a facet. Take $n$ aligning with $\xi_3$ for example (hence $\frac{\partial \xi_1}{\partial n} = \frac{\partial \xi_2}{\partial n} = 0$), the term $\frac{\partial f}{\partial n}$ writes
\begin{align}\label{eq:dfdn_3d_appendix}
    \frac{\partial f}{\partial n}|_{n\to\xi_3^+} &= \frac{\partial v(\xi_1, \xi_2, \xi_3)}{\partial\xi_3}\frac{\partial\xi_3}{\partial n} \nonumber\\
    &= \frac{\partial[\phi_i(\xi_1)\phi_j(\xi_2)\phi_k(\xi_3)]}{\partial\xi_3}\frac{\partial\xi_3}{\partial n} \nonumber\\
    &= \phi_i(\xi_1)\phi_j(\xi_2)\frac{\partial \phi_k(\xi_3)}{\partial\xi_3}\frac{\partial\xi_3}{\partial n} \nonumber\\
    &= \phi_i(\xi_1)\phi_j(\xi_2)\frac{\partial \phi_k(\xi_3)}{\partial n}.
\end{align}
Note that $\xi_1$ and $\xi_2$ are the tangential direction on the facet in this example in the reference frame $(\xi_1,\xi_2,\xi_3)$ and $\xi_3$ is the normal direction. Therefore, Equation (\ref{eq:dfdn_3d_appendix}) shows the multiplication of the 2 polynomials on the tangential direction and the derivative of the rest 1 polynomial on the normal direction. Similar operations could be done for other facets, and thus this fully explains how we reach $\phi_{t_1, i} \phi_{t_2, j} \frac{\partial \phi_{n, k}}{\partial n}$ in Equation (\ref{eq:GJP_RHS}).

\bibliography{ref}

@article{BREUER2009433,
	author = {M. Breuer and N. Peller and Ch. Rapp and M. Manhart},
	doi = {10.1016/j.compfluid.2008.05.002},
	issn = {0045-7930},
	journal = {Computers \& Fluids},
	number = {2},
	pages = {433–457},
	title = {Flow over periodic hills – Numerical and experimental study in a wide range of Reynolds numbers},
	url = {https://www.sciencedirect.com/science/article/pii/S0045793008001126},
	volume = {38},
	year = {2009}
}

@article{Chatterjee2017,
    author = {Chatterjee, Tanmoy and Peet, Yulia T.},
    title = {Effect of artificial length scales in large eddy simulation of a neutral atmospheric boundary layer flow: A simple solution to log-layer mismatch},
    doi = {10.1063/1.4994603},
    journal = {Physics of Fluids},
    volume = {29},
    number = {7},
    pages = {075105},
    year = {2017},
    month = {07},
    issn = {1070-6631}
}

@article{Fischer2001,
title = {Filter-based stabilization of spectral element methods},
journal = {Comptes Rendus de l'Académie des Sciences - Series I - Mathematics},
volume = {332},
number = {3},
pages = {265-270},
year = {2001},
issn = {0764-4442},
doi = {https://doi.org/10.1016/S0764-4442(00)01763-8},
url = {https://www.sciencedirect.com/science/article/pii/S0764444200017638},
author = {Paul Fischer and Julia Mullen}
}

@article{Gillyns2022,
    author = {Gillyns, Emmanuel and Buckingham, Sophia and Winckelmans, Grégoire},
    title = {Implementation and Validation of an Algebraic Wall Model for LES in Nek5000},
    journal = {Flow, Turbulence and Combustion},
    year = {2022},
    volume = {109},
    pages = {1111-1131},
    doi = {10.1007/s10494-022-00378-y}
}

@article{Gloerfelt2019,
	risfield_0_da = {2019/06/01},
	author = {Gloerfelt, Xavier and Cinnella, Paola},
	doi = {10.1007/s10494-018-0005-5},
	issn = {1573-1987},
	journal = {Flow, Turbulence and Combustion},
	number = {1},
	pages = {55–91},
	title = {Large Eddy Simulation Requirements for the Flow over Periodic Hills},
	volume = {103},
	year = {2019}
}

@article{Hoyas2006,
    author = {Hoyas, Sergio and Jiménez, Javier},
    title = {Scaling of the velocity fluctuations in turbulent channels up to ${R}e_\tau=2003$},
    journal = {Physics of Fluids},
    volume = {18},
    number = {1},
    pages = {011702},
    year = {2006},
    month = {01},
    doi = {10.1063/1.2162185}
}

@article{Huang_2022,
    author = {Huang, Bohua and Wang, Rui and Wu, Feng and Xu, Hui},
    title = {Applications of wall-models to implicit large eddy simulations in the spectral/hp element method},
    journal = {Physics of Fluids},
    volume = {34},
    number = {10},
    pages = {105129},
    year = {2022},
    month = {10},
    doi = {10.1063/5.0117432},
    issn = {1070-6631}
}

@article{Huusko2025,
	author = {Huusko, L. and Mukha, T. and Donati, L. L. and Sullivan, P. P. and Schlatter, P. and Svensson, G.},
	doi = {10.1029/2025MS005233},
	journal = {Journal of Advances in Modeling Earth Systems},
	number = {10},
	pages = {e2025MS005233},
	title = {Large Eddy Simulation of Canonical Atmospheric Boundary Layer Flows With the Spectral Element Method in Nek5000},
	volume = {17},
	year = {2025}
}

@article{Kirby2006,
title = {Stabilisation of spectral/hp element methods through spectral vanishing viscosity: Application to fluid mechanics modelling},
journal = {Computer Methods in Applied Mechanics and Engineering},
volume = {195},
number = {23},
pages = {3128-3144},
year = {2006},
issn = {0045-7825},
author = {Robert M. Kirby and Spencer J. Sherwin},
doi = {10.1016/j.cma.2004.09.019}
}

@article{Kou2023,
    title = {Jump penalty stabilization techniques for under-resolved turbulence in discontinuous {G}alerkin schemes},
    journal = {Journal of Computational Physics},
    volume = {491},
    pages = {112399},
    year = {2023},
    issn = {0021-9991},
    author = {Jiaqing Kou and Oscar A. Marino and Esteban Ferrer},
    doi = {10.1016/j.jcp.2023.112399}
}

@article{Lilly1966,
    title={The representation of small-scale turbulence in numerical simulation experiments},
    author={Lilly, D. K.},
    year={1966},
    doi={10.5065/D62R3PMM}
}

@article{Manzanero2020,
	author = {Juan Manzanero and Esteban Ferrer and Gonzalo Rubio and Eusebio Valero},
	issn = {0045-7930},
	journal = {Computers \& Fluids},
	pages = {104440},
	title = {Design of a Smagorinsky spectral Vanishing Viscosity turbulence model for discontinuous Galerkin methods},
	volume = {200},
	year = {2020},
        doi = {10.1016/j.compfluid.2020.104440}
}

@InProceedings{Massaro2023,
author="Massaro, Daniele
and Peplinski, Adam
and Schlatter, Philipp",
editor="Melenk, Jens M.
and Perugia, Ilaria
and Sch{\"o}berl, Joachim
and Schwab, Christoph",
title="Interface Discontinuities in Spectral-Element Simulations with Adaptive Mesh Refinement",
booktitle="Spectral and High Order Methods for Partial Differential Equations ICOSAHOM 2020+1",
year="2023",
publisher="Springer International Publishing",
address="Cham",
pages="375--386",
isbn="978-3-031-20432-6"
}

@article{Moura2022GJP,
	author = {Rodrigo C. Moura and Andrea Cassinelli and André F.C. {da Silva} and Erik Burman and Spencer J. Sherwin},
	issn = {0045-7825},
	journal = {Computer Methods in Applied Mechanics and Engineering},
	pages = {114200},
	title = {Gradient jump penalty stabilisation of spectral/hp element discretisation for under-resolved turbulence simulations},
	volume = {388},
	year = {2022},
        doi = {10.1016/j.cma.2021.114200}
}

@misc{Mukha2024,
      title={Wall-modeled large-eddy simulation based on spectral-element discretization}, 
      author={Timofey Mukha and Philipp Schlatter},
      year={2024},
      eprint={2404.05378},
      archivePrefix={arXiv},
      doi = {10.48550/arXiv.2404.05378},
      primaryClass={physics.flu-dyn}
}

@article{Neko,
	author = {Niclas Jansson and Martin Karp and Artur Podobas and Stefano Markidis and Philipp Schlatter},
	issn = {0045-7930},
	journal = {Computers \& Fluids},
	pages = {106243},
	title = {Neko: A modern, portable, and scalable framework for high-fidelity computational fluid dynamics},
        doi = {10.1016/j.compfluid.2024.106243},
	volume = {275},
	year = {2024}
}

@article{Nicoud2011,
    author = {Nicoud, Franck and Toda, Hubert Baya and Cabrit, Olivier and Bose, Sanjeeb and Lee, Jungil},
    title = {Using singular values to build a subgrid-scale model for large eddy simulations},
    journal = {Physics of Fluids},
    doi = {10.1063/1.3623274},
    volume = {23},
    number = {8},
    pages = {085106},
    year = {2011},
    month = {08},
    issn = {1070-6631}
}

@article{Ntoukas2025,
	author = {Gerasimos Ntoukas and Gonzalo Rubio and Oscar Marino and Alexandra Liosi and Francesco Bottone and Julien Hoessler and Esteban Ferrer},
	issn = {2590-1230},
	journal = {Results in Engineering},
	pages = {104425},
	title = {A comparative study of explicit and implicit Large Eddy simulations using a high-order discontinuous Galerkin solver: Application to a Formula 1 front wing},
        doi = {10.1016/j.rineng.2025.104425},
	volume = {25},
	year = {2025}
}

@misc{PySEMTools,
      title={PySEMTools: A library for post-processing hexahedral spectral element data}, 
      author={Adalberto Perez and Siavash Toosi and Tim Felle Olsen and Stefano Markidis and Philipp Schlatter},
      year={2025},
      eprint={2504.12301},
      archivePrefix={arXiv},
      primaryClass={physics.comp-ph},
      doi = {10.48550/arXiv.2504.12301}
}

@article{Schlatter2004,
	author = {Philipp Schlatter and Steffen Stolz and Leonhard Kleiser},
	doi = {10.1016/j.ijheatfluidflow.2004.02.020},
	issn = {0142-727X},
	journal = {International Journal of Heat and Fluid Flow},
	note = {Turbulence and Shear Flow Phenomena (TSFP-3)},
	number = {3},
	pages = {549–558},
	title = {LES of transitional flows using the approximate deconvolution model},
	url = {https://www.sciencedirect.com/science/article/pii/S0142727X04000281},
	volume = {25},
	year = {2004}
}

@phdthesis{Schlatter2005_thesis,
	address = {Z\"urich},
	author = {Schlatter, Philipp Christian},
	copyright = {In Copyright - Non-Commercial Use Permitted},
	language = {en},
	publisher = {ETH Z\"urich},
	school = {ETH Zurich},
	size = {1 v.},
	title = {Large-eddy simulation of transition and turbulence in wall-bounded shear flow},
	type = {Doctoral Thesis},
        doi = {10.3929/ethz-a-004945776},
	year = {2005}
}

@article{Smagorinsky1963,
	address = "Boston MA, USA",
	author = "J. Smagorinsky",
	doi = "10.1175/1520-0493(1963)0912.3.CO;2",
	journal = "Monthly Weather Review",
	number = "3",
	pages = "99–164",
	publisher = "American Meteorological Society",
	title = "GENERAL CIRCULATION EXPERIMENTS WITH THE PRIMITIVE EQUATIONS: I. THE BASIC EXPERIMENT",
	url = "https://journals.ametsoc.org/view/journals/mwre/91/3/1520-0493_1963_091_0099_gcewtp_2_3_co_2.xml",
	volume = "91",
	year = "1963"
}

@article{Vreman2004,
    author = {Vreman, A. W.},
    title = {An eddy-viscosity subgrid-scale model for turbulent shear flow: Algebraic theory and applications},
    doi = {10.1063/1.1785131},
    journal = {Physics of Fluids},
    volume = {16},
    number = {10},
    pages = {3670-3681},
    year = {2004},
    month = {10},
    issn = {1070-6631}
}

@article{Wang2021,
    author = {Wang, Rui and Wu, Feng and Xu, Hui and Sherwin, Spencer J.},
    title = {Implicit large-eddy simulations of turbulent flow in a channel via spectral/hp element methods},
    doi = {10.1063/5.0040845},
    journal = {Physics of Fluids},
    volume = {33},
    number = {3},
    pages = {035130},
    year = {2021},
    month = {03},
    issn = {1070-6631}
}

@article{Yilmaz2015,
    author = {Yilmaz, Ilyas and Davidson, Lars},
    title = {Comparison of SGS models in Large-Eddy Simulation for transition to turbulence in Taylor–Green flow},
    journal = {In The 16th International Conference on Fluid Flow Technologies CMFF 2015, Budapest, Hungary},
    year = {2015}
}

@misc{Rapp2009Phill,
    author = {Rapp, Christoph and Breuer, Michael and Manhart, Michael and Peller, Nikolaus},
    title = {Abstr:2D Periodic Hill Flow},
    url  = {https://www.kbwiki.ercoftac.org/w/index.php/Abstr:2D_Periodic_Hill_Flow},
    addendum = {accessed: 01.06.2025}
}

@book{Karniadakis2005_hpbook,
author = {Karniadakis, George. and Sherwin, Spencer J.},
address = {New York},
booktitle = {Spectral/hp element methods for computational fluid dynamics},
edition = {2nd ed.},
isbn = {1-4294-2205-X},
language = {eng},
publisher = {Oxford University Press},
series = {Numerical mathematics and scientific computation},
title = {Spectral/hp element methods for computational fluid dynamics },
year = {2005},
}

@article{Boussinesq_1877,
    author = {Boussinesq, Joseph},
    title = {Essai sur la théorie des eaux courantes},
    journal = {Mémoires présentés par divers savants à l'Académie des Sciences},
    year = {1977},
    volume = {23},
    number = {1},
    page = {1-680}
}

@inbook{Fröhlich_Rodi_2002,
    address = {Cambridge},
    title = {Introduction to Large Eddy Simulation of Turbulent Flows},
    booktitle = {Closure Strategies for Turbulent and Transitional Flows},
    publisher = {Cambridge University Press},
    author = {Fr{\"o}hlich, J. and Rodi, W.},
    year = {2002}, 
    pages = {267–298}
}

\end{document}